\pdfoutput=1
\documentclass[11pt]{article}

\usepackage{amsmath,amsthm,amssymb}
\usepackage{mathtools}
\usepackage{mathrsfs}
\usepackage{graphicx}
\usepackage{float}
\usepackage{lmodern,microtype}
\usepackage{enumitem}
\usepackage[margin=1in]{geometry}
\usepackage[round]{natbib}
\usepackage{setspace}
\usepackage{booktabs,tabularx}
\usepackage{longtable}
\usepackage{tikz}
\usetikzlibrary{positioning}
\usepackage[hidelinks]{hyperref}
\usepackage{xurl}  
\definecolor{codegray}{gray}{0.45}  

\usepackage{euler}

\newcommand{\Ahat}{\widehat{\mathbf{A}}}
\newcommand{\Aaug}{\widetilde{\mathbf{A}}}

\newcommand{\Fset}[1]{\mathcal{F}_{#1}}
\newcommand{\onestep}{\widehat{\mathbf m}}

\theoremstyle{plain}
\newtheorem{proposition}{Proposition}
\newtheorem{lemma}{Lemma}

\newtheorem{definition}{Definition}

\newtheoremstyle{boldremark}%
  {0.5\baselineskip}
  {0.5\baselineskip}
  {\itshape}
  {}
  {\bfseries}
  {.}
  { }
  {}
\theoremstyle{boldremark}

\graphicspath{ {./Images/} }

\title{\Large{\textbf{Reconstructing Large-Scale Production Networks}}}

\author{Ashwin Bhattathiripad\thanks{Economics Area, Indian Institute of Management Kozhikode, Kerala 673 570, India.} \and  Vipin P. Veetil\protect\footnotemark[1]}

\date{\small\today}
\begin{document}
{\renewcommand{\thefootnote}{\fnsymbol{footnote}}\maketitle}
\setcounter{footnote}{0}
\begin{abstract}
Firm-to-firm production networks matter for aggregate propagation, but they are rarely observed.
This paper reconstructs national-scale, weighted firm-to-firm networks from two public objects:
a sectoral input--output table and the distribution of firm sizes by sector. The algorithm first
draws a binary buyer-seller backbone from a sector-aware gravity model and then assigns weights
by a minimum-energy program. A Markov closure makes the reconstructed network primitive, so it has a unique stationary distribution. The weighting program keeps one-step firm balances and sectoral flows close to the data; the stationary money vector is then checked ex post and remains close in aggregate. For the United States we reconstruct a network with about
6.5 million firms and 340 million links in roughly four hours on a single workstation. We
also reconstruct the networks of Japan, the United Kingdom, Australia, Finland, and Denmark. The Japanese
reconstruction, built without any link data, reproduces the heavy-tailed degree regime
documented in the country's observed production network. The reconstructed networks exhibit customer tails heavier than supplier tails, though the algorithm treats the two sides symmetrically. We also run computational experiments on the reconstructed networks to assess the systemic risk posed by the failure of individual firms. These experiments show that neither firm size nor degree nor sectoral position is a good proxy for the aggregate losses generated by a firm's failure. For such questions, there is no good substitute for the complete weighted buyer-seller network that we reconstruct.  We release the generated networks and the reconstruction code as a Python library.
\end{abstract}
\vspace{0.2cm}

\newpage
\setstretch{1.25}
\section{Introduction}

Which firms buy from which firms? For most of the world economy, no one knows. What we do know
is that this network of buyer-seller relations between firms shapes economic
phenomena ranging from how granular shocks generate business cycles to how an earthquake in
one part of the world affects production in another.\footnote{See
\citet{AcemogluEtAl2012_Econometrica,CarvalhoNireiSaitoTahbazSalehi2021,JRC2023FloodStress,PalepuClarkFrancisGarcia2025_Cascading,PelosiMandel2025}.} 
We cannot know how a shock at one firm, plant, or region becomes an aggregate event without
tracing the chain of suppliers and customers through which it travels. Natural disasters, pandemics,
climate events, and monetary disturbances all pose this problem.  It is here that we hit a wall. Firm-to-firm data
do not exist for most countries. Where they exist, they are not public. And where they are observed by private vendors, they are often censored, with a few major partners standing in for the full spectrum of a firm's buyer-seller
relations.\footnote{Near-universe firm-to-firm production-network data exist for Belgium
\citep{DhyneKikkawaMogstadTintelnot2021_REStud}, Hungary \citep{DiemEtAl2022_SciRep},
Ecuador \citep{BacilieriEtAl2023_INET}, Chile \citep{HuneeusKroftLim2021_NBER}, Japan
\citep{MizunoSoumaWatanabe2014_PLOSONE}, and the Dominican Republic
\citep{CardozaGrigoliPierriRuane2024_REStud}. Large but sub-national coverage exists for India
\citep{Panigrahi2022_JMP}. Though these datasets exist, they are not publicly available. So most
researchers cannot use them to estimate the empirical consequences of their theoretical insights.} Simply put, one of the most significant data objects we need for economic analysis is not available.

The production network, however, casts two public shadows. Many countries periodically publish
input--output tables, which are the firm buyer-seller network summed into sectors. Statistical
agencies also publish the distribution of firm sizes by sector, which is the network's size margin read
off one node at a time. This paper asks how much of the missing firm-to-firm network can be
recovered from these two shadows alone. We develop an algorithm that reconstructs a national-scale production network from these two objects. The reconstruction proceeds in
two stages. The first stage generates a binary, unweighted backbone that determines which firms
transact with one another. We model the probability that two firms are connected with an augmented
gravity model. The probability that a buyer links to a seller rises with the sizes of the two firms and with the flow from the buyer's sector to the seller's sector in the input--output table, and a single density scalar sets the number of links per firm. We estimate the sector-pair parameters so that the share of links the model places in each pair of sectors matches that pair's share of the flows recorded in the input--output table, subject to a band on the expected mean degree and a cap on the average error over the worst-fitting sector pairs.  The aim is not simply to generate a graph with the
right average density. It is to generate a graph whose sectoral pattern of links reflects the sectoral
pattern of production.

The reconstructed backbone is ultimately read as a Markov chain of money flows. It must therefore be irreducible and aperiodic. We first join the strongly connected pieces of the drawn graph with the smallest number of component-level connections. Each pair of pieces receives a number of ordinary firm-level links that grows with the size of the smaller piece, placed where they least disturb the sectoral pattern of the graph. If the resulting graph is periodic, we add one link between two firms that breaks the period, again chosen to disturb the sectoral pattern least. Because the degree floor gives every firm at least two suppliers, some cyclic class of the closed backbone always contains two firms, so an admissible edge between distinct firms always exists. 

The second stage turns the binary backbone into a weighted network. Because the procedure uses
no information on the intensity of individual buyer-seller relationships, we assign weights by a
minimum-energy criterion. Among all weightings consistent with the data, we select the one that
imposes the least additional structure. Concretely, we minimize the sum of squared edge weights, which spreads each firm's purchases as evenly across its suppliers as the caps on firm balances and sectoral totals allow. For small buyers the caps are slack and the weights are close to uniform. For large buyers the caps decide the weights, and their spending concentrates on large sellers. The result is the most uniform, or equivalently least informative, set of weights that remains
consistent with the firms' sizes and the sectoral flows.\footnote{Minimum energy is itself a maximum-entropy principle. Minimizing the sum of squared weights subject to linear constraints maximizes the quadratic entropy, which is one minus that sum \citep{Tsallis1988}. And for a firm with a given number of links, the chi-square divergence of its spending allocation from the uniform allocation across those links equals the number of links times its sum of squared weights, minus one. The objective is therefore the sum over firms of the chi-square divergence from the uniform allocation, the zero-structure benchmark, weighted by the inverse of the firm's link count.}

The algorithm is designed to scale to a large economy. Estimating the gravity parameters is inexpensive because it is done on representative firms for narrow sector-size bins. Under the sparse-sampling conditions stated in Appendix~\ref{sec:compu}, the degree-floor repair, the Markov closure, and each iteration of the weighting step run in time proportional to the number of firms and realized links. The Bernoulli draw remains the genuinely quadratic stage, since it makes one decision for every ordered pair of firms, but the decisions are independent and are run in parallel on the GPU. We reconstruct the complete production network of the United States,
with 6.46 million firms and 340 million weighted edges, using BEA sectoral flows and
SUSB firm-size-by-sector data. The reconstruction takes roughly four hours on a Mac Studio (M1)
with 128 GB of RAM. We also reconstruct the production networks of Japan, the United Kingdom,
Australia, Finland, and Denmark. The largest of these is Japan, with about 3.5 million firms, and
the smallest is Denmark, with an order of magnitude fewer firms. These are, to our knowledge, the
largest production networks reconstructed so far.

The natural test of a reconstruction built from aggregates alone is whether it reproduces the
heavy-tailed degree signature of a real-world economy for which granular data are available. Japan is the one large economy whose firm-to-firm production network has been mapped nearly in full and studied at length. We reconstruct Japan's network from publicly available information on
sectoral flows and firm-size distributions by sector. With no observed links as input, the
reconstruction lands in the same degree regime as the real network. The customer-count distribution
is heavy-tailed, with a tail index such that the mean is finite but the variance diverges. Furthermore,
as in the observed network, the customer tail is heavier than the supplier tail. The agreement is also quantitative. We measure the customer tail, the distribution of the number of customers per firm, as the empirical literature measures it, by a counter-cumulative Hill index. So measured, the reconstruction gives a tail exponent between 1.2 and 1.5 at the top 10 to 20 percent of firms, against an observed exponent of about 1.3
\citep{fujiwara2010large,CarvalhoNireiSaitoTahbazSalehi2021}. Shown only a marginal size
distribution and a sector-level input--output table, the algorithm thus recovers the degree
tails of a network it never observed.

Beyond this out-of-sample test against Japan's observed network, the reconstruction is also internally consistent. Summed back to the sector level, the reconstructed firm-to-firm flows remain close to the input--output target. The one-step firm-balance residuals are small by construction. Running each fitted money chain to stationarity relocates only a few percent of total money in all six economies. For the full US network, the detailed firm-level check also gives a median stationary-to-census ratio of about 1.04. Thus the stationary closeness is an empirical self-consistency result, not a direct consequence of the weighting caps.

The reconstructed networks also show what a thresholded dataset loses. In every economy, links below one percent of a buyer's spending carry between 88 and 99 percent of second-order reach, the customers of customers, and more than 96 percent of third-order reach. A network that records only major suppliers therefore keeps the first neighbourhood of each firm and loses the paths through which shocks travel beyond it.

We illustrate the usefulness of the reconstructed networks through a canonical propagation experiment, the systemic risk posed by the failure of individual firms.  Across all six economies the distribution of systemic risk is heavy-tailed. The failure of most firms costs the economy little, while the failure of a few firms costs it a great deal. Our experiments show that what matters for systemic risk is a firm's position in the network, not its size alone. They also show that a network with fewer firms, or one that keeps only the major connections, overstates systemic risk, by an order of magnitude or more in the tail. For questions about the propagation of granular shocks, there may be no good substitute for the full weighted production networks we reconstruct.

\subsection{Code, Library, and GUI}
\label{sec:code}

The reconstruction toolkit is released as an open-source Python package, \texttt{prodnet-generator}, distributed under the MIT license. The source is at \href{https://bitbucket.org/VipinVeetil/prodnet-generator}{\textcolor{codegray}{\nolinkurl{https://bitbucket.org/VipinVeetil/prodnet-generator}}}. A built distribution is on PyPI at \href{https://pypi.org/project/prodnet-generator/}{\textcolor{codegray}{\nolinkurl{https://pypi.org/project/prodnet-generator/}}}. A single \texttt{pip install prodnet-generator} pulls the package and its dependencies and makes it importable at once, with no manual build or configuration. The aim is not merely to release code but to make the reconstructed economies operational. From a national firm-size distribution and an input--output table alone, the package builds the binary backbone, imposes the degree floor and Markov closure, and solves the minimum-energy weighting, returning a firm-level weighted production network ready to use. On that network the user can threshold links, knock out firms, and run the systemic-risk cascades and other counterfactuals reported in this paper. The user thus recreates, modifies, and simulates forward the very objects behind the empirical results without re-implementing any of the algorithm.\footnote{The implementation uses the available hardware automatically. The independent pairwise link decisions of the Bernoulli draw are sent to the GPU when a compatible device is present, while the sparser downstream stages (graph repair, closure, weighting, and counterfactual simulation) run on the CPU with sparse data structures.} Reconstructions are fully deterministic given a random seed, so every network and experiment reported here is exactly reproducible. A graphical interface exposes the main controls: economy, network scale, mean degree, degree-tail calibration, link threshold, and shock specification. These controls turn the reconstruction into a small laboratory for production networks.

\subsection{Relation to Existing Reconstruction Methods}
\label{sec:literature}

The problem of reconstructing a production network arises because the object needed for many questions, the complete weighted firm-to-firm graph, is rarely observed. Existing reconstruction methods can be classified according to which
 objects they take as primitive inputs.  One class of methods begins with observed firm-to-firm links. The observed network is typically
partial and unweighted. It is interpreted as a censored view of the true production network, and the
task is to infer the missing part. There are several ways to do this. One approach estimates the
observable correlates of a link, such as firm size, sector, geography, or other firm attributes. The
fitted model is then used to predict additional links, which are added to the observed ones. Another
approach combines the observed links with aggregate accounting restrictions. Since the sum of
firm-to-firm flows must reproduce the sectoral input--output table, the missing links are chosen so that the completed network aggregates back to the observed sectoral flows \citep{HooijmaaijersBuiten2019,BuitenEtAl2021CBS,WelburnEtAl2023_GlobalFirmIO}.\footnote{The missing part may be filled in by a deterministic algorithm, returning a single completed graph, or by a maximum-entropy procedure describing an ensemble consistent with the observed constraints \citep{Squartini2017}. Weights are then assigned by an optimization that brings firm or sectoral flows close to their observed values \citep{Bacilieri2023}.}

A second class of methods does not require observed firm-to-firm links. Instead, they use
information such as firm sizes, sector labels, and
input--output flows to generate a network from scratch \citep{Ialongo2022,DiVece2023}. Some versions also use
firm-level degrees or strengths \citep{Mastrandrea2014,DevetakMandel2026}. The principal strength of this
class of methods is that they can be applied to economies for which no buyer-seller network has been
mapped. The problem is that the model must infer all firm-to-firm links from aggregate information and
firm characteristics alone.

Our reconstruction belongs to the second class. In this respect, our closest starting point is the
``stripes-corrected'' gravity model of \citet{Ialongo2022}, which rules out links between sector pairs
with no input--output flow. We generalize that idea by estimating a multiplier for
each sector pair, so that the model-implied distribution of links across sector blocks matches the
flow shares in the input--output table. And with this generalization, we separate the role of sectoral flows from the role
of firm size. The sectoral multipliers determine where links go across sectors. The firm-size fitness
determines how links are distributed across firms within those sectors.  This separation matters because the main difficulty is not merely to reproduce aggregate flows.
That can be done in many ways. The harder task is to reproduce aggregate flows while also
generating the heavy-tailed degree distributions observed in production networks. In most
maximum-entropy and fitness-based reconstructions, a single object, the firm's observed strength, carries both constraints. That strength is used to reproduce firm or sectoral flows, and the induced degree
distribution is left as a by-product. The result is often a degree tail that is too light
\citep{Bacilieri2023,DevetakMandel2026}. The model fits the accounting constraints, but it does not
generate the granular network structure that is central for propagation. Our construction assigns these tasks to different parameters. The sectoral flow pattern is carried by the block multipliers, one for each pair of sectors, estimated so that the share of links in each sector block tracks its share of input--output flows. The degree-tail shape is carried by a size elasticity parameter in the fitness map, which the flow shares do not identify. Changing that parameter merely redistributes link probability among firms within a block without altering the target block shares, so it can be disciplined separately to reproduce the observed tail regime.\footnote{A heavy-tailed firm-size distribution, passed through the fitness map, generates a heavy-tailed expected-degree distribution, and the sectoral multipliers place those links in the correct parts of the input--output table.} Fat tails and sectoral flows are together matched because the model does not ask a single parameter to explain both.

Our algorithm also differs from previous reconstruction efforts in its ability to scale.  Previous reconstructions have generally operated on networks with
roughly ten thousand to a hundred thousand firms. We reconstruct the full US firm sector, about
6.46 million firms, and repeat the exercise for five additional economies. Scale matters because many network quantities are scale-dependent. A small graph places too much economic mass on too few nodes, and a thresholded graph removes weak links that may carry higher-order reach. A sectoral graph removes the identity of the firm altogether, which is the object on which the motivating counterfactuals depend. Table~\ref{tab:add_links_final_compact} compares existing methods not only by their
information requirements, but also by the scale at which they operate and the objects they are able
to preserve. From an economic point of view, the relevant question is not only whether a method can complete a network. It is
whether it can produce a full-scale weighted graph on which firm-level propagation can be measured.

\begin{table}[H]
\centering
\caption{Papers that reconstruct production networks}
\label{tab:add_links_final_compact}
\setstretch{1.0}\footnotesize
\setlength{\tabcolsep}{4pt}
\renewcommand{\arraystretch}{1.25}
\begin{tabularx}{\textwidth}{@{}
  >{\raggedright\arraybackslash\hsize=.80\hsize}X
  r r r r
  >{\raggedright\arraybackslash\hsize=1.20\hsize}X @{}}
\toprule
\textbf{Paper} & \textbf{$N$} & \textbf{$E_{\text{obs}}$} & \textbf{$\Delta E$} & \textbf{$E_{\text{tot}}$} & \textbf{Region / Notes} \\
\midrule
\citet{WelburnEtAl2023_GlobalFirmIO} & 24{,}879 & 137{,}811 & 218{,}259 & 356{,}070 & Global (FactSet); ensemble thresholding \\
\citet{ReischEtAl2022_SciRep} & 89{,}000+ & - & 235{,}000 & 235{,}000 & Hungary (VAT benchmark); reconstruction from mobile-phone data \\
\citet{Mungo2023}, Compustat & 915 & NR & NR & NR & U.S. (public firms); link-prediction study \\
\citet{Mungo2023}, FactSet & 6{,}714 & NR & NR & NR & Global (listed); link-prediction study \\
\citet{Mungo2023}, Ecuador & 10{,}000 & NR & NR & NR & Ecuador (VAT); link-prediction study \\
\citet{Ialongo2022} & NR & NR & NR & NR & Netherlands (two bank datasets); generalized MaxEnt \\
\citet{BuitenEtAl2021CBS} & NR & NR & NR & NR & Netherlands (CBS method); deterministic pipeline \\
\citet{Mastrandrea2014} & - & - & - & NR & ECM (MaxEnt with degrees and strengths); no fixed support \\
\citet{DevetakMandel2026} & $2.0\times10^5$ & - & $1.8\times10^6$ & $1.8\times10^6$ & Hungary (production network); MaxEnt with input--output constraints \\
\citet{DiVece2023} & - & - & - & NR & Integrated and conditional MaxEnt (ITN); no fixed support \\
\citet{IalongoBangma2024} & $10^5$ & NR & $10^6$ & $10^6$ & Netherlands (multi-scale embeddings); method paper \\
This paper & $6.46\times10^6$ & 0 & $3.40\times10^8$ & $3.40\times10^8$ & United States \\
\bottomrule
\end{tabularx}

\vspace{0.5ex}
{\setstretch{1.0}\raggedright\footnotesize
Notes: $E_{\text{obs}}$ = observed links in the starting (partial) network. $\Delta E$ = links added. $E_{\text{tot}}=E_{\text{obs}}+\Delta E$.
`-' = not applicable (from-scratch methods don't start from a fixed $E_{\text{obs}}$). NR = not reported publicly in the cited sources.\par}
\end{table}

\subsection{Organization of the Paper}

The rest of the paper is organized as follows. Section~\ref{sec:model} describes the reconstruction
algorithm. It specifies the logistic-gravity model for link probabilities, the estimation of sector-pair
multipliers, the Bernoulli draw of the binary backbone, the Markov closure that makes the graph
irreducible and aperiodic, and the minimum-energy program that assigns weights.

Section~\ref{sec:US} applies the algorithm to the United States. It reconstructs the full US firm
sector, compares the result with the largest available sample of the observed US firm-to-firm
network, and checks that the reconstructed network aggregates back to the input--output table.

Section~\ref{app:countries} provides external validation and cross-country evidence. It compares the
Japanese reconstruction with the observed Japanese production network, and then applies the same
pipeline to Australia, the United Kingdom, Finland, and Denmark. This section documents which
regularities recur across all six economies and which features the independent-edge backbone does
not reproduce.

Section~\ref{sec:weaklinks} examines the weighted structure of the reconstructed graph. It shows that the network's higher-order reach, the set of firms reachable in two or three steps, is carried mainly by the many weak links rather than by the few strong ones. Thresholding a network to its major suppliers, as observed datasets do, therefore destroys most of the multi-hop propagation while making the surviving graph look more concentrated than it is.

Section~\ref{sec:esri} asks whether the full graph changes economic measurement. The application is
firm-level systemic risk: the aggregate output loss caused by the failure of a single firm. The section
shows that this object cannot be recovered from sectoral aggregation, firm size alone, degree alone,
or a thresholded map of major links. It requires the full weighted graph.

Section~\ref{sec:conclusion} concludes the paper. Appendix~\ref{app:notation} collects notation.
Appendix~\ref{appendix} states and proves the formal results. The main result shows how heavy
degree tails are inherited from the firm-size distribution through the fitness map, and it records the form the minimum-energy weights take on every realized edge. We also show that, conditional on a fixed probability matrix and before the deterministic repairs, a single Bernoulli draw concentrates around its model-implied edge count and binned degree distribution.
Appendix~\ref{app:robustness} proves the stationary-money bound and reports computational
robustness checks. Appendix~\ref{app:factory} describes how the reconstruction can be extended to
firms with multiple geographically located production units.

\section{The Model}
\label{sec:model}
Firms are indexed by \(i,j\) and sectors by \(k,\ell\), with \(\pi(i)\) firm \(i\)'s sector and \(m_i\in(0,1]\) its normalized size.  A directed edge \(i\to j\) means that firm \(i\) buys from
firm \(j\), and the corresponding weight is the share of firm \(i\)'s spending allocated to firm \(j\). The construction proceeds in two stages. The first stage builds the binary support of the network. A
sector-aware gravity model assigns to every ordered firm pair a probability of trading. An
independent Bernoulli draw realizes the corresponding adjacency matrix. The draw is then regularized:
we impose a minimum-degree floor, connect the strongly connected components, and add an
aperiodizing edge if needed. The second stage assigns weights to the realized support. The weights are chosen by a minimum-energy program that keeps one-step firm balances and sectoral flows close to the data.

Table~\ref{tab:algo} summarizes the algorithm.

\begin{table}[H]
\centering
\caption{The reconstruction algorithm. Inputs: the input--output table \(\mathbf{IO}\) and the firm
census \(\{(m_i,\pi(i))\}_{i=1}^{N_F}\). Output: the weighted network \(\mathbf W\) of
Definition~\ref{def:target}.}
\label{tab:algo}
\setstretch{1.05}\small
\setlength{\tabcolsep}{4pt}
\begin{tabularx}{\textwidth}{@{}l >{\raggedright\arraybackslash}X c@{}}
\toprule
\textbf{Step} & \textbf{Action} & \textbf{Output} \\
\midrule
\multicolumn{3}{@{}l}{\textbf{Stage I. Binary support}}\\
\addlinespace[3pt]
1. Gravity
&
Estimate the sector-pair multipliers \(\lambda\) from input--output flow shares; calibrate the
density scalar \(z\); fix the fitness shape \((a,\eta,m^\star)\); form
\(p_{ij}=x_{ij}/(1+x_{ij})\), where
\(x_{ij}=z\lambda_{\pi(i)\pi(j)}g(m_i)g(m_j)\).
&
\(\mathbf P\)
\\
\addlinespace[2pt]
2. Bernoulli draw
&
Draw \(a_{ij}\sim\mathrm{Bernoulli}(p_{ij})\) independently across ordered firm pairs.
&
\(\mathbf A\)
\\
\addlinespace[2pt]
3. Degree floor
&
Repair firms below the floor so that every firm has at least two suppliers and two customers.
&
\(\mathbf A\) with \(d\ge2\)
\\
\addlinespace[2pt]
4. Irreducibility
&
Connect the strongly connected components through a minimum component-level augmentation, and place the size-adaptive firm-level links using the sectoral placement rule.
&
\(\Ahat\)
\\
\addlinespace[2pt]
5. Aperiodicity
&
Add one aperiodizing edge between distinct firms, if needed, so that the backbone is primitive.
&
\(\Aaug\)
\\
\addlinespace[5pt]
\multicolumn{3}{@{}l}{\textbf{Stage II. Weights}}\\
\addlinespace[3pt]
6. Weighting
&
Solve the minimum-energy weighting problem on the realized support, subject to firm-size and
sectoral-flow restrictions.
&
\(\mathbf W\)
\\
\bottomrule
\end{tabularx}
\end{table}

\begin{definition}[Reconstruction target]
\label{def:target}
A reconstructed network is a row-stochastic matrix \(\mathbf W=(w_{ij})\), \(\mathbf W\mathbf 1=\mathbf 1\), supported on the final binary backbone \(\Aaug=(a_{ij})\): \(w_{ij}=0\) if \(a_{ij}=0\) and \(w_{ij}>0\) if \(a_{ij}=1\). The backbone \(\Aaug\) is irreducible and aperiodic, so \(\mathbf W\) is primitive and admits a unique stationary distribution.
\end{definition}

\subsection{Sector-Aware Logistic Gravity}
\label{subsec:gravity}

The first step specifies the probability that one firm buys from another. The probability depends on
three objects: the size of the buyer, the size of the seller, and the intensity of directional trade between their
sectors. The sectoral input--output table disciplines the third object.

We orient the sectoral input--output table \(\mathbf{IO}\in\mathbb R_+^{N_S\times N_S}\) with buyer sectors in rows and seller sectors in columns, so \(IO_{k\ell}\) is spending by sector \(k\) on sector \(\ell\). Two normalizations are used below,
\[
I_{k\ell}:=\frac{IO_{k\ell}}{\sum_{\ell'}IO_{k\ell'}},
\qquad
\sigma_{k\ell}:=\frac{IO_{k\ell}}{\sum_{k',\ell'}IO_{k'\ell'}}
\]
The row-stochastic expenditure matrix \(\mathbf I\) gives sector \(k\)'s share of intermediate spending on sector \(\ell\).\footnote{Normalizing an input--output table as a Markov transition matrix of money flows is standard \citep{Augusztinovics1965MoneyCirculation,LeontiefBrody1993MoneyFlow}.} The block flow shares \(\sigma_{k\ell}\), which sum to one, are the empirical target for the sectoral composition of links.

For any firm pair \((i,j)\), let \(k=\pi(i)\) and \(\ell=\pi(j)\). We call \((k,\ell)\) the sector block
of the ordered pair. Firm size enters through a scalar fitness function,\footnote{This is a hidden-variable fitness specification with firm size as the hidden variable \citep{GarlaschelliLoffredo2004}. For small firms \(g(m)\approx m^a\), and for very large firms the denominator dampens the growth of fitness. The parameter \(a\) governs the size elasticity, \(m^\star\) the saturation point, and \(\eta\) how sharply the saturation occurs.}
\begin{equation}
\label{eq:fitness}
        g(m)
        :=
        \frac{m^a}{\bigl(1+(m/m^\star)^a\bigr)^{1-\eta}},
        \qquad
        a>0,\quad m^\star\in(0,1],\quad \eta\in[0,1]
\end{equation}

A directed link \(i\to j\) means that firm \(i\) buys from firm \(j\). Conditional on the probability
matrix, links are drawn independently. The intensity of link \(i\to j\) is
\[
        x_{ij}
        :=
        z\,\lambda_{k\ell}\,g(m_i)g(m_j),
        \qquad k=\pi(i),\ \ell=\pi(j)
\]
and the corresponding probability is
\begin{equation}
\label{eq:kernel}
        p_{ij}
        :=
        \frac{x_{ij}}{1+x_{ij}},
        \qquad i\neq j,
        \qquad
        p_{ii}=0
\end{equation}
Let \(\mathbf P=(p_{ij})_{i\neq j}\) denote the resulting probability matrix.

The specification separates three roles: the scalar \(z>0\) fixes overall density, the multipliers \(\lambda=\{\lambda_{k\ell}\}\) allocate link intensity across sector blocks, and the fitness map \(g(\cdot)\) controls the degree tail within a block. The multipliers lie on the unit simplex, \(\lambda_{k\ell}\ge0\) and \(\sum_{k,\ell}\lambda_{k\ell}=1\), which removes the redundancy \((\lambda,z)\mapsto(c\lambda,z/c)\) that leaves every \(x_{ij}\) unchanged. The fitness parameters \((a,\eta,m^\star)\) are not identified from the sectoral flow moments that pin down \(\lambda\) and \(z\), exactly so in the unsaturated limit, where the closed form for \(\lambda\) absorbs any value of \(a\), and only weakly otherwise, so we fix them by economy.

\subsubsection{The sector-pair multipliers}

The sector-pair multipliers are chosen so that the expected distribution of links across sector pairs
tracks the input--output distribution of flows across sector pairs. The premise is that relationships follow money: a sector pair carrying twice the flow share should carry roughly twice the share of buyer-seller links, with the intensity of individual links left to the weighting stage. For a given \(\lambda\), define the
expected number of links in block \((k,\ell)\) by
\[
        P_{k\ell}(\lambda)
        :=
        \sum_{i\in\Fset{k}}\sum_{j\in\Fset{\ell}}p_{ij}
\]
The model-implied edge share of block \((k,\ell)\) is
\[
        \varphi_{k\ell}(\lambda)
        :=
        \frac{P_{k\ell}(\lambda)}
             {\sum_{k',\ell'}P_{k'\ell'}(\lambda)}
\]
We estimate \(\lambda\) by solving
\begin{equation}
\label{eq:lambda-program}
\begin{aligned}
\min_{\lambda\in\Delta}\quad
& \sum_{(k,\ell)}
  \bigl(\varphi_{k\ell}(\lambda)-\sigma_{k\ell}\bigr)^2 \\[2pt]
\text{s.t.}\quad
& \left|
\frac{1}{N_F\bar d}\sum_{k,\ell}P_{k\ell}(\lambda)-1
\right|
\le
\delta_{\mathrm{den}}, \\[2pt]
& \frac{1}{|\mathcal B_{\mathrm{tail}}|}
\sum_{(k,\ell)\in\mathcal B_{\mathrm{tail}}}
\bigl(\varphi_{k\ell}(\lambda)-\sigma_{k\ell}\bigr)^2
\le
\tau^2_{\mathrm{tail}}
\end{aligned}
\end{equation}
where
\[
        \Delta=\{\lambda\ge0:\sum_{k,\ell}\lambda_{k\ell}=1\}
\]
The first constraint keeps the expected mean degree within a band around the target \(\bar d\). The block counts \(P_{k\ell}(\lambda)\) depend on the density scalar \(z\) through \(p_{ij}\), so we solve \eqref{eq:lambda-program} over the pair \((\lambda,z)\), the density band holding the pair near the target, and with \(\lambda\) fixed at the solution we then replace the band by the exact mean-degree calibration \eqref{eq:zcalib}. The
second constraint limits the average error among the worst-fitting sector blocks. Its role is to
prevent small or sparse sector pairs from being sacrificed to improve the aggregate least-squares fit.\footnote{The estimation is finite-dimensional, one multiplier per non-empty block. In the unsaturated limit it has the closed form \(\lambda_{k\ell}\propto\sigma_{k\ell}/(G_kG_\ell)\), with \(G_k=\sum_{i\in\Fset{k}}g(m_i)\), which we use as the starting value for the numerical solution of \eqref{eq:lambda-program}. We estimate \(\lambda\) on a representative sector-size-bin collapse of the firm population, so its cost scales with the number of sectors and bins rather than with \(N_F\), and we hold it fixed across network scales for the same economy, re-solving only the density scalar \(z\) when \(N_F\) changes.}

\subsubsection{Density and degree-tail calibration}

The multipliers \(\lambda\) fix the shape of the probability matrix across sector pairs, and the scalar \(z\) fixes its level. Holding \(\lambda\) and \((a,\eta,m^\star)\) fixed, \(z\) is chosen so that the expected mean degree equals the target\footnote{The left-hand side is continuous and strictly increasing in \(z\) and equals zero at \(z=0\). As \(z\to\infty\) it tends to the number of active ordered pairs divided by \(N_F\), which exceeds \(\bar d\), so the root is unique. We compute it by bisection over size bins, still accounting for logistic saturation among the largest firm pairs. The calibration is repeated for every firm population, since the same \(\lambda\) spread over more or fewer firms needs a different absolute density.},
\begin{equation}
\label{eq:zcalib}
        \frac{1}{N_F}\sum_{i\neq j}p_{ij}(z,\lambda)
        =
        \bar d
\end{equation}

As noted, the block flow shares and the mean-degree condition do not identify the fitness parameters \((a,\eta,m^\star)\); we fix them by economy, disciplined by the tail-inheritance result of Appendix~\ref{appendix}. Over the unsaturated range in which \(g(m)\) behaves like \(m^a\), a firm-size counter-cumulative tail index \(\beta\) induces the degree-tail index \(\chi=\beta/a\), so the calibration relation is \(a=\beta/\chi\). The mechanism is transparent: a firm's expected degree scales as a power of its size, and raising a heavy-tailed size to the power \(a\) divides the tail index by \(a\), while the Bernoulli draw around those means preserves the tail. With a saturation knee and \(0<\eta<1\), the strict far-tail index is instead \(\beta/(a\eta)\); at \(\eta=0\), no finite strict-tail power-law index is asserted. Thus \(\beta/a\) is the calibration benchmark for the empirically relevant unsaturated range, while \(\eta\) and \(m^\star\) govern the transition and the extreme tail.

\subsection{The Bernoulli Draw}
\label{subsec:bernoulli}

Given the probability matrix \(\mathbf P=(p_{ij})_{i\neq j}\), we draw one independent Bernoulli variable for each ordered firm pair,
\[
a_{ij}\sim\mathrm{Bernoulli}(p_{ij}),\qquad i\neq j,\qquad a_{ii}=0
\]
producing a single realized adjacency matrix \(\mathbf A=(a_{ij})\) whose edges are independent conditional on \(\mathbf P\). We use one draw rather than the full ensemble because, conditional on the fixed probability matrix, the pre-repair edge count and binned degree distribution concentrate at national scale; this requires independence and a growing total variance, not identical \(p_{ij}\). Appendix~\ref{appendix} states and proves the relevant central limit theorems (Propositions~\ref{lem:edges_mean_degree} and~\ref{rem:degree_distr_CLT}) and finite-sample bounds. These results justify using one Bernoulli draw for the stochastic backbone. They do not describe the deterministic degree-floor and Markov-closure additions applied afterward.

\subsection{A Minimum-Degree Floor}
\label{subsec:degreefloor}

A raw Bernoulli draw can leave small firms with too few trading partners. We therefore impose the smallest nondegenerate degree floor: every firm must have at least two suppliers and two customers, \(d_i^{\mathrm O}\ge2\) and \(d_i^{\mathrm I}\ge2\). A firm with a single outgoing edge would pass all of its spending to one counterparty, a degenerate row of the money-flow matrix, so the floor is also what keeps \(\mathbf W\) free of point-mass rows. Deficient rows are repaired by drawing additional absent edges from the gravity-weighted conditional law in Appendix~\ref{app:degfloor}, with the largest-\(p_{ij}\) absent entries used only as a deterministic feasibility fallback. The same operation is then applied to the transpose. Since each pass only adds edges, the first pass is not undone by the second.

The floor is a lower-tail correction, not a new source of economic mass. Firm sizes and sector labels are unchanged, and only deficient firms are touched. If the aggregate expected repair count is \(O(N_F)\), the number of added edges remains linear and the network remains sparse; the calibrated \(z\) is not re-solved after the repair. Under sparse non-materialized sampling with bounded expected work per added edge, the expected repair cost is then \(O(N_F+E)\). Appendix~\ref{app:degfloor} gives the tilted conditional law, its deterministic feasibility fallback, and the coupling properties of the repair.

\subsection{Markov Regularity}
\label{subsec:markov}

The weighted network is interpreted as a Markov chain of money flows, so the binary support must be primitive. Irreducibility lets money eventually move between any two firms and pins down a unique stationary distribution. Aperiodicity makes the chain converge to that distribution from any start. The repaired Bernoulli graph need not be strongly connected, and even a strongly connected graph may be periodic. We therefore add a minimum component-level augmentation rather than claiming a minimum number of firm-level links.

First, we collapse the graph into strongly connected components and construct a minimum source--sink pairing that makes the condensation graph strongly connected. Each pairing receives a size-adaptive number of firm-level links. The links are placed by the relaxed sectoral-placement surrogate and final group-wise top-\(k\) rule of Appendix~\ref{app:closure}; no global binary optimality is claimed. Second, if the resulting graph is periodic, we add one admissible ordinary edge between distinct firms that changes the period to one, chosen to make the smallest available change in the placement score. Self-loops are never used. Because the degree floor gives every firm at least two suppliers, some cyclic class always contains two firms, and an admissible edge between distinct firms therefore always exists (Lemma~\ref{lem:noexception}). Appendix~\ref{app:closure} gives the pairing, placement, candidate-screening, and gcd rules.

\subsection{Weights}
\label{subsec:weights}

The previous steps produce a primitive binary backbone. The final step assigns weights to its
edges. The result is a row-stochastic matrix \(\mathbf W\) supported on \(\Aaug\). The entry
\(w_{ij}\) is the share of firm \(i\)'s spending that goes to firm \(j\). The weights must do two things. First, they must respect the support: non-edges receive zero
weight, and realized edges receive positive weight. Second, they must make the money-flow economy
close to the empirical firm-size and sector-size targets.

Let \(\mathbf m=(m_i)\) be the vector of firm sizes. For a candidate weight matrix \(\mathbf W\), the one-step size\footnote{We impose the one-step condition \(\mathbf W^\top\mathbf m\approx\mathbf m\) rather than force the stationary distribution of \(\mathbf W\) to equal \(\mathbf m\) exactly, which a fixed sparse support need not admit. We do not prove that the stationary vector is close to \(\mathbf m\) in general. The guarantees in Appendix~\ref{app:robustness} are conditional. Proposition~\ref{rem:stationary_vs_onestep} bounds the stationary gap by the one-step error, times the ergodic sum \(\Theta(\mathbf W)=\sum_{t\ge0}\tau(\mathbf W^t)\) of the Dobrushin coefficients of the powers of \(\mathbf W\), which is finite for every primitive \(\mathbf W\) but which we cannot bound a priori for a sparse non-normal \(\mathbf W\). The local sensitivity result, Proposition~\ref{rem:error_bound_W}, holds only inside a neighborhood of an exactly money-preserving weighting: a ball that need not contain our fitted point and outside which the map need not be smooth. Neither result certifies that the reconstruction lands in the favorable regime, so we check the gap computationally. In all six economies, running \(\mathbf W\) to stationarity relocates only a few percent of total money mass. The fitted firm-size vector is therefore close to the stationary money distribution used in the propagation exercises.} inherited by firm \(j\) is
\[
        (\mathbf W^\top\mathbf m)_j
\]
The relative one-step firm-size error is
\[
        e_j(\mathbf W)
        =
        \frac{|(\mathbf W^\top\mathbf m)_j-m_j|}{m_j}
\]
Let \(M=\mathbf 1^\top\mathbf m\) and \(\mu_j=m_j/M\). Define the size-weighted root-mean-square
firm error by
\[
        J(\mathbf W)
        =
        \left(\sum_{j=1}^{N_F}\mu_j e_j(\mathbf W)^2\right)^{1/2}
\]

We also track sectoral sizes. For sector \(\ell\), define
\[
        \widehat s_\ell(\mathbf W)
        =
        \sum_{i=1}^{N_F}m_i
        \sum_{j\in\Fset{\ell}}w_{ij},
        \qquad
        s_\ell=\sum_{j\in\Fset{\ell}}m_j
\]
The sectoral relative error is
\[
        h_\ell(\mathbf W)
        =
        \frac{\widehat s_\ell(\mathbf W)-s_\ell}{s_\ell}
\]

Among all feasible weights, we choose the minimum-energy assignment:
\begin{equation}
\label{eq:weight-program}
        \min_{\mathbf W}
        \sum_{i=1}^{N_F}\sum_{j=1}^{N_F}w_{ij}^2
\end{equation}
subject to row-stochasticity,
\[
        \sum_{j=1}^{N_F}w_{ij}=1,
        \qquad i=1,\ldots,N_F
\]
support and positivity,
\[
        \varepsilon_{\mathrm{floor}}a_{ij}
        \le
        w_{ij}
        \le
        a_{ij},
        \qquad i,j=1,\ldots,N_F
\]
the firm-size caps,
\[
        J(\mathbf W)\le\delta^{\mathrm{firm}},
        \qquad
        \mathrm{RMS}_q(\mathbf e(\mathbf W))\le\delta^{\mathrm{tail}}
\]
and the sector-size caps,
\[
        \mathrm{RMS}(\mathbf h(\mathbf W))\le\delta^{\mathrm{sec}},
        \qquad
        \mathrm{RMS}_q(\mathbf h(\mathbf W))\le\delta^{\mathrm{sec}}_{\mathrm{tail}}
\]
Here \(\mathrm{RMS}_q\) denotes the root-mean-square over the worst-fitting fraction \(q\). For firms,
it is computed over the largest \(\lceil qN_F\rceil\) values of \(e_j\). For sectors, it is computed
over the worst-fitting fraction of \(h_\ell\) values.

The objective spreads weight as evenly as possible across the realized links: a minimum-structure rule that does not concentrate trade on a few edges unless the caps require it. Row-stochasticity allocates each firm's spending, and the support constraints zero out non-edges while the positivity floor keeps every realized edge active. The two firm caps control the typical and worst-tail firm-size errors, and the sector caps keep the sectoral totals from drifting as the weights fit firm sizes. These caps are one-step conditions, but they carry the long-run discipline: a matrix that nearly preserves every firm's money in a single step leaves drift little room to accumulate, unless small errors compound along long cycles. Whether they do is what the ex-post stationary check measures.  The program is convex, so whenever the feasible set is nonempty the minimizer is unique.\footnote{The
worst-\(q\) constraints are convex because the sum of the largest \(k\) squared deviations is a
pointwise maximum over all subsets of size \(k\) of sums of convex functions. The positivity floor
keeps every backbone edge active, so \(\mathbf W\) inherits irreducibility and aperiodicity from
\(\Aaug\).}

We now record the form the solution takes on every realized edge.

\begin{proposition}[Structure of the minimum-energy weights]
\label{prop:kkt}
Let \(\mathbf W\) solve \eqref{eq:weight-program} and suppose the caps admit a strictly feasible weighting. Then there are scalars \(\alpha_i\), one per buyer, and \(\gamma_j\), one per seller, such that \(w_{ij}=\max\{\varepsilon_{\mathrm{floor}},\,\alpha_i+m_i\gamma_j\}\) on every realized edge.
\end{proposition}
\noindent Proof in Appendix~\ref{app:weights_structure}.

The seller term \(\gamma_j\) is a nonnegative combination of two one-step shortfalls. The first is that of seller \(j\), \(m_j-(\mathbf W^\top\mathbf m)_j\), and the second is that of its sector, \(s_\ell-\widehat s_\ell(\mathbf W)\). The coefficients vanish for every slack cap. Row \(i\) of \(\mathbf W\) is the Euclidean projection of the vector \((m_i\gamma_j)_j\) onto the rows that sum to one and respect the floor. Two regimes follow. A buyer whose size satisfies \(m_i\max_j|\gamma_j|\ll1/d_i^{\mathrm O}\) receives a row that is uniform up to a small correction, so for small buyers the objective does spread spending evenly. A buyer with \(m_i\max_j|\gamma_j|\gg1/d_i^{\mathrm O}\) receives the positive part of the seller profile \(\gamma_j+\alpha_i/m_i\), so for large buyers the caps, not the objective, decide the row. When only the firm-size cap binds, \(\gamma_j\) has the sign of seller \(j\)'s one-step shortfall. Under flat weights the expected inflow of a seller is proportional to its fitness \(g(m_j)\) rather than to its size (Appendix~\ref{app:weights_structure}), so the shortfall grows with seller size and the sellers with the highest \(\gamma_j\) are the large ones. Large buyers therefore concentrate their spending on large sellers, and the program places the balance adjustment on the rows that have the most room to absorb it.

\section{Reconstruction of the Complete US Production Network}
\label{sec:US}

We now apply the algorithm to reconstruct the full US economy with \(6.46\times10^6\) firms. The section proceeds in five steps. Section~\ref{subsec:data} describes the public data used as
inputs. Section~\ref{subsec:interfirm} converts book receipts into the inter-firm sizes that the
network must reproduce. Section~\ref{subsec:values} reports the estimation: the sector-pair shape
\(\lambda\) is estimated on the simplex from input--output flow shares, the density scalar \(z\) is
calibrated to a mean degree of \(50\), and the degree-tail parameters
\((a,\eta,m^\star)\) are fixed. Section~\ref{subsec:plots} compares the reconstructed network with
the largest available sample of the observed US firm-to-firm network. Section~\ref{subsec:fidelity}
then aggregates the reconstructed firm network back to sectors and checks that it recovers the
input--output table from which it was built.

\subsection{Data}
\label{subsec:data}

Two public datasets carry the reconstruction. The first fixes the number and size of firms. The
second fixes how sectors trade with one another.

The firm data come from the Census Bureau's Statistics of U.S. Businesses (SUSB). SUSB covers
the near-universe of US employer firms. It reports firm counts by NAICS industry and enterprise receipt-size class. We aggregate its industry detail to a two-digit, \(24\)-sector system. Because the receipt-size tabulation is released only for years ending in \(2\) and \(7\) (Economic Census years), the counts are drawn from such a year. SUSB does not report exact firm sizes.
We therefore instantiate every counted firm and assign it a size inside its own reported bin.\footnote{
The finite-bin breakpoints, in dollars, are
\([0,10^5)\), \([10^5,5\times10^5)\), \([5\times10^5,10^6)\),
\([10^6,2.5\times10^6)\), \([2.5\times10^6,5\times10^6)\),
\([5\times10^6,10^7)\), \([10^7,5\times10^7)\),
\([5\times10^7,2.5\times10^8)\), and an open top bin
\([2.5\times10^8,\infty)\).}

Within a finite bin, firm size is drawn uniformly between the lower and upper bin edges. The top
bin is open-ended, so there we draw from a truncated Pareto distribution with counter-cumulative
tail exponent \(\alpha=1\), a Zipf law. This gives the upper tail the empirically relevant shape and
avoids understating the largest firms.\footnote{We use the relatively heavy exponent \(\alpha=1\)
because the upper cap below removes the very largest draws, and the heavy tail restores part of the
mass that the cap would otherwise remove. This choice does not drive the degree-tail results. In
the United States, the top bin contains \(66{,}847\) firms, about \(1\%\) of the \(6.46\)-million firm
population. The degree tails are estimated over the top \(5\)--\(10\%\) of firms, roughly \(323{,}000\)
to \(646{,}000\) firms. Most firms in the estimated degree tail therefore lie outside the open top
size bin. Their sizes are drawn from finite bins, and their degrees are generated by the
gravity--Bernoulli reconstruction rather than by the top-bin sampling rule.}

We impose two bounds on the size support. The lower bound is one-tenth of the smallest bin's
upper edge, so that the smallest firms do not approach zero. The upper bound on the open top bin
is one thousand times that bin's lower edge. Both bounds are relative, so they are invariant to
currency units. The synthetic census is generated once under a fixed seed and then held fixed for
all downstream estimation and simulation.

The sectoral flow data come from the Bureau of Economic Analysis input--output table. We
aggregate the table to the same two-digit, NAICS-based sector system as the firm census, giving a
\(24\times24\) matrix of sector-to-sector flows. Because the two datasets use the same sector
classification, no concordance is required. Each synthetic firm has a sector, a receipt-size bin, and,
through its sector, a row and column in the input--output table. The resulting US firm population
has $N_F=6{,}462{,}423$ firms. This is the population on which the reconstruction is run.

\subsection{From Book Receipts to Inter-Firm Sizes}
\label{subsec:interfirm}

Before estimation, the firm census and the input--output table must be put on the same accounting
basis. SUSB reports total firm receipts. The production network we reconstruct carries only
inter-firm transactions. These are not the same thing. The difference matters most for consumer-facing sectors. SUSB receipts include sales to other
firms, but also sales to households, government, investment, and exports. The reconstructed network
should carry only the intermediate part of these sales. For sectors that mostly sell to other firms,
such as wholesale, professional-service inputs, and basic manufacturing, total receipts and inter-firm
sales are close. For sectors that sell mainly to final demand, they are far apart. Retail trade, health
care, accommodation and food services, much of real estate, and part of utilities have book receipts
that can exceed their inter-firm sales by a large factor.

If we ignored this distinction, the reconstruction would be forced to absorb final-demand sales into
firm-to-firm trade. A consumer-facing sector would enter the gravity model with firm sizes too large
for its intermediate-flow position in the input--output table. The algorithm would then either create
spurious firm-to-firm links or leave large firms with sales that cannot be routed to intermediate
buyers. That would undermine the one-step firm-balance condition imposed in the weighting stage.

We therefore convert book receipts into inter-firm receipts before estimation. For each sector
\(\ell\), define the inter-firm share
\[
\kappa_\ell
:=
1-
\frac{\text{final-demand sales of sector }\ell}
     {\text{total output of sector }\ell}
\in(0,1]
\]
This share is read from the final-demand columns of the BEA Use table
\citep{MillerBlair2009}. It is close to one for intermediate-input sectors and small for
consumer-facing sectors. For example, \(\kappa_\ell\approx0.15\) for health care,
\(\kappa_\ell\approx0.20\) for retail trade, and \(\kappa_\ell\approx0.10\) for accommodation and
food services.

Two sectors require special treatment. For real estate, sector \(53\), we separate commercial rents
paid by businesses from residential rents paid by households. The former are inter-firm payments. The latter are final demand. We use the business-rent share to define \(\kappa_{53}\). The imputed
rent of owner-occupied housing does not appear as a firm receipt and is therefore excluded
automatically.\footnote{This is why we anchor \(\kappa_\ell\) in firm receipts rather than in
input--output gross output. Gross output includes the imputed housing services of owner-occupiers,
which have no firm on either side and would bias \(\kappa_{53}\) toward final demand.} For utilities,
sector \(22\), we similarly separate sales to businesses from sales to households and government to
obtain \(\kappa_{22}\).

There is no firm-level measure of final-demand exposure. We therefore apply \(\kappa_\ell\) uniformly
to all firms in sector \(\ell\). This changes the level of each sector's inter-firm activity but leaves the
within-sector size ranking unchanged.

Let \(r_i\) be the book receipts of firm \(i\), and let
\[
        R_\ell=\sum_{i\in\Fset{\ell}}r_i
\]
be total book receipts in sector \(\ell\). We define inter-firm receipts by
\[
        \tilde r_i
        :=
        \kappa_{\pi(i)}r_i,
        \qquad
        M_\ell
        :=
        \kappa_\ell R_\ell
        =
        \sum_{i\in\Fset{\ell}}\tilde r_i
\]
The normalized firm size \(m_i\) used in the gravity model is the normalization of
\(\tilde r_i\), not of \(r_i\).

This operation only shrinks sectors. Health care falls to roughly one-seventh of book receipts,
accommodation and food services to roughly one-tenth, and real estate and utilities to their
firm-receipt inter-firm levels. Intermediate-input sectors, whose \(\kappa_\ell\) is close to one, barely
move. Nothing is scaled up.\footnote{One further guard is applied to the rescaled sizes, and it binds
only for the smallest economies. Because the open top receipt bin is drawn from a truncated
Zipf-Pareto distribution, a small synthetic census can occasionally generate a ``lonely giant'': one
firm with an implausibly large share of the entire inter-firm economy. Such a firm would distort the
density calibration and isolate much of the network. We therefore cap any one firm's inter-firm size
at three percent of the economy-wide inter-firm total, iterating the clip to a fixed point. In the United
States, Japan, the United Kingdom, Australia, and Denmark, the largest firm's inter-firm size is
already below this threshold. The cap leaves them untouched and binds only in the synthetic small-economy
case where the heavy top bin creates an artificial giant.} We write the rescaled sizes to a separate
dataset, leaving the original synthetic census intact, so book-receipt and inter-firm runs remain
comparable.

The final step is to construct a sectoral inter-firm flow target consistent with these rescaled firm
sizes. The input--output table supplies the pattern of flows across sectors. The rescaled census
supplies the level of inter-firm activity in each sector. We combine the two using RAS or iterative
proportional fitting \citep{bacharach1970biproportional}. Starting from the intermediate-flow structure of the input--output table, we compute a balanced
matrix
\[
        \mathbf F^\star\in\mathbb R_+^{N_S\times N_S}
\]
whose row and column margins both equal the inter-firm sector totals \(M\):
\[
        \sum_{\ell}F^\star_{k\ell}
        =
        \sum_{\ell}F^\star_{\ell k}
        =
        M_k,
        \qquad
        k=1,\ldots,N_S
\]
The equality of row and column margins is deliberate. It imposes at the sector level the same
closure condition later imposed at the firm level: inter-firm purchases equal inter-firm sales. This is
the condition needed for the firm-size vector to be close to stationary under the reconstructed
money-flow matrix. The iteration converges to such a matrix whenever some nonnegative matrix with the seed's zero pattern has these margins \citep{bacharach1970biproportional}, which holds here because the diagonal matrix with entries \(M_k\) has the required margins and its support lies in the seed's whenever every own-sector flow is positive.

The block flow-share target used in the gravity estimation is then read from this balanced
inter-firm table:
\[
        \sigma_{k\ell}
        =
        \frac{F^\star_{k\ell}}
             {\sum_{k',\ell'}F^\star_{k'\ell'}}
\]
We also read the balanced expenditure shares from the same table, \(I^\star_{k\ell}:=F^\star_{k\ell}/\sum_{\ell'}F^\star_{k\ell'}\), which satisfy \(\sum_kM_kI^\star_{k\ell}=M_\ell\) by the margin condition. In sum, the estimation uses two mutually consistent inter-firm objects: firm sizes \(\{m_i\}\) stripped
of final demand, and sectoral flow shares \(\{\sigma_{k\ell}\}\) balanced to the same inter-firm
margins. With these in place, the estimation in Section~\ref{subsec:values} proceeds unchanged.

\subsection{Estimation and Parameter Values}
\label{subsec:values}

We estimate the US reconstruction in a fixed sequence. First we build the synthetic firm population.
Then we estimate the sector-pair multipliers \(\lambda\), calibrate the density scalar \(z\), draw the
Bernoulli backbone, impose Markov regularity, and assign weights. Table~\ref{tab:gravity-params}
reports the parameter values.

The fitness shape is not estimated from the sectoral flow moments. For the identification reason
given in Section~\ref{subsec:gravity}, we hold
\[
        (a,\eta,m^\star)
\]
fixed. In the US reconstruction we set
\[
        (a,\eta,m^\star)=(0.6,\,0.7,\,q_{98})
\]
where \(q_{98}\) is the \(98\)th percentile of the firm-size distribution. The saturation knee is
therefore placed among the largest firms.

\subsubsection{Building the firm population}

We begin by turning the binned SUSB counts into individual firms. Each counted firm becomes one
synthetic firm. Its size is drawn according to the sampling rule in Section~\ref{subsec:data}. For
the full US economy this produces a fixed census of $N_F=6{,}462{,}423$ firms. Each firm has an identifier, a sector, and a normalized size $m_i\in(0,1]$, where sizes are normalized by the largest firm.

For the scale experiments in Section~\ref{subsec:esri-scale}, we draw size-stratified subsamples
from this same census. Firms are retained independently with probability \(r\). Thus a cell with
\(c\) firms contributes \(rc\) firms in expectation. The underlying synthetic population is unchanged;
only the scale at which it is sampled changes.

\subsubsection{Estimating \(\lambda\) and calibrating \(z\)}

The gravity model has two free objects: the shape \(\lambda\) and the level \(z\). They play different
roles and are obtained differently.

The shape \(\lambda\) is the vector of sector-pair multipliers. We estimate it on the simplex by
minimizing the mismatch between model-implied block flow shares and the input--output flow
shares, subject to the density band and the worst-tail cap described in Section~\ref{subsec:gravity}.
The fit is tight. The residual root-mean-square error is of order \(10^{-3}\). We estimate
\(\lambda\) once on the full US census and then hold it fixed. Since \(\lambda\) is the scale-free
sectoral shape, the scale experiments reuse the same \(\lambda\). This is exact in the unsaturated limit. Retaining firms independently with probability \(r\) multiplies every fitness mass \(G_k\) by \(r\) up to sampling noise, which leaves the model-implied block shares \(\varphi_{k\ell}(\lambda)\), and hence the minimizer of \eqref{eq:lambda-program}, unchanged. Only \(z\) is recomputed as \(N_F\)
changes. For the other economies in Section~\ref{app:countries}, the same procedure is applied once
per country.

\begin{figure}[H]
  \centering
  \includegraphics[width=.75\linewidth]{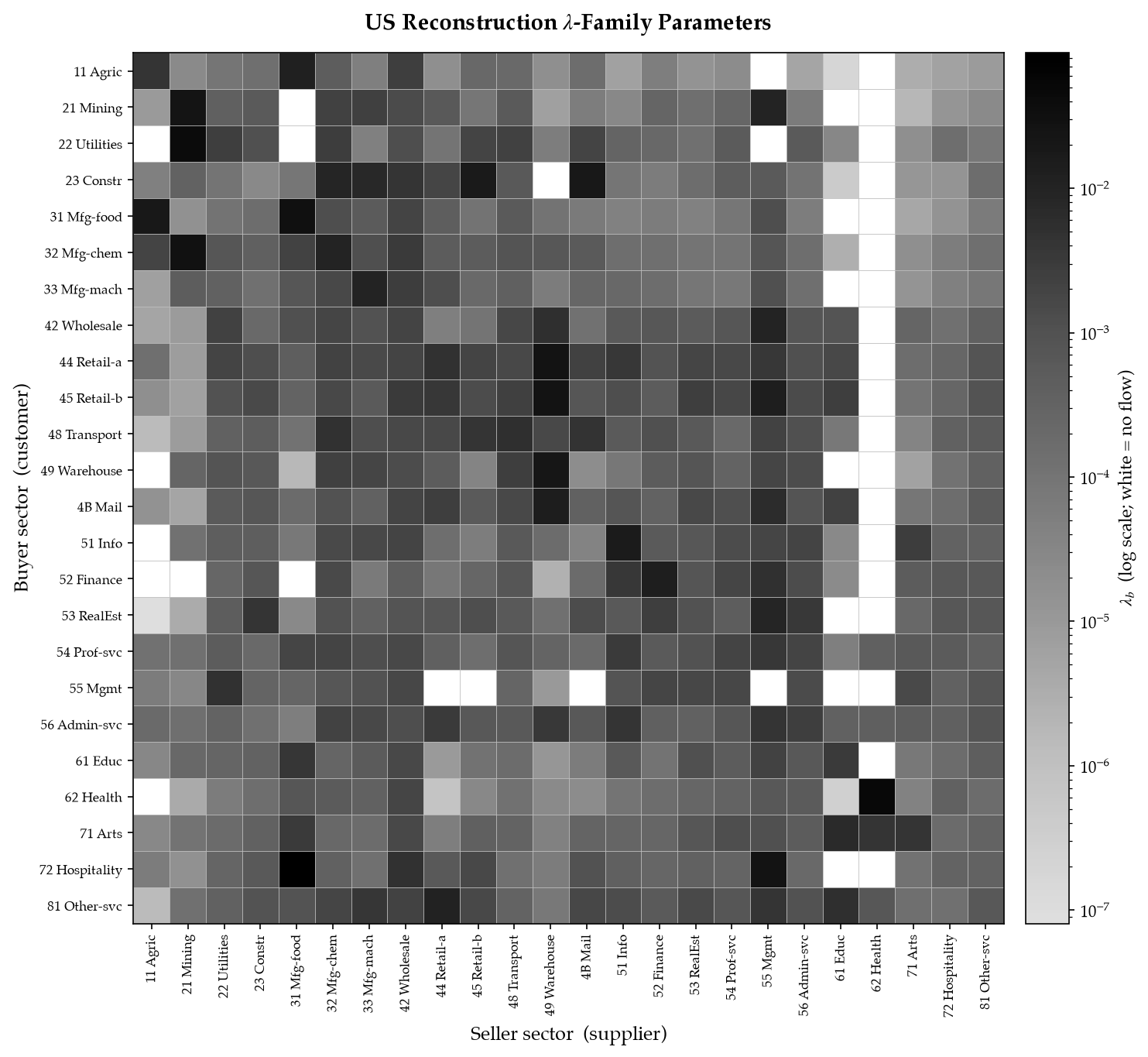}
  \caption{Estimated sector-pair multipliers \(\{\lambda_{k\ell}\}\). Rows are seller sectors and
  columns are buyer sectors, at the two-digit NAICS level. Color gives the multiplier on a logarithmic
  scale. There are \(534\) active blocks out of \(576\); blocks with no inter-sector flow are masked.}
  \label{fig:lambda_heatmap}
\end{figure}

The scalar \(z\) fixes the density of the network. Holding \(\lambda\) and the fitness shape fixed, we
choose \(z\) to solve the mean-degree condition $\frac{1}{N_F}\sum_{i\neq j}p_{ij}=50$. This root is unique, and it is recomputed for every firm population because density is genuinely
scale-dependent. The resulting US value is reported in Table~\ref{tab:gravity-params}. There is one
positive multiplier for each non-empty sector pair, about \(534\) out of the \(24^2=576\) possible
blocks. Figure~\ref{fig:lambda_heatmap} shows the estimated multiplier matrix. The bright seller
columns are the hub sectors: wholesale, management of companies, and professional services.\footnote{
All runs were carried out on a Mac Studio (M1, 128 GB RAM). The multiplier fit runs in minutes.
The binding cost is the weight solve in Section~\ref{subsec:comput_weights}.}

\begin{table}[H]
\centering
\begin{tabular}{llc}
\hline
Parameter & Role & Value \\
\hline
\(a\) & Fitness size elasticity (frozen) & \(0.6\) \\
\(\eta\) & Fitness knee sharpness (frozen) & \(0.7\) \\
\(m^\star\) & Fitness saturation knee (frozen) & \(98\)th size pct.\ \\
\(z\) & Global density (calibrated to \(\bar d=50\)) & \(\approx 2.75\times10^{2}\) \\
\(\lambda\) & Sector-pair multipliers (estimated on simplex) & \(\sim534\) blocks \\
\(\bar d\) & Target mean degree & \(50\) \\
\hline
\end{tabular}
\caption{Parameter values for the US reconstruction, \(N_F=6{,}462{,}423\) firms.}
\label{tab:gravity-params}
\end{table}

\subsubsection{Drawing the backbone}

Once \(\lambda\) and \(z\) are fixed, we draw the binary backbone. For every ordered firm pair
\((i,j)\), we draw an edge independently with probability \(p_{ij}\). At full US scale this means about $N_F(N_F-1)\approx4\times10^{13}$ Bernoulli decisions. The GPU implementation described in Appendix~\ref{sec:compu} makes this
step feasible by assigning those decisions to parallel threads. The output is not a dense
\(N_F\times N_F\) matrix, but the realized edge list.

We then apply the minimum-degree floor and the Markov closure from Section~\ref{sec:model}. The
result is an irreducible and aperiodic directed backbone. This is the support on which weights are
assigned.

\subsubsection{Weighting caps and achieved fidelity}
\label{subsec:comput_weights}

The weighting step solves the minimum-energy program in Section~\ref{subsec:weights}. There is
one variable per realized edge, about $3.40\times10^8$ variables at full scale. The consensus-ADMM solver and its \(O(E)\)-per-iteration cost are described in Appendix~\ref{sec:compu}. This is the slowest stage in wall-clock time, taking roughly four hours
at full scale.

The run uses two kinds of caps. Firm-size caps are hard constraints on the one-step balance residuals. Sector caps are soft targets. The firm caps do not directly constrain the stationary distribution; stationary closeness is checked ex post in Appendix~\ref{app:robustness}. Table~\ref{tab:caps} reports the caps and the achieved one-step deviations. All four targets are met. The hard firm caps are met with room to spare in the tail.

\begin{table}[H]\centering
\caption{Weighting constraints and achieved values, full US network.}
\label{tab:caps}
\begin{tabular}{lccc}
\toprule
Constraint & Symbol & Cap & Achieved \\
\midrule
Firm size-RMS & \(\delta^{\mathrm{firm}}\) & 0.05 & 0.05 \\
Firm tail-RMS (worst \(10\%\)) & \(\delta^{\mathrm{tail}}\) & 0.20 & 0.16 \\
Sector size-RMS & \(\delta^{\mathrm{sec}}\) & 0.10 & 0.07 \\
Sector tail-RMS (worst \(10\%\)) & \(\delta^{\mathrm{sec}}_{\mathrm{tail}}\) & 0.25 & 0.18 \\
\midrule
Tail fraction & \(q\) & \multicolumn{2}{c}{\(0.10\)} \\
Per-edge floor & \(\varepsilon_{\mathrm{floor}}\) & \multicolumn{2}{c}{\(10^{-7}\)} \\
Mean degree & \(\bar d\) & \multicolumn{2}{c}{\(50\)} \\
\bottomrule
\end{tabular}
\end{table}

\subsection{The Reconstructed US Network}
\label{subsec:plots}

We now describe the network produced by the algorithm.  We use the goods-flow convention. A firm's in-degree is its number of suppliers, and its out-degree
is its number of customers.\footnote{The stored matrix uses money-flow edges buyer\(\to\)seller,
with weights equal to the buyer's spending shares. Goods flow in the opposite direction. Throughout
the paper we report degrees in the goods-flow convention.}

Table~\ref{tab:recon_vs_real_stats} compares the reconstructed network with the largest available
sample of the observed US firm-to-firm network: \(78{,}020\) nodes and \(172{,}636\) directed edges
from S\&P Capital IQ, as reported in \citet{MandelVeetil2025}. The comparison is qualitative. The
reconstruction is a full-economy object. The observed network is a selected sample. Their mean
degrees therefore need not coincide. The question is whether they belong to the same structural
regime.

They do. Both networks are extremely sparse. Both have low clustering and low reciprocity.
Average clustering is \(0.01\) in the reconstruction and \(0.01\) in the sample. Reciprocity is lower
in the reconstruction, as expected from an independent-edge backbone, but remains small in both.
Degree assortativity is weakly negative in both networks. The reconstruction has \(-0.01\), while
the sample lies between \(-0.02\) and \(-0.09\). We do not put weight on the exact sign. Pearson
assortativity is fragile in heavy-tailed networks \citep{LitvakvanderHofstad2013}. The important
point is that both networks show weak degree mixing, not a strong rich-club or strong hub-avoidance
pattern.\footnote{This is a qualitative comparison. Conditional on \(\mathbf P\), edges are drawn
independently, so the reconstruction cannot generate much transitivity or reciprocity by chance.
The agreement places the reconstruction in the low-clustering, low-reciprocity regime of the
observed sample. It does not imply that an independent-edge model can reproduce the higher
clustering and reciprocity found in fully mapped administrative networks.}

\begin{table}[H]\centering
\caption{Summary statistics: reconstructed network versus original US sample. In-degree \(=\) suppliers. Out-degree \(=\) customers, using the goods-flow convention. Degree assortativity is the
Pearson correlation of endpoint total degree across edges. Statistics are for the full
\(6.46\times10^6\) network, except local clustering, which is estimated on the \(10^6\)-firm
reconstruction.}
\label{tab:recon_vs_real_stats}
\begin{tabular}{lcc}
\toprule
\textbf{Statistic} &
\textbf{Reconstructed network} &
\textbf{Small US sample} \\
\midrule
Firms \(N_F\)                     & \(6{,}462{,}423\)     & \(78{,}020\)       \\
Edges \(E\)                       & \(340{,}400{,}000\)   & \(172{,}636\)      \\
Density                         & \(8.2\times 10^{-6}\) & \(2.8\times 10^{-5}\) \\
Mean degree                     & \(\approx 52.7\)      & \(\approx 2.2\)    \\
Reciprocity                     & \(0.0009\)            & \(0.03\)           \\
Average clustering              & \(0.013\)             & \(0.01\)           \\
Degree assortativity            & \(-0.009\)            & \(-0.02\) to \(-0.09\) \\
Max suppliers (in-degree)       & \(11{,}044\)          & n/a              \\
Max customers (out-degree)      & \(38{,}358\)          & n/a              \\
Isolated firms                  & \(0\%\)               & n/a              \\
\bottomrule
\end{tabular}
\end{table}

\begin{figure}[H]
  \centering
  \includegraphics[width=.7\linewidth]{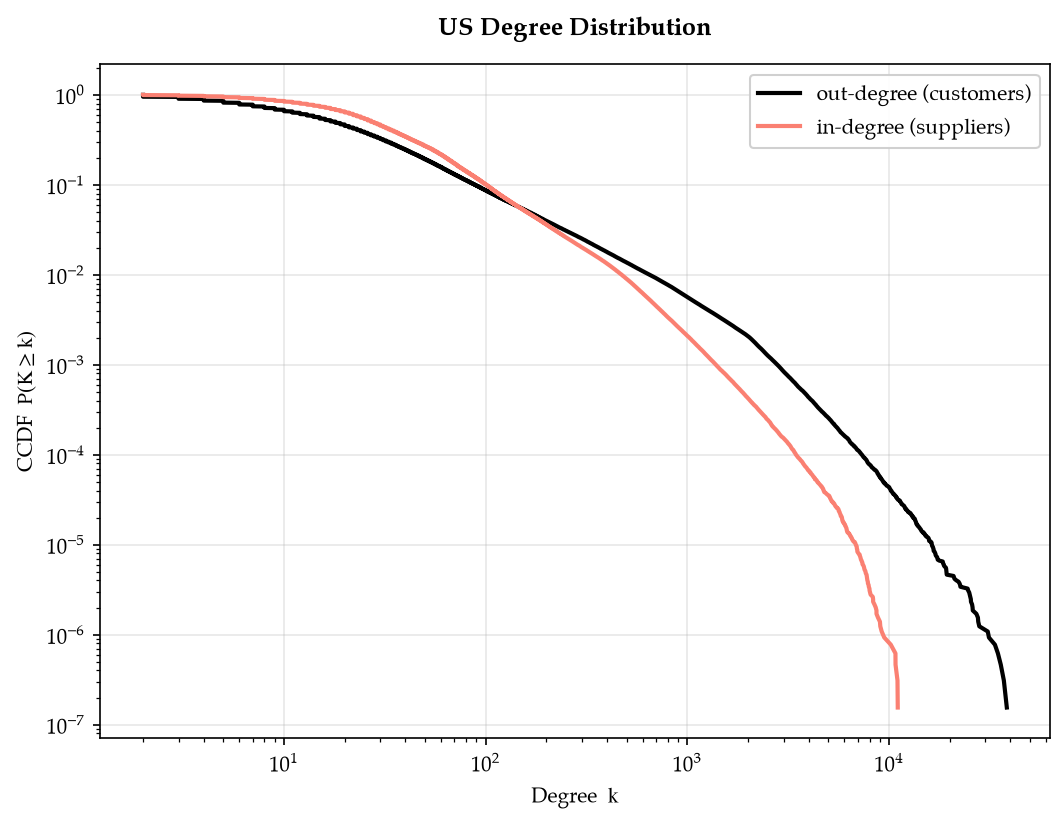}
  \caption{Complementary CDF \(P(K\ge k)\) of the reconstructed US production network on
  log-log axes. Out-degree is customers. In-degree is suppliers. The empirical survival functions are
  plotted without binning.}
  \label{fig:degree_dist}
\end{figure}

The degree distribution is the central diagnostic. Figure~\ref{fig:degree_dist} plots the
complementary CDF of supplier and customer counts on log-log axes. Both tails are heavy. They are
also asymmetric. The customer tail is fatter than the supplier tail, as in observed production
networks. A few firms sell to tens of thousands of customers. The largest customer count is close to
\(4\times10^4\). The largest supplier count is much smaller, about \(10^4\).

Hill estimates at the top decile give a counter-cumulative customer-tail exponent of about $\chi\approx1.2$
and a supplier-tail exponent of about $\chi\approx1.5$ (Table~\ref{tab:country_powerlaw}). Both are below two. This is the important economic range:
the mean degree exists, but the variance does not. 

The customer-heavy asymmetry is not imposed by the firm-size kernel. A firm enters the gravity
term through the same fitness \(g\) whether it is a buyer or a seller. There is no separate seller-side
exponent that mechanically fattens the customer tail. The asymmetry comes from the sectoral flow
structure. Selling is more concentrated than buying. Hub seller sectors, such as wholesale,
management of companies, and professional services, sell into many buyer sectors. The estimated
multipliers in Figure~\ref{fig:lambda_heatmap} transmit that sectoral concentration to the firm
level. Firms in those sectors acquire many customers, and the customer tail stretches farther than
the supplier tail.

Proposition~\ref{prop:tail_inheritance} gives the formal mechanism. Both degree tails inherit the
firm-size tail through the fitness map, with the same index \(\beta/a\). The sectoral masses enter only the prefactors, as \(\sum_k\pi_k\widetilde S_k^{\beta/a}\) on the customer side and \(\sum_k\pi_k\widetilde B_k^{\beta/a}\) on the supplier side, where \(\pi_k\) is the share of firms in sector \(k\) (Proposition~\ref{prop:sparse_constants}). The two families of masses have the same mean, and the seller-side masses are the more dispersed, because selling is concentrated in the hub sectors, so by Jensen's inequality the customer prefactor is the larger and the customer tail extends further at every tail probability. The supplier side also reaches the saturation knee sooner. The local tail index of the degree distribution at size \(m\) is \(\beta/\epsilon(m)\), where \(\epsilon(m)=a[1-(1-\eta)(m/m^\star)^{a}/(1+(m/m^\star)^{a})]\) is the local size elasticity of the fitness map, which falls from \(a\) below the knee to \(a\eta\) above it. Because the buyer-side masses are nearly flat across sectors, the top decile of supplier counts is populated by the largest firms, which sit above the knee, whereas the top decile of customer counts is populated by hub-sector firms of moderate size, which sit below it. The supplier tail therefore appears steeper at
the same Hill depth.

The heavy customer tail is stable across scale. Table~\ref{tab:us_scale_degree} repeats the
tail-exponent estimates at four reconstruction sizes. The customer-tail exponent moves from \(1.32\)
at \(10^4\) firms to \(1.18\) at full scale. The supplier tail is similarly stable once the network is
large. Thus the heavy customer tail is a full-economy property, not a small-sample artifact.

\begin{table}[H]\centering
\caption{US firm degree-tail exponents across reconstruction scale. The table reports
counter-cumulative Hill estimates at top \(10\%\) and top \(20\%\) depths. In-degree \(=\) suppliers. Out-degree \(=\) customers. The \(10^4\) and \(10^5\) values are means over \(100\) networks.}
\label{tab:us_scale_degree}
\setlength{\tabcolsep}{7pt}
\begin{tabular}{lcccc}
\toprule
 & \multicolumn{2}{c}{In-degree (suppliers)} & \multicolumn{2}{c}{Out-degree (customers)} \\
\cmidrule(lr){2-3}\cmidrule(lr){4-5}
Scale & Hill \(10\%\) & Hill \(20\%\) & Hill \(10\%\) & Hill \(20\%\) \\
\midrule
\(10^4\)  & 1.86 & 1.75 & 1.32 & 1.26 \\
\(10^5\) & 1.62 & 1.58 & 1.20 & 1.18 \\
\(10^6\)   & 1.55 & 1.54 & 1.17 & 1.17 \\
Full   & 1.55 & 1.54 & 1.18 & 1.15 \\
\bottomrule
\end{tabular}
\end{table}

\subsection{Reconstruction Fidelity}
\label{subsec:fidelity}

Does the reconstructed firm network return the sectoral table it was built from? To run this basic  internal-consistency test, we aggregate the reconstructed network to the two-digit sectoral level and compare it to the  RAS-balanced input--output target \(\mathbf F^\star\) from which it was created. We test aggregate consistency along two dimensions. The first is concentration: we compute the tail exponent \(\zeta\) of the sector Domar-weight
distribution, using the Gabaix--Ibragimov rank-\(1/2\) estimator on sector sales shares
\citep{gabaix2011rank}. The estimate is \(0.85\) for the RAS-balanced input--output table and
\(0.83\) for the reconstructed firm network aggregated back to sectors. Thus the reconstruction
preserves the concentration of sectoral sales shares. 

The second dimension is the sector-to-sector flows themselves, a more demanding benchmark.  We run the reconstructed money-flow matrix to its
closed-market stationary state and then aggregate the resulting flows back to sectors. These
stationary sectoral flows are close to \(\mathbf F^\star\): the Pearson correlation is \(0.995\), cosine
similarity is \(0.996\), and total-variation distance is \(0.06\). In other words, only about \(6\%\) of
flow share is misplaced. The largest single-cell residual is \(0.5\%\) of total flow.

\begin{figure}[H]
  \centering
  \includegraphics[width=\linewidth]{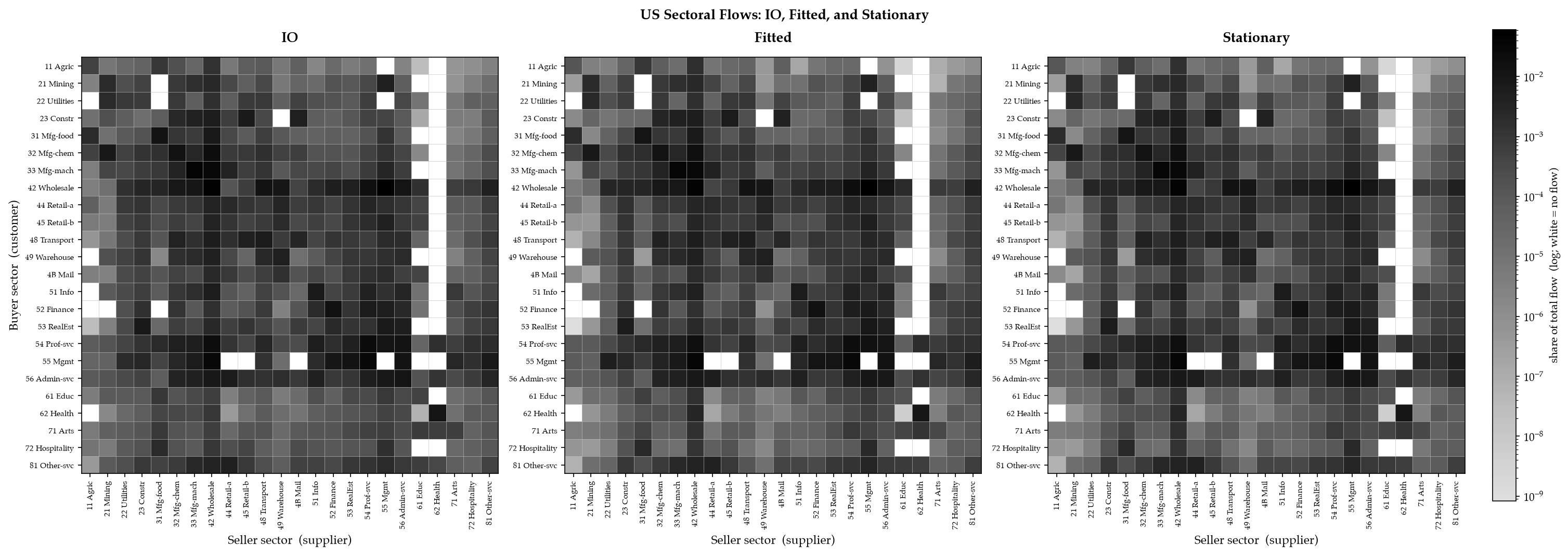}
  \caption{Sectoral flow matrices, buyer sector to seller sector, for the full US network. Entries are
  flow shares, shown on a shared logarithmic color scale. The panels show the RAS-balanced
  input--output target \(\mathbf F^\star\), the flows implied by the fitted weights, and the flows at the
  closed-market stationary state.}
  \label{fig:sector_flow_compare}
\end{figure}

This comparison is not just a test of the fitted weights. It is a test of the reconstructed economy
after money has circulated to stationarity. The fitted flows are even closer by construction. The
stationary flows ask whether the closed-market dynamics preserve the sectoral object that the
network was built to match. The remaining residual is small and systematic. Stationarity routes slightly more flow into the
management-of-companies hub sector, whose seller share rises from \(12.2\%\) to \(13.4\%\). It also
reduces own-sector trade along the diagonal, from \(19.8\%\) to \(18.1\%\).  Figure~\ref{fig:sector_flow_compare}
shows the three matrices side by side: the target \(\mathbf F^\star\), the fitted flows, and the
stationary flows. Table~\ref{tab:sector_flow_div} reports the divergence between stationary flows
and the target. The signed residuals are reported in Appendix~\ref{app:robustness}.

\begin{table}[H]\centering
\caption{Sectoral flow divergence between the US network's closed-market stationary flows
\(\mathbf F_{\mathrm{eq}}\) and the RAS-balanced input--output target \(\mathbf F^\star\), measured as
shares, full US network.}
\label{tab:sector_flow_div}
\begin{tabular}{lc}
\toprule
Metric & Value \\
\midrule
Pearson correlation (active cells) & \(0.995\) \\
Cosine similarity & \(0.996\) \\
Total-variation distance \(\tfrac12\sum|\mathbf F_{\mathrm{eq}}-\mathbf F^\star|\) & \(0.057\) \\
Max single-cell \(|\text{residual}|\) & \(0.45\%\) of flow \\
Within-sector (diagonal) share: \(\mathbf F^\star\to\mathbf F_{\mathrm{eq}}\) & \(19.8\%\to18.1\%\) \\
Management (\(55\))-as-seller share: \(\mathbf F^\star\to\mathbf F_{\mathrm{eq}}\) & \(12.2\%\to13.4\%\) \\
\bottomrule
\end{tabular}
\end{table}

The reconstruction also passes a firm-level self-consistency check. The census firm-size vector is
close to a fixed point of the network's own money dynamics. Running the row-stochastic weight
matrix to stationarity relocates only about \(3\%\) of total money share, and the median firm moves
by about \(4\%\). Appendix~\ref{app:robustness} reports the full census-versus-stationary comparison.

\section{Validation and Verification}
\label{app:countries}

This section does two things. First, it validates the reconstruction against the one economy in our
sample whose firm-to-firm production network has been mapped: Japan. This gives a strict
out-of-sample test. The algorithm is shown only aggregate data. It is not shown any of the observed
firm-to-firm links. The question is whether it nevertheless recovers the degree-tail regime of the
real network.

Second, the section asks whether the structural patterns found for the United States are specific to
the United States. We therefore run the same pipeline on five more economies: Japan, Australia,
the United Kingdom, Finland, and Denmark. Four of these economies have no mapped national
firm-to-firm network. They cannot be used for external validation in the strict sense, but they can
be used to ask whether the same structural regularities recur across different economies.

Only a few objects are re-estimated or recalibrated country by country. For each economy we
estimate the sector-pair shape \(\lambda\) from that country's input--output flow shares. We recalibrate
the density scalar \(z\) to the country's firm count. And we fix the degree-tail parameters
\((a,\eta,m^\star)\) to match that economy's observed or targeted degree-tail regime, while avoiding
a fit that starves either tail of the firm-size distribution (Table~\ref{tab:country_tailparams}).
Everything else is held fixed: the mean-degree target of \(50\), the degree floor, the Markov closure,
and the weighting caps in Table~\ref{tab:caps}.

\begin{table}[H]\centering
\caption{Frozen degree-tail parameters \((a,\eta,m^\star)\) by economy. The saturation knee
\(m^\star\) is quoted as the firm-size percentile at which it sits. It is the \(98\)th percentile in every
economy. The elasticity \(a\) is anchored by the unsaturated-range calibration relation \(a=\beta/\chi\) in Proposition~\ref{prop:tail_inheritance}.}
\label{tab:country_tailparams}
\begin{tabular}{lccc}
\toprule
Economy & \(a\) & \(\eta\) & \(m^\star\) (size pct.) \\
\midrule
United States  & 0.6 & 0.7 & 98 \\
Japan          & 0.5 & 0.5 & 98 \\
United Kingdom & 0.6 & 0.6 & 98 \\
Australia      & 0.5 & 0.5 & 98 \\
Finland        & 0.5 & 0.5 & 98 \\
Denmark        & 0.6 & 0.6 & 98 \\
\bottomrule
\end{tabular}
\end{table}

\subsection{External Validation against the Observed Japanese Network}
\label{app:japan}

Japan is the natural validation case. Its production network has been mapped from credit-registry
data covering close to the universe of firms \citep{fujiwara2010large,MizunoSoumaWatanabe2014_PLOSONE}.
No comparable benchmark exists for the other economies in our sample. We therefore ask a narrow
question: when the reconstruction is built from aggregates alone, does it recover the heavy-tailed
and asymmetric degree regime of the mapped Japanese network?

The comparison is not one of density. The observed network is a partial map. The credit
registry records major supplier-customer relationships, not the full universe of transactions. Its mean
degree, about four to five, therefore substantially understates true connectivity. The reconstruction targets a mean
degree of about fifty on a larger synthetic census. The two networks differ in density by construction
(Table~\ref{tab:japan_compare}). The relevant comparison is the shape of the degree distribution,
not the level of the mean degree.

\begin{figure}[H]
  \centering
  \includegraphics[width=.7\linewidth]{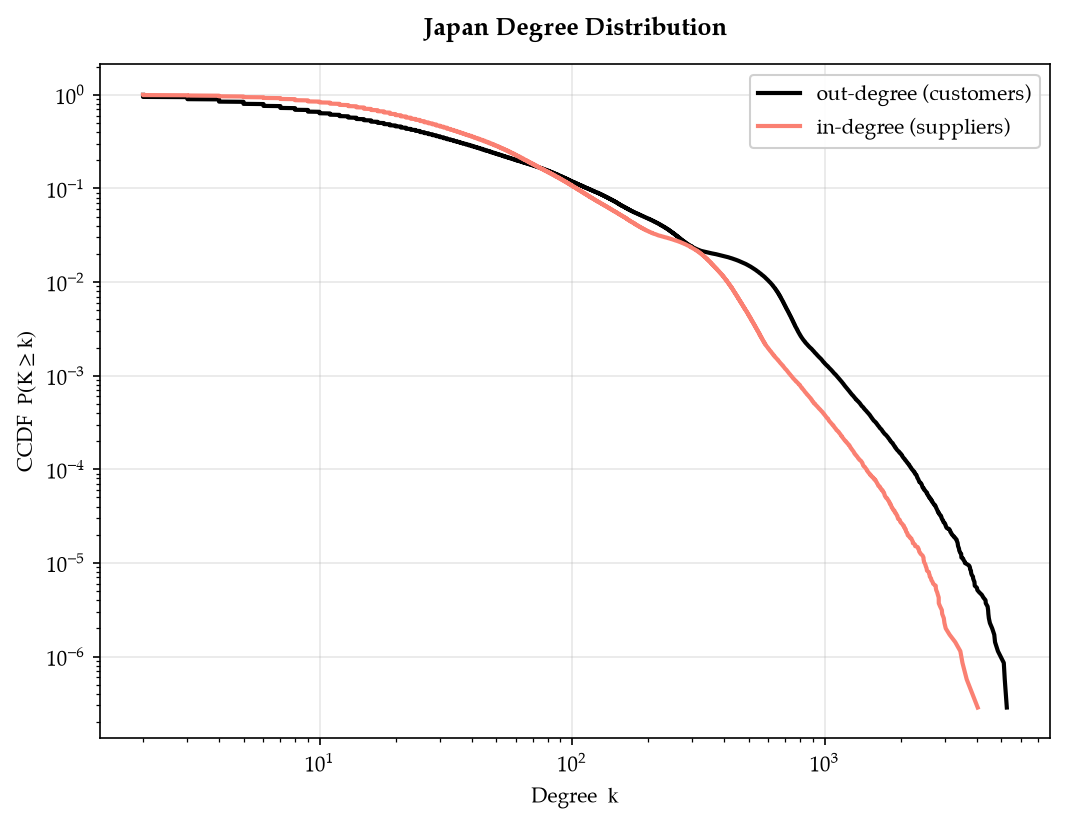}
  \caption{Degree CCDF \(P(K\ge k)\) of the reconstructed Japanese production network:
  out-degree (customers, black) versus in-degree (suppliers, salmon), log-log axes.}
  \label{fig:japan_ccdf}
\end{figure}

On that dimension, the reconstruction does well. Figure~\ref{fig:japan_ccdf} plots the degree CCDFs
of the reconstructed Japanese network. The qualitative signature matches the observed network.
Both tails are heavy. And in both networks the customer tail is fatter than the supplier tail.
\citet{fujiwara2010large} estimate the observed degree distributions over the range \(10\) to \(10^3\)
and report counter-cumulative exponents of \(1.35\pm0.02\) for suppliers and \(1.26\pm0.02\) for
customers. The reconstruction gives \(\chi\approx1.67\) for suppliers and \(1.46\) for customers. These
values lie in the same heavy-tailed band, \(\chi\in(1,2)\), where the mean exists but the variance
diverges. They also reproduce the same customer-heavier-than-supplier asymmetry. Thus, using only
aggregate data, the reconstruction lands in the granular degree regime of the real Japanese network.

The reconstruction does not match higher-order correlations, and this is expected. The backbone is an independent-edge model: conditional on the probability matrix \(\mathbf P\), edges are drawn independently, so reciprocity, clustering, and degree-degree correlations can arise only by chance. At the densities considered here, that chance rate is small. The Japanese comparison therefore draws a clear boundary around the validation exercise. The reconstructed network recovers the heavy-tailed, variance-divergent degree regime and the customer-versus-supplier tail asymmetry. But it is essentially degree-neutral, with assortativity \(+0.006\), compared with mild disassortativity of \(-0.075\) in the observed full network and about \(-0.21\) in listed-firm subsamples. Its reciprocity is \(0.0003\), compared with about \(0.046\) in the observed network, and its local clustering is of order \(10^{-2}\) or below, compared with about \(0.10\) in the observed network. These gaps are not accidental. They follow from the independence assumption. Real firms form reciprocated pairs, triangles, and correlated trading neighborhoods for reasons not contained in a model conditioned only on firm sizes and sector pairs. The validation is therefore positive but limited: for the one economy in which the true network is observable, the reconstruction matches the degree-tail objects it is designed to recover, but it does not reproduce disassortativity, reciprocity, or clustering. Those objects require correlated edge draws, and we do not claim to recover them.

\begin{table}[H]\centering
\caption{The reconstructed Japanese network versus the observed Japanese production network.
Degree-tail exponents are counter-cumulative CCDF indices, listed as the unordered
\(\{\)supplier, customer\(\}\) pair.}
\label{tab:japan_compare}
\begin{tabular}{lcc}
\toprule
Quantity & Observed network & Reconstruction \\
\midrule
Firms \(N\)                 & \(\approx 1.0\times10^{6}\) & \(3.51\times10^{6}\) \\
Links \(E\)                 & \(\approx 5\times10^{6}\)   & \(1.76\times10^{8}\) \\
Mean degree               & \(\approx 4\)--\(5\)          & \(50.2\) \\
Degree-tail exponents \(\chi\) & \(1.26,\ 1.35\)          & \(1.46,\ 1.67\) \\
Degree assortativity      & \(-0.075\) (full), \(-0.21\) (listed) & \(+0.006\) \\
Reciprocity               & \(\approx 0.046\)           & \(0.0003\) \\
Local clustering          & \(\approx 0.10\)            & \(\lesssim 0.01\) \\
\bottomrule
\end{tabular}

\vspace{0.5ex}
\raggedright\footnotesize Notes: Observed statistics from \citet{fujiwara2010large}
(degree-tail exponents as cumulative exponents \(1.35,1.26\), and full-network assortativity)
and \citet{MizunoSoumaWatanabe2014_PLOSONE}; cross-study reciprocity, clustering, and
listed-firm assortativity are compiled by \citet{BacilieriEtAl2023_INET}. Reconstruction statistics
come from Tables~\ref{tab:country_summary}--\ref{tab:country_powerlaw}, and tail exponents are
Hill estimates at top-\(10\%\) depth. The reconstruction matches the observed heavy-tailed degree
regime and tail asymmetry, but, as an independent-edge model, not the observed disassortativity,
reciprocity, or clustering.
\end{table}

\subsection{Cross-Country Regularities}
\label{app:crosscountry}

We next compare the six reconstructed economies. Across all of them, the same structural regime
appears: sparse networks, heavy-tailed degree distributions, customer tails fatter than supplier tails,
near-zero degree assortativity, and negligible reciprocity and clustering. This is the signature of the
fitness-based independent-edge backbone. It is not a peculiarity of the US data.\footnote{There is
substantial cross-country heterogeneity in the frequency, sectoral detail, and valuation of official
input--output releases. Many advanced economies publish annual Supply-Use Tables and less
frequent benchmark input--output tables, often every five years, with varying sectoral resolution
and import/price conventions. A number of low- and middle-income countries release input--output
tables more sporadically and at coarser levels of aggregation. Data on firm sizes are also widely
available. Statistical agencies in the United States, the European Union, Japan, and Australia
regularly release industry-by-size tables. Comparable series are available for several large developing
economies, including India, though usually at lower frequency and with less standardized binning.}

\begin{figure}[H]
  \centering
  \includegraphics[width=\linewidth]{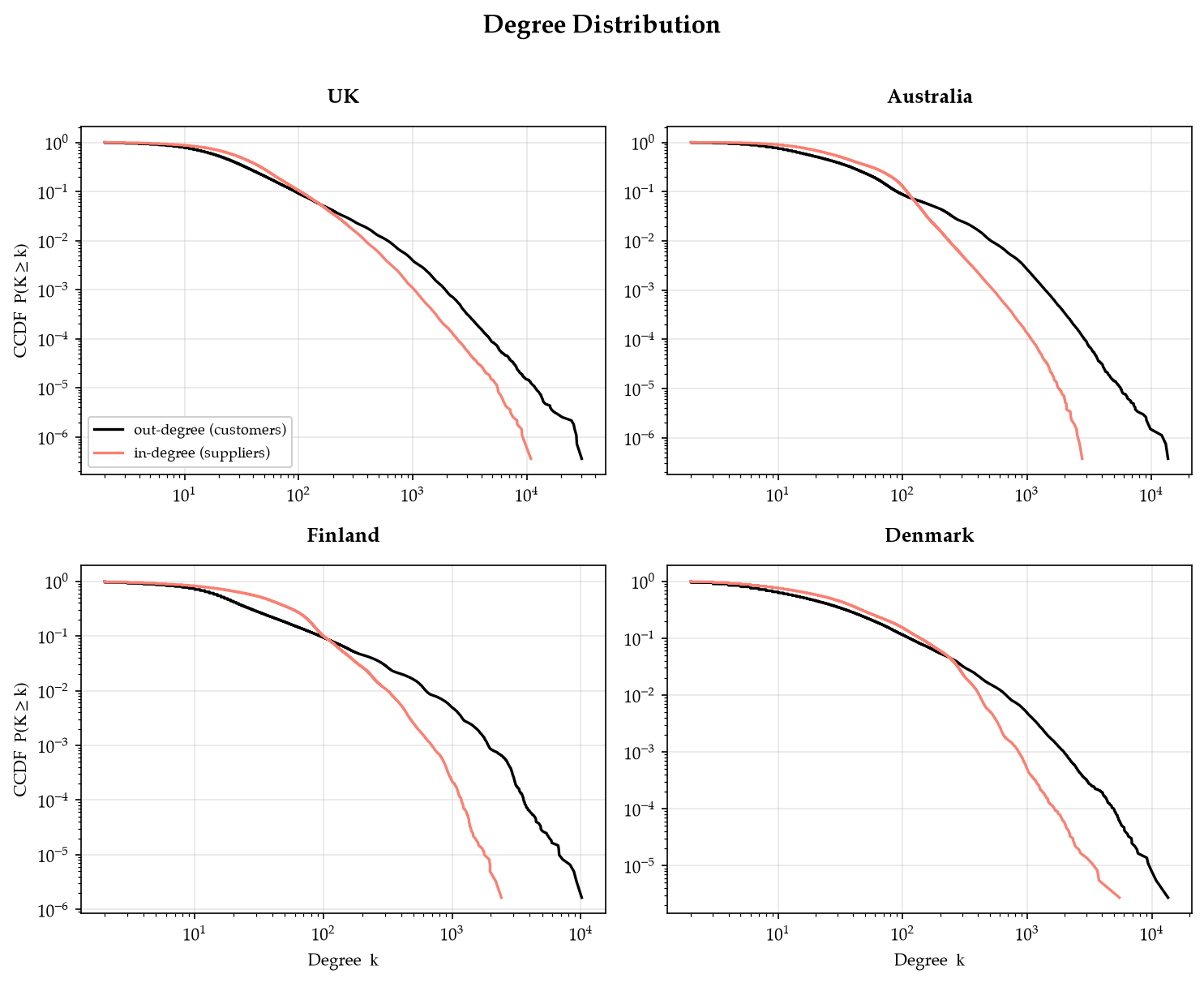}
  \caption{Directed degree CCDFs \(P(K\ge k)\) for the United Kingdom, Australia, Finland, and
  Denmark: out-degree (customers, black) versus in-degree (suppliers, salmon), log-log axes.}
  \label{fig:cc_directed}
\end{figure}

Figure~\ref{fig:cc_directed} shows the tail asymmetry for the United Kingdom, Australia, Finland,
and Denmark. The United States and Japan are shown separately in Figures~\ref{fig:degree_dist}
and~\ref{fig:japan_ccdf}. In every economy the out-degree tail, the customer side, lies above the
in-degree tail, the supplier side.

Table~\ref{tab:country_summary} reports network statistics for all six economies, and
Table~\ref{tab:country_powerlaw} reports the degree-tail exponents. Two facts stand out. First, the networks are uniformly degree-neutral. Reciprocity stays below \(0.002\), clustering is of
order \(10^{-2}\) or below, and degree assortativity lies between \(-0.009\) and \(+0.006\). These are
not empirical discoveries about all production networks. They are what one should expect from an
independent-edge fitness backbone. Second, the customer tails are consistently heavy. The out-degree counter-cumulative exponents
cluster around \(\chi\approx1.2\)--\(1.5\), again in the variance-divergent band below two. The supplier tails are lighter. This customer-side fat tail is the economically
important regularity, because it implies that a small number of firms sell to very many buyers and
can therefore occupy large positions in the propagation structure.

\begin{table}[H]\centering
\caption{Network summary statistics of the reconstructed full-economy networks. Goods
convention: in-degree \(=\) suppliers, out-degree \(=\) customers. All statistics are unweighted and
computed on a single full-economy draw per country. Degree assortativity is the Pearson correlation
of endpoint total degree across edges.}
\label{tab:country_summary}
\setstretch{1.0}\small
\setlength{\tabcolsep}{5pt}
\renewcommand{\arraystretch}{1.2}
\begin{tabular}{lrrrrrr}
\toprule
Statistic & US & Japan & Australia & UK & Finland & Denmark \\
\midrule
Firms \(N\) & \(6.46\times10^{6}\) & \(3.51\times10^{6}\) & \(2.66\times10^{6}\) & \(2.77\times10^{6}\) & \(6.08\times10^{5}\) & \(3.63\times10^{5}\) \\
Edges \(E\) & \(3.40\times10^{8}\) & \(1.76\times10^{8}\) & \(1.33\times10^{8}\) & \(1.44\times10^{8}\) & \(3.06\times10^{7}\) & \(2.03\times10^{7}\) \\
Mean degree & 52.7 & 50.2 & 50.1 & 51.9 & 50.3 & 55.8 \\
Density & \(8.2\text{e-}6\) & \(1.4\text{e-}5\) & \(1.9\text{e-}5\) & \(1.9\text{e-}5\) & \(8.3\text{e-}5\) & \(1.5\text{e-}4\) \\
Reciprocity & 0.0009 & 0.0003 & 0.0001 & 0.0006 & 0.0011 & 0.0013 \\
Deg.\ assortativity & \(-0.009\) & \(+0.006\) & \(-0.005\) & \(+0.002\) & \(+0.006\) & \(-0.003\) \\
Max suppliers & 11{,}044 & 4{,}043 & 2{,}780 & 10{,}861 & 2{,}411 & 5{,}492 \\
Max customers & 38{,}358 & 5{,}272 & 13{,}610 & 30{,}329 & 10{,}155 & 13{,}503 \\
Isolated \% & 0 & 0 & 0 & 0 & 0 & 0 \\
\bottomrule
\end{tabular}
\end{table}

\begin{table}[H]\centering
\caption{Firm degree-distribution tail exponents. The counter-cumulative index \(\chi\) is defined
by \(P(K\ge k)\propto k^{-\chi}\) and estimated by the Hill estimator at the top \(10\%\) and
top \(20\%\) depths. In-degree \(=\) suppliers. Out-degree \(=\) customers. The final column gives the unsaturated-range benchmark \(\beta/a\) from Proposition~\ref{prop:tail_inheritance}, using census \(\beta\) from Table~\ref{tab:eqsize} and frozen \(a\) from Table~\ref{tab:country_tailparams}.}
\label{tab:country_powerlaw}
\setlength{\tabcolsep}{7pt}
\begin{tabular}{lccccc}
\toprule
 & \multicolumn{2}{c}{In-degree (suppliers)} & \multicolumn{2}{c}{Out-degree (customers)} & \\
\cmidrule(lr){2-3}\cmidrule(lr){4-5}
Economy & Hill \(10\%\) & Hill \(20\%\) & Hill \(10\%\) & Hill \(20\%\) & \(\beta/a\) \\
\midrule
United States  & 1.55 & 1.54 & 1.18 & 1.15 & 1.20 \\
Japan          & 1.67 & 1.59 & 1.46 & 1.21 & 1.40 \\
United Kingdom & 1.67 & 1.52 & 1.24 & 1.22 & 1.33 \\
Australia      & 3.06 & 2.58 & 1.22 & 1.37 & 2.28 \\
Finland        & 2.00 & 2.25 & 1.16 & 1.04 & 1.32 \\
Denmark        & 1.83 & 1.42 & 1.23 & 1.14 & 1.40 \\
\bottomrule
\end{tabular}
\end{table}

The heavy tails are not confined to degree. Table~\ref{tab:eqsize} reports the tail exponent of the
firm-size distribution for both the census input and the stationary firm sizes obtained by iterating
the money-circulation chain. In most economies the census size distribution is close to Zipf, with a
counter-cumulative exponent below one. This is the regime in which firm granularity matters
\citep{Axtell2001,Gabaix2011Granular}. The stationary tail is almost identical to the census tail in
every economy. The closed-market dynamics move some money across firms, but they barely change
the shape of the size distribution. Thus the granular propagation objects computed on the
stationary network are not artifacts of a changed size distribution.

\begin{table}[H]\centering
\caption{Firm-size tail exponent, counter-cumulative Hill index at top-\(10\%\): census input versus
network-equilibrium stationary firm sizes.}
\label{tab:eqsize}
\begin{tabular}{lccc}
\toprule
Economy & \(N\) (millions) & Census \(\beta\) & Stationary \(\beta\) \\
\midrule
United States  & 6.46 & 0.72 & 0.73 \\
Japan          & 3.51 & 0.70 & 0.71 \\
United Kingdom & 2.77 & 0.80 & 0.81 \\
Australia      & 2.66 & 1.14 & 1.14 \\
Finland        & 0.61 & 0.66 & 0.66 \\
Denmark        & 0.36 & 0.84 & 0.83 \\
\bottomrule
\end{tabular}
\end{table}

The reconstruction also preserves the sectoral concentration of the input--output table.
Table~\ref{tab:recon_fidelity} reports the tail exponent \(\zeta\) of the sector Domar-weight
distribution for the original input--output table and for each reconstructed network aggregated
back to sectors. In every economy,
\[
        \zeta_{\mathrm{recon}}\approx \zeta_{\mathrm{orig}}
\]
with differences no larger than \(0.02\). The realized sector flows also correlate with the target at
\(0.965\)--\(0.999\). The reconstruction therefore reproduces not only the average sectoral flows but
also the concentration of sectoral weight.

\begin{table}[H]\centering
\caption{Reconstruction fidelity: tail exponent \(\zeta\) of the sector Domar-weight
(sales-share) distribution, original input--output table versus reconstructed firm network aggregated
to \(S\) sectors.}
\label{tab:recon_fidelity}
\begin{tabular}{lccc}
\toprule
Economy & \(S\) & \(\zeta_{\mathrm{orig}}\) & \(\zeta_{\mathrm{recon}}\) \\
\midrule
United States & \(24\) & \(0.85\) & \(0.83\) \\
Japan & \(18\) & \(0.54\) & \(0.54\) \\
United Kingdom & \(17\) & \(0.73\) & \(0.72\) \\
Australia & \(19\) & \(0.62\) & \(0.62\) \\
Finland & \(18\) & \(0.53\) & \(0.53\) \\
Denmark & \(19\) & \(0.46\) & \(0.46\) \\
\bottomrule
\end{tabular}
\end{table}

Two conclusions follow. First, the recurrence of the same heavy-tailed, degree-neutral,
customer-heavier regime across six economies is not driven by peculiarities in the US data. The six economies
differ by more than an order of magnitude in firm count and have different sectoral concentration
profiles, yet the same basic structure appears. Second, the method succeeds and fails in predictable places. It succeeds on the heavy-tailed degree
structure. The customer-tail exponents, about \(1.2\)--\(1.5\), lie in the range documented in mapped
production networks \citep{BacilieriEtAl2023_INET,Cimini2021}. It fails on the higher-order
correlations for the same reason it fails there in Japan.

\section{Weak Links Carry the Higher-Order Reach}
\label{sec:weaklinks}

We have so far focused on degrees. We now turn to a related but different question: how many other firms can be reached through the weighted graph in two or three steps, and which links make that reach possible. This distinction matters because many observed firm-to-firm datasets record only major trading partners. If discarded small links are economically irrelevant, then a thresholded network may still be sufficient for propagation exercises. If they carry higher-order reach, however, then a major-supplier-only graph removes precisely the links through which shocks travel beyond the first neighborhood. We find the latter. Across all six economies, higher-order reach is carried mainly by the many small links, not by the few large ones. Weak links are individually small, but collectively they connect firms to distant parts of the production network. These are exactly the links that thresholded datasets discard.

Start with the reach itself. In the US reconstruction, the largest hubs reach almost the entire
\(6.46\)-million-firm economy within two hops. Direct links are only the first layer. Most of the
network's reach appears in the second-order kernel: customers of customers, or suppliers of
suppliers.\footnote{The two-hop reach is read from the second-order out-degree distribution. The
number of customers-of-customers a US firm reaches extends to \(\sim 10^7\), nearly the full firm
count.}

\begin{figure}[H]
  \centering
  \includegraphics[width=0.75\linewidth]{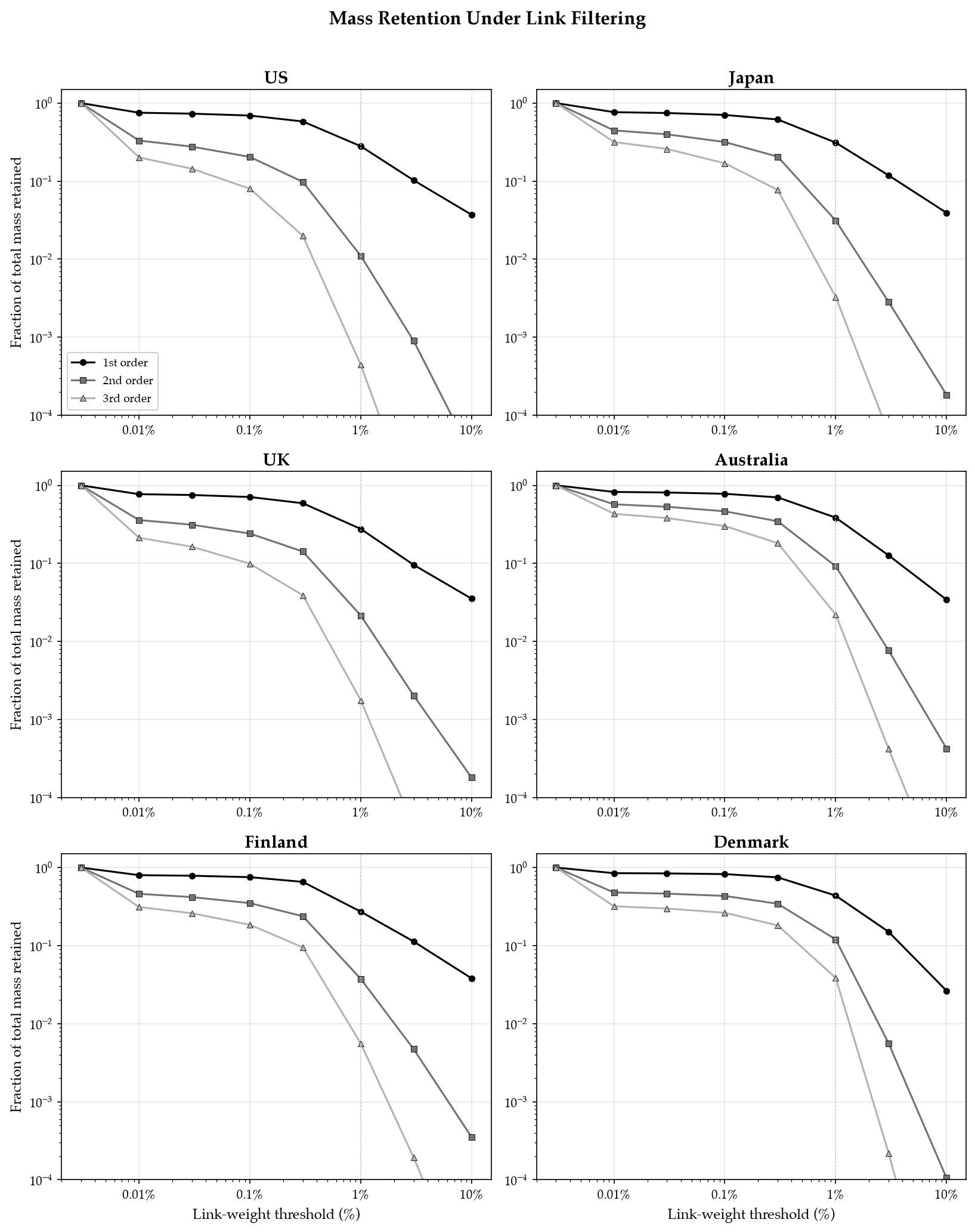}
  \caption{Fraction of total out-reach mass retained when links below a weight threshold are
  deleted, one panel per economy. The horizontal axis is the threshold, measured as the link's share
  of the buyer's spending, on a logarithmic scale from \(0.01\%\) to \(10\%\). The vertical axis is the
  retained fraction of reach mass, for reach orders \(1\), \(2\), and \(3\).}
  \label{fig:order_retention}
\end{figure}

We then ask which links carry this reach. For each economy, we delete every link below a weight
threshold, where the weight is the link's share of the buyer's spending. We then measure how much
first-, second-, and third-order out-reach survives. Figure~\ref{fig:order_retention} gives the full
threshold path.

The result is sharp. At the \(1\%\)-of-spending cutoff, first-order reach survives moderately:
between \(27\%\) and \(44\%\) across economies. But second-order reach falls to \(1\%\)--\(12\%\),
and third-order reach falls to \(0.04\%\)--\(3.9\%\). Read the other way, links below \(1\%\) of buyer
spending carry about \(88\%\)--\(99\%\) of second-order reach and \(96\%\)--\(99.96\%\) of third-order
reach. The pattern appears in every economy (Table~\ref{tab:link_filter}).

\begin{table}[H]\centering
\caption{Link-weight filtering: reach retained at the \(\ge1\%\)-of-spend cutoff, and the norm
\(\|\mathbf v\|_2\) of the stationary money distribution recomputed on the filtered network.
Higher-order reach collapses under the cutoff, while \(\|\mathbf v\|_2\) rises. Thresholding destroys
higher-order propagation but overstates first-order concentration.}
\label{tab:link_filter}
\setlength{\tabcolsep}{5pt}\small
\begin{tabular}{lrrrrcc}
\toprule
 & Edges & \multicolumn{3}{c}{Reach retained at \(\ge1\%\) cut} & \(\|\mathbf v\|_2\) & filtered \(\|\mathbf v\|_2\) \\
\cmidrule(lr){3-5}
Economy & (M) & 1st & 2nd & 3rd & full & (ratio) \\
\midrule
United States  & 340 & \(28\%\) & \(1.1\%\) & \(0.04\%\) & 0.0142 & 0.0355 (\(2.5\times\)) \\
Japan          & 176 & \(31\%\) & \(3.1\%\) & \(0.33\%\) & 0.0172 & 0.0330 (\(1.9\times\)) \\
United Kingdom & 144 & \(28\%\) & \(2.2\%\) & \(0.18\%\) & 0.0475 & 0.0850 (\(1.8\times\)) \\
Australia      & 133 & \(39\%\) & \(9.2\%\) & \(2.2\%\)  & 0.0343 & 0.0503 (\(1.5\times\)) \\
Finland        & 31  & \(27\%\) & \(3.7\%\) & \(0.56\%\) & 0.0576 & 0.0853 (\(1.5\times\)) \\
Denmark        & 20  & \(44\%\) & \(12\%\)  & \(3.9\%\)  & 0.0375 & 0.0680 (\(1.8\times\)) \\
\bottomrule
\end{tabular}
\end{table}

Thresholding therefore creates two distortions at once. First, it removes the weak links that carry
multi-hop propagation. Second, it makes the surviving graph look more concentrated than the full
graph. The last two columns of Table~\ref{tab:link_filter} show this directly. The norm \(\|\mathbf v\|_2\) of the stationary money distribution rises by a factor of \(1.5\) to \(2.5\).\footnote{The reported numbers are based on renormalized buyer weights. The results are nearly identical without the renormalization.} The mechanics are simple: once the small links are deleted, each buyer's surviving spending shares are renormalized upward, so stationary money piles onto the remaining hubs.

The order-by-order pattern is not a quirk of the \(1\%\) cutoff. At every threshold, second- and
third-order reach fall much faster than first-order reach. This is a composition effect. If a fraction
of direct links is retained, the fraction of two-hop paths that survives is roughly the square of that
fraction, and the fraction of three-hop paths is roughly the cube. Thus even a moderate loss of
direct links becomes a large loss of higher-order reach.\footnote{The retention percentages are
conditional on the calibrated density. At \(\bar d=50\), the minimum-energy weighting spreads a
buyer's budget over roughly fifty links, so a typical spending share is of order
\(1/\bar d\approx2\%\). The \(1\%\) cutoff is therefore near the middle of the weight distribution. A
sparser calibration would raise average link shares and soften the first-order cutoff. The retention
levels in Table~\ref{tab:link_filter} should therefore be read together with \(\bar d\). The same
exercise can be repeated at alternative density targets.}

\begin{figure}[H]
  \centering
  \includegraphics[width=0.75\linewidth]{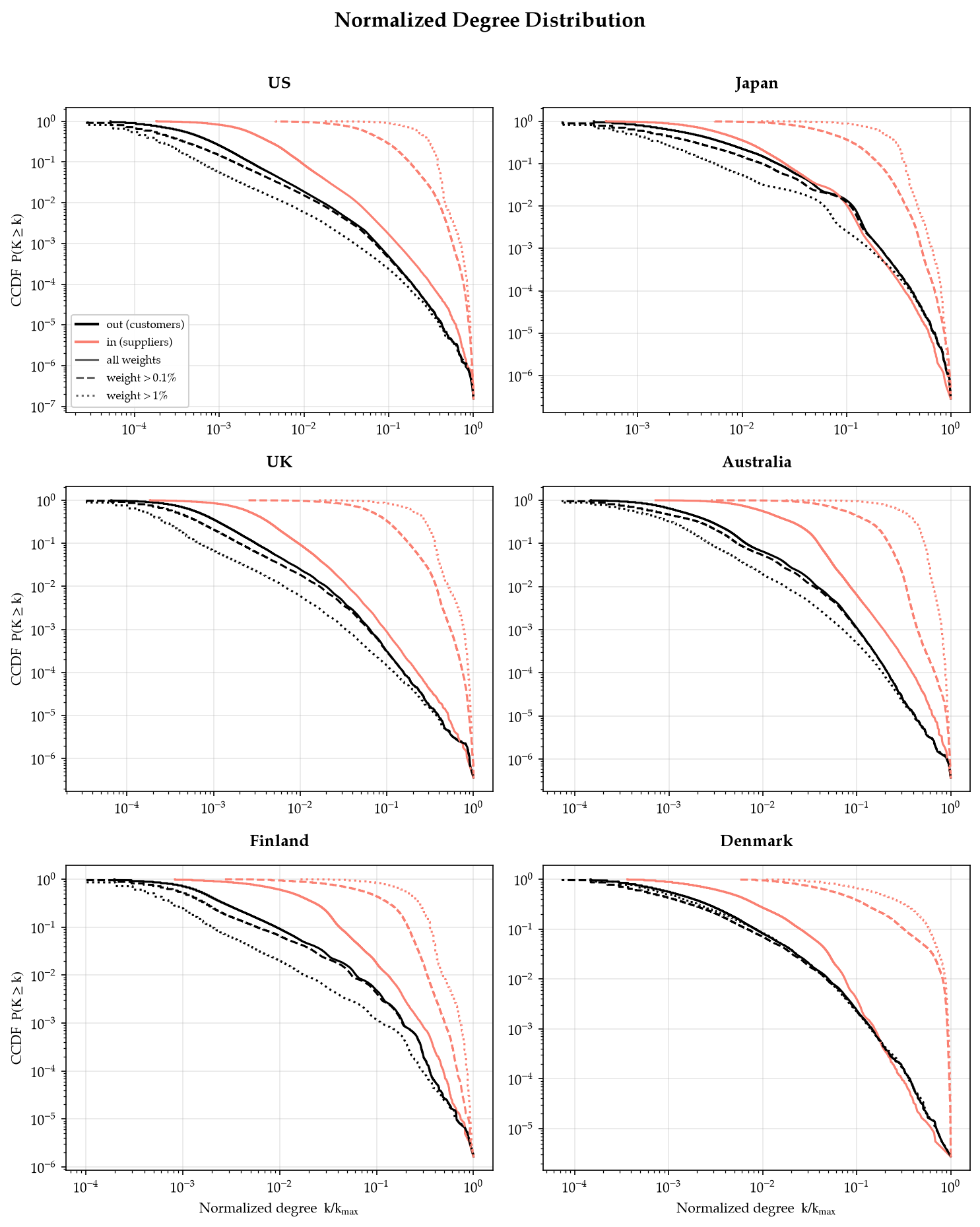}
  \caption{Normalized degree CCDF \(P(K\ge k)\) of the reconstructed networks, one panel per
  economy, on log-log axes. Each series' degree is rescaled by its own maximum, \(k/k_{\max}\), so
  that the figure compares tail shape rather than scale. Out-degree is customers. In-degree is suppliers. Line style marks the weight filter: all links, \(>0.1\%\), and \(>1\%\) of buyer spending.}
  \label{fig:us_ccdf_norm}
\end{figure}

\begin{figure}[H]
  \centering
  \includegraphics[width=0.75\linewidth]{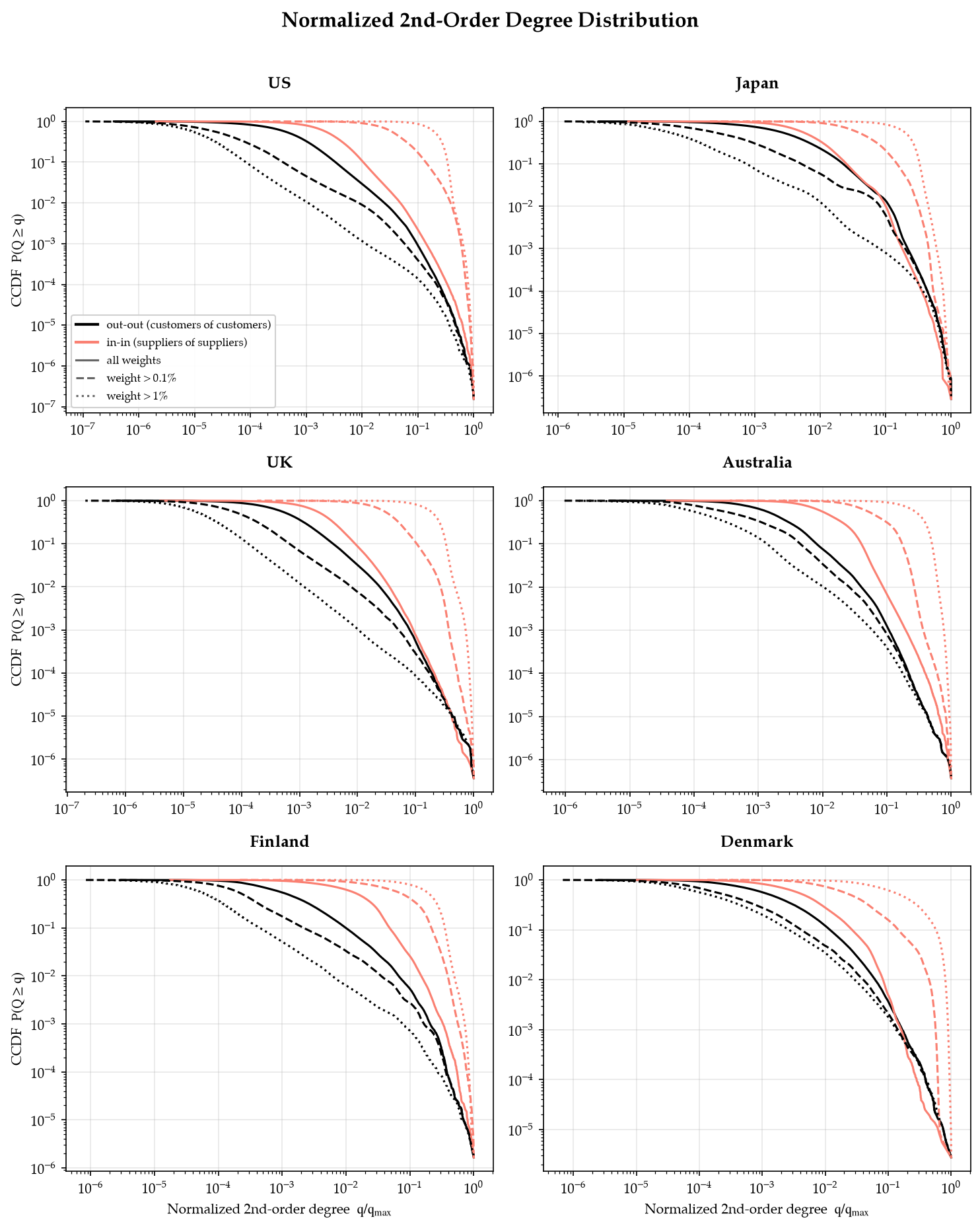}
  \caption{Normalized second-order degree CCDF \(P(Q\ge q)\) of the reconstructed networks,
  one panel per economy, on log-log axes. Each series' two-hop degree is rescaled by its own
  maximum, \(q/q_{\max}\). Out-out means customers of customers. In-in means suppliers of
  suppliers. Line style marks the weight filter, and second-order degrees are recomputed on each
  filtered subgraph.}
  \label{fig:us_ccdf_2nd_norm}
\end{figure}

The same conclusion appears in the shape of the degree distributions. Figure~\ref{fig:us_ccdf_norm}
rescales each first-order degree distribution by its own maximum degree, so that the comparison is
about tail shape rather than scale. The first-order customer tail is fairly stable across filters. Hubs
remain hubs even after links below \(1\%\) of spending are removed. The first-order heavy tail is
therefore not simply an artifact of many tiny links. The second-order distribution behaves differently. Figure~\ref{fig:us_ccdf_2nd_norm} shows that
the customers-of-customers tail collapses under the same filters. The \(>1\%\) curve lies one to two
orders of magnitude below the all-links curve in every economy. This is because two-hop reach is
built by chaining through weak links. Removing those links reshapes the second-order tail much
more than the first-order tail.

Table~\ref{tab:tail-exponent-filtering} gives the same result numerically. Under the \(>1\%\) filter,
the first-order customer-tail exponent changes by \(-22\%\) on average. The second-order
customer-tail exponent changes by \(-42\%\) on average. The second-order distortion is therefore
about twice as large. The Kolmogorov--Smirnov distances in Table~\ref{tab:ks-filtering} give the
same message: filtering moves the second-order shape about \(1.7\) times as much as the first-order
shape. In the United States and Finland, the filter even pushes the second-order tail exponent below
one, a concentration regime that the first-order distribution never enters.

\begin{table}[H]
\centering
\caption{Change in the out-degree, or customer-side, tail exponent under link filtering: all weights
versus keeping only links above \(1\%\) of buyer spending. The table compares first- and
second-order degree distributions.}
\label{tab:tail-exponent-filtering}
\begin{tabular}{@{}l cc cc@{}}
\toprule
 & \multicolumn{2}{c}{1st order (customers)} & \multicolumn{2}{c}{2nd order (customers of customers)} \\
\cmidrule(lr){2-3}\cmidrule(lr){4-5}
Economy & exponent (all\(\to{>}1\%\)) & \% change & exponent (all\(\to{>}1\%\)) & \% change \\
\midrule
United States  & \(1.53 \to 1.19\) & \(-22\%\) & \(1.59 \to 0.93\) & \(-42\%\) \\
Japan          & \(2.43 \to 1.61\) & \(-34\%\) & \(2.42 \to 1.10\) & \(-54\%\) \\
United Kingdom & \(1.77 \to 1.33\) & \(-25\%\) & \(1.79 \to 1.06\) & \(-40\%\) \\
Australia      & \(2.05 \to 1.59\) & \(-23\%\) & \(2.11 \to 1.33\) & \(-37\%\) \\
Finland        & \(1.79 \to 1.25\) & \(-30\%\) & \(1.79 \to 0.98\) & \(-45\%\) \\
Denmark        & \(1.84 \to 1.85\) & \(+1\%\)  & \(1.97 \to 1.33\) & \(-33\%\) \\
\midrule
Mean           & & \(-22\%\) & & \(-42\%\) \\
\bottomrule
\end{tabular}
\end{table}

\begin{table}[H]
\centering
\caption{Kolmogorov--Smirnov distance between the all-weights and \(>1\%\)-filtered normalized
degree CCDFs, for first- and second-order degree distributions. A larger KS distance means the
shape moves more under filtering.}
\label{tab:ks-filtering}
\begin{tabular}{@{}l ccc cc@{}}
\toprule
 & \multicolumn{3}{c}{out-degree (customers)} & \multicolumn{2}{c}{in-degree (suppliers)} \\
\cmidrule(lr){2-4}\cmidrule(lr){5-6}
Economy & 1st order & 2nd order & ratio & 1st order & 2nd order \\
\midrule
United States   & 0.44 & 0.75 & 1.7 & 0.98 & 0.97 \\
Japan           & 0.40 & 0.71 & 1.8 & 0.95 & 0.93 \\
United Kingdom  & 0.54 & 0.75 & 1.4 & 0.98 & 0.97 \\
Australia       & 0.38 & 0.58 & 1.5 & 0.96 & 0.94 \\
Finland         & 0.49 & 0.70 & 1.4 & 0.89 & 0.84 \\
Denmark         & 0.08 & 0.42 & 5.0 & 0.86 & 0.79 \\
\midrule
Mean            & 0.39 & 0.65 & 1.7 & 0.94 & 0.91 \\
\bottomrule
\end{tabular}
\end{table}

This is the edge-completeness case for the full weighted network. It matters because real
firm-to-firm datasets are often thresholded by design. The Japanese production network from Tokyo
Shoko Research and RIETI records each firm's principal trading partners
\citep{fujiwara2010large,OhnishiTakayasuTakayasu2010,BernardMoxnesSaito2019}. The Belgian
value-added-tax network censors transactions below a reporting floor
\citep{DhyneKikkawaMogstadTintelnot2021_REStud,DhyneMagermanRubinova2015_NBB}. The links
these datasets do not see are exactly the links that carry higher-order reach. The exercises in this section show what the full reconstruction is for. It is not only a way to recover
degree tails or match sectoral aggregates. It is a way to recover the weak links and higher-order
paths through which shocks travel.

\section{Economic Systemic Risk: Why the Full Graph Has No Substitute}\label{sec:esri}
 Suppose one wants to know which firms are systemic, in the simple sense that their failure causes a large fall in aggregate output. Can this be read from firm size? Can it be read from sectoral input--output linkages? Can it be read from a small or thresholded version of the firm network? Or does one need the reconstructed national-scale weighted graph? We answer this question by running a single counterfactual for every firm. Firm $i$ is forced to produce zero, the production cascade is allowed to settle, and we record the resulting fall in aggregate output. This number is the firm's economic systemic risk index, or ESRI \citep{DiemEtAl2022_SciRep,ElliottGolubJackson2014}. We compute it on the full reconstructed network and then ask how much of it can be recovered from reduced views of the same economy.

The exercise is deliberately simple. It is not meant to be a general theory of firm failure. It is a test of what the full graph buys. If a sectoral table, a size ranking, a fitted tail, a small reconstruction, or a thresholded network can reproduce the ESRI distribution and the ESRI ranking, then the full graph is unnecessary for this purpose. If they cannot, then the reconstruction is doing economic work that no reduced object can do.

The cascade runs on the reconstructed network read as a closed inter-firm economy. Write $\mathbf W$ for the row-stochastic matrix of input shares, with $w_{jr}\ge0$ the fraction of buyer $j$'s intermediate spend sourced from seller $r$. Let $m_j>0$ be firm $j$'s size (its equilibrium sales) and $s\in(0,1)$ the intermediate-input share of output, so that $1-s$ is value added (canonically $s=0.8$). Each firm carries a \emph{normalized output}, or health, $h_j\in[0,1]$, with $h_j=1$ the undisturbed state and $h_j=0$ complete shutdown. One propagation step limits each firm by the worse of a supply term set by its inputs and a demand term set by its customers, clipped to $[0,1]$. Appendix~\ref{app:esri-impl} gives the map and both channels. Aggregate output is the size-weighted average of firm health, $Y_t=\sum_j m_j h_{j,t}/\sum_j m_j$, equal to one at the undisturbed fixed point $\mathbf h=\mathbf 1$. The vector $\mathbf v=\mathbf W^\top\mathbf v$ ($\mathbf 1^\top\mathbf v=1$) is the left Perron vector of $\mathbf W$, the stationary distribution of the money chain. Its entries are the firms' sales shares: the Domar weights of Section~\ref{subsec:weights}, normalized to sum to one.

The knock is an output constraint, not a deletion. Firm $i$'s health is held at $h_i\equiv0$ while its links, input shares, size, and sales flows stay unchanged, so its customers see a dead supplier through the supply term and its suppliers a dead customer through the demand term. Holding the relationships fixed isolates the question of interest: how a given failure propagates through the existing web of relations, not how the web would eventually rewire around it. The supply term encodes the substitution assumption: when a supplier's output falls, how far can its customers re-source? We answer with a constant-elasticity-of-substitution aggregator that makes each edge independently \emph{essential} with probability $p$, splitting each buyer's suppliers into a substitutable set and an essential set (Appendix~\ref{app:ces-esri}). The per-edge essentiality draws are frozen, so a sweep over $p$ is nested. Two limits bracket the CES channel. As $p\to0$ the supply term is linear, the perfectly substitutable Hulten case. As $\rho\to-\infty$ the essential-input aggregate becomes the minimum health among the essential suppliers, damped by the value-added envelope of Appendix~\ref{app:ces-esri}, so a dead essential input removes the essential-input contribution to output but leaves the firm producing. The undamped Leontief rule of Appendix~\ref{app:leontief-esri}, in which a single dead essential input idles the firm, is a separate mechanism that we run alongside the CES channel. Our headline sets $\rho=-1$ and $p=0.01$: gross complements, but only mildly, an elasticity close to the Cobb--Douglas ($\rho=0$) case in which Hulten's theorem is exact.\footnote{Estimated elasticities of substitution across intermediate inputs are low---inputs are closer to complements than to substitutes \citep{Atalay2017}---so both the headline $\rho=-1$ and the Leontief limit fall in the economically relevant region.}

Pinning firm \(i\) and re-imposing the knock at each step gives a monotone, continuous update map. It has a greatest fixed point \(\mathbf h^{\star}(i)\) by the Knaster--Tarski theorem, and iteration from the healthy economy with entry \(i\) zeroed converges to that point under the continuity conditions stated in Appendix~\ref{app:esri-impl}. Settling at the greatest fixed point is the conservative choice: it attributes to the failure only the contraction the network forces, not deeper collapses that the same map would also sustain. The economic systemic risk of firm \(i\) is the settled, size-weighted output loss,
\begin{equation}\label{eq:esri}
\mathrm{ESRI}_i:=1-\frac{Y_\infty(i)}{Y_0}=\frac{\sum_j m_j\big(1-h^{\star}_j(i)\big)}{\sum_j m_j}\ \in[0,1]
\end{equation}
Since firm $i$'s own health is held at zero, $\mathrm{ESRI}_i$ is at least firm $i$'s own share of aggregate output, $m_i/\sum_jm_j$, with equality only if the failure reaches no other firm, and the ratio of $\mathrm{ESRI}_i$ to that share measures the network amplification of the failure. The two mechanisms feed this one construction different supply terms---the knock, the iteration, and the measure are identical---so CES and Leontief are compared on an equal footing and bracket the substitution assumption. The parameter $p$ is the one that matters. Economically it sets how much of the network is genuinely complementary: a fraction $p$ of edges are essential, with no substitute, and the rest are substitutable.\footnote{The parameter also governs cost, and the two mechanisms sit at opposite ends. Leontief is a threshold rule: supply health is the minimum over the essential inputs whose spend share clears $\theta=0.05$. Because a buyer's shares sum to one, only a handful of suppliers qualify---just $6.7\%$ of the full US network's $3.4\times10^8$ edges---so each step is a cheap scatter-$\min$ over the surviving $\sim\!23$ million. The CES map instead keeps every edge live: the substitutable fraction $1-p$ enters a sparse matrix--vector product over essentially the whole graph, and the essential fraction $p$ a CES power-mean that materializes a dense buffer. That buffer grows linearly in $p$, from a few million rows at $p=0.01$ to about $170\text{M}\times B$ for a batch of $B$ knocked firms at $p=0.5$. At that point it exceeds the GPU's $\sim\!86$~GB limit at any useful batch, so the engine must shrink $B$ and runtime rises faster than linearly. This is why the full-economy granular runs are pinned to small $p$.}

\begin{figure}[H]
  \centering
  \includegraphics[width=\linewidth]{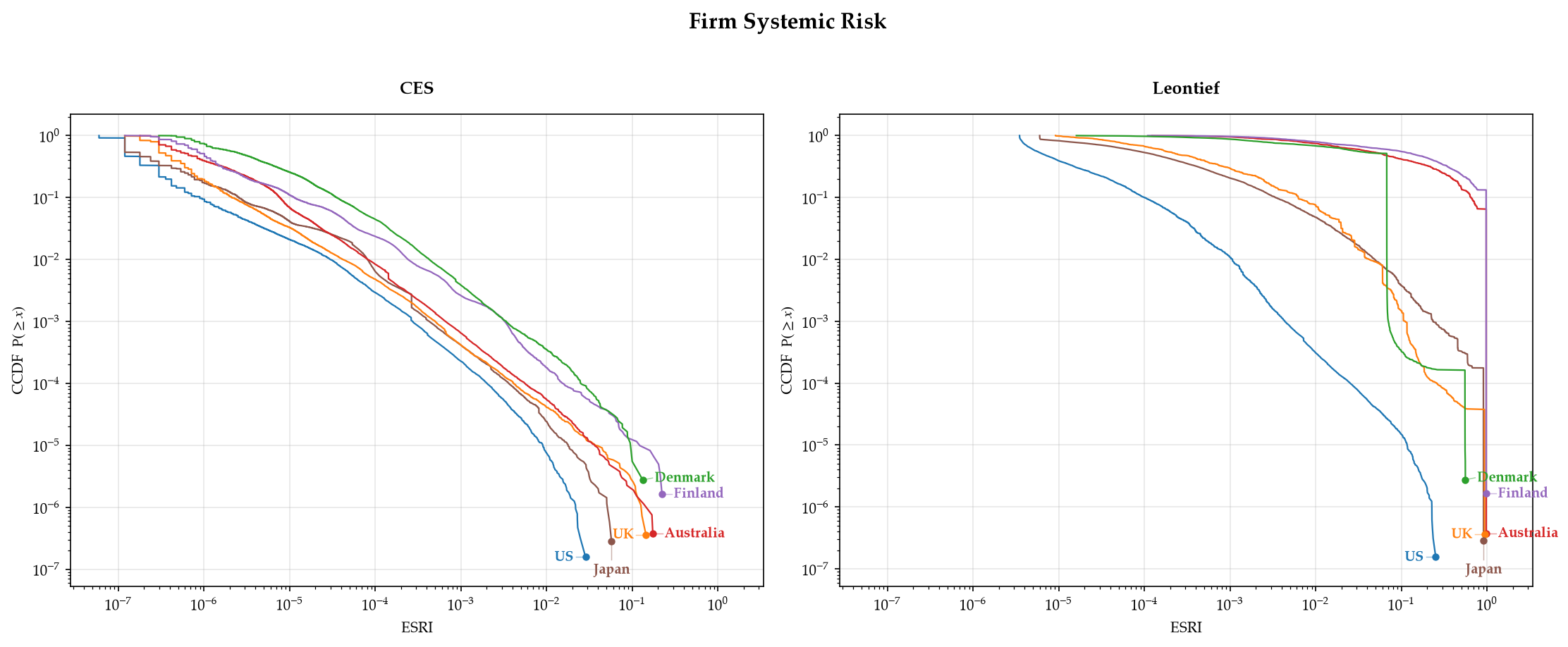}
  \caption{Counter-cumulative distribution of firm-level economic systemic risk $\mathrm{ESRI}_i$, one curve per economy, under CES input substitution (left, $\rho=-1$) and Leontief no-substitution (right). Under CES systemic risk is heavy-tailed and its tail orders the economies by granularity. Under Leontief it rises sharply and the economies separate: Australia and Finland approach a near-degenerate regime in which a large fraction of firms become critical, while in the United States risk stays concentrated in a heavier tail. Log axes.}
  \label{fig:esri_ccdf}
\end{figure}

Figure~\ref{fig:esri_ccdf} plots the distribution of $\mathrm{ESRI}$ across firms, and it shows how the two mechanisms differ at a glance. Under CES the distribution is heavy-tailed across several orders of magnitude: the overwhelming majority of firms carry negligible systemic risk, while the firms in a thin tail each subtract a substantial fraction of output on failing. The tail broadly orders the economies by their granularity (Table~\ref{tab:esri_tails}). Most firms are replaceable. A few are not, and the network decides which. The Leontief panel is the same experiment with the substitution cushion removed. With no substitution a single dead essential input can idle a firm, so systemic risk rises sharply and spreads beyond the thin CES tail. Removing substitution also pulls apart the six economies, which under CES had looked alike and uniformly resilient. Some, notably Australia and Finland, approach a near-degenerate regime in which a large fraction of firms become critical; in others, including the largest, the United States, risk stays concentrated in a heavier tail (Table~\ref{tab:esri_tails}). Substitutability, not the country's wiring, is thus the master variable of systemic fragility. A real economy lies between the two bounds: the same graph, read with and without the ability to substitute, produces both regimes.\footnote{Systemic risk stays most contained in the United States, whose sheer scale leaves even the no-substitution economy with enough redundancy that few firms are critical.} What such a failure looks like over time is traced in Appendix~\ref{app:esri} (Figures~\ref{fig:esri_dist}--\ref{fig:esri_ces_leontief_irf}).

\begin{table}[H]\centering
\caption{Mean firm-level systemic risk $\mathrm{ESRI}_i$ in the worst-$X\%$ tail (size-weighted), by economy and cascade mechanism. $\mathrm{ESRI}$ is the share of aggregate output lost when a firm fails. The worst-$X\%$ are the $X\%$ of firms with the highest systemic risk. Numeric companion to Figure~\ref{fig:esri_ccdf}.}
\label{tab:esri_tails}
\setlength{\tabcolsep}{6pt}\small
\begin{tabular}{@{}lrrrr@{}}
\toprule
Economy & mean & worst $10\%$ & worst $1\%$ & worst $0.1\%$ \\
\midrule
\multicolumn{5}{@{}l}{\emph{CES ($\rho=-1$)}}\\
United States  & $2.5\mathrm{e}{-6}$ & $1.9\mathrm{e}{-5}$ & $1.5\mathrm{e}{-4}$ & $9.2\mathrm{e}{-4}$ \\
Japan          & $4.5\mathrm{e}{-6}$ & $4.0\mathrm{e}{-5}$ & $2.7\mathrm{e}{-4}$ & $1.6\mathrm{e}{-3}$ \\
United Kingdom & $5.2\mathrm{e}{-6}$ & $4.3\mathrm{e}{-5}$ & $3.5\mathrm{e}{-4}$ & $2.5\mathrm{e}{-3}$ \\
Australia      & $7.4\mathrm{e}{-6}$ & $6.0\mathrm{e}{-5}$ & $4.5\mathrm{e}{-4}$ & $3.0\mathrm{e}{-3}$ \\
Finland        & $2.2\mathrm{e}{-5}$ & $2.0\mathrm{e}{-4}$ & $1.6\mathrm{e}{-3}$ & $9.5\mathrm{e}{-3}$ \\
Denmark        & $3.6\mathrm{e}{-5}$ & $3.0\mathrm{e}{-4}$ & $2.1\mathrm{e}{-3}$ & $1.2\mathrm{e}{-2}$ \\
\addlinespace
\multicolumn{5}{@{}l}{\emph{Leontief (no substitution)}}\\
United States  & $6.8\mathrm{e}{-5}$ & $5.5\mathrm{e}{-4}$ & $2.9\mathrm{e}{-3}$ & $1.3\mathrm{e}{-2}$ \\
Japan          & $2.8\mathrm{e}{-3}$ & $2.4\mathrm{e}{-2}$ & $1.3\mathrm{e}{-1}$ & $4.8\mathrm{e}{-1}$ \\
United Kingdom & $2.8\mathrm{e}{-3}$ & $2.2\mathrm{e}{-2}$ & $7.7\mathrm{e}{-2}$ & $1.8\mathrm{e}{-1}$ \\
Australia      & $2.0\mathrm{e}{-1}$ & $8.9\mathrm{e}{-1}$ & $9.8\mathrm{e}{-1}$ & $9.8\mathrm{e}{-1}$ \\
Finland        & $2.9\mathrm{e}{-1}$ & $9.8\mathrm{e}{-1}$ & $9.8\mathrm{e}{-1}$ & $9.8\mathrm{e}{-1}$ \\
Denmark        & $4.0\mathrm{e}{-2}$ & $6.9\mathrm{e}{-2}$ & $7.8\mathrm{e}{-2}$ & $1.7\mathrm{e}{-1}$ \\
\bottomrule
\end{tabular}
\end{table}

\subsection{Why Smaller or Major-Supplier-Only Networks Overstate Losses}
\label{subsec:esri-scale}
The first question is whether the size of the reconstructed graph matters. A skeptical answer would be no. If the object of interest is systemic risk, perhaps a network with $10^4$ or $10^5$ firms is already large enough. Or perhaps it is enough to observe the major suppliers of each firm and ignore the small links. Here we show that neither view is correct. Small networks and major-supplier-only networks both make the economy look more fragile than it is.

There are two reasons for this. The first is scale. The second is thresholding. These two margins are useful to separate because they correspond to the two ways in which real firm-to-firm data are usually incomplete. Many datasets observe only a subset of firms. And even for the firms they observe, they record only the large buyer-seller relationships. If either omission changes the distribution of economic systemic risk, then the full graph is not a luxury. It is part of the measurement.

\begin{figure}[H]
  \centering
  \includegraphics[width=\linewidth]{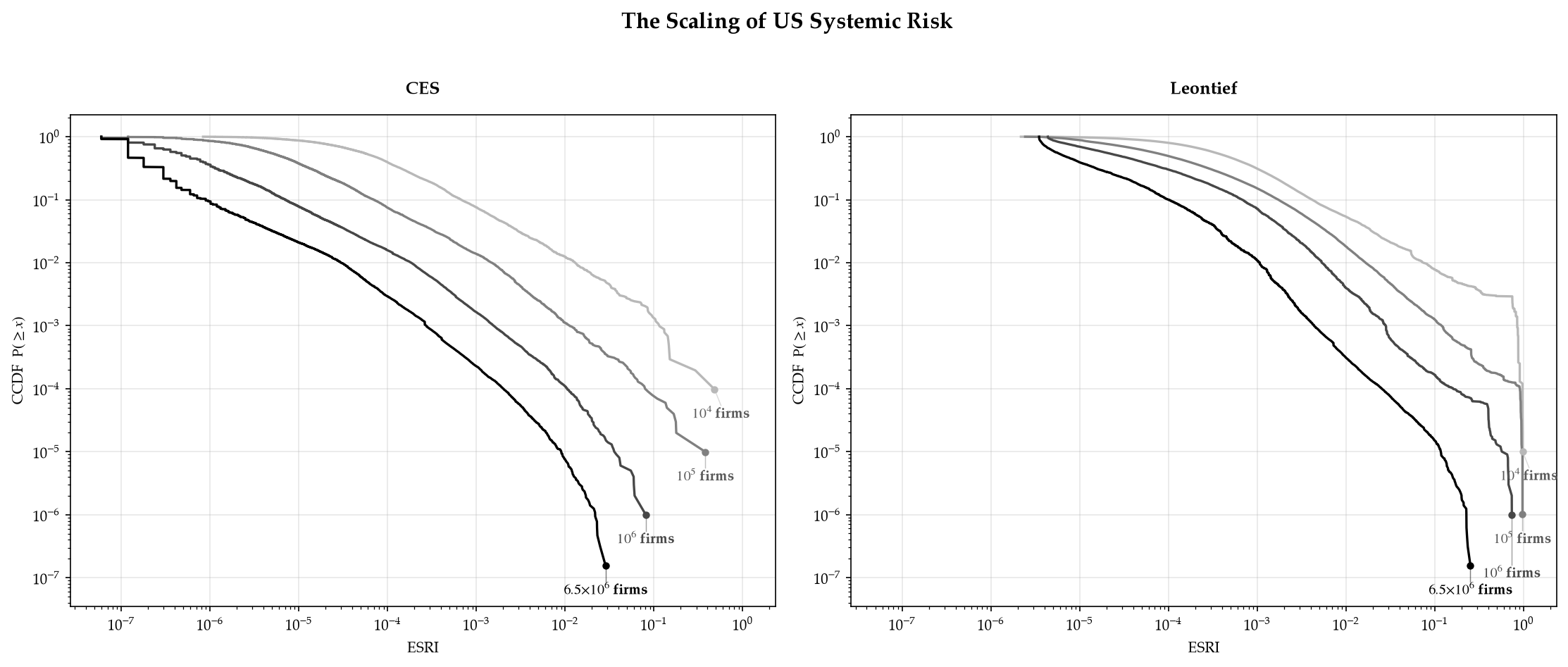}
  \caption{The scaling of US firm-level systemic risk with reconstruction size. The counter-cumulative $\mathrm{ESRI}$ distribution is shown at $N_F\in\{10^4,10^5,10^6,6.46\times10^6\}$ under CES (left) and Leontief (right). The distribution shifts left, toward lower systemic risk, as the economy is resolved to more firms, under both mechanisms: a coarse or sub-sampled network overstates per-firm systemic risk. Log axes.}
  \label{fig:esri_scaling}
\end{figure}

We begin with scale. Figure~\ref{fig:esri_scaling} recomputes the US ESRI distribution at four reconstruction sizes, $N_F=10^4,10^5,10^6$, and $6.46\times 10^6$. The result is simple. As the network is resolved into more firms, the ESRI distribution shifts left. This happens under both CES and Leontief. A small reconstruction packs the economy into too few nodes. Each node then carries too much output, and removing one of the large nodes appears to destroy too much of the economy. Adding the missing firms spreads the same economy over many more production units. The typical firm then matters less, and the upper tail of systemic risk becomes lighter.

Thus, the problem with a small reconstruction is not merely that it misses some firms. It changes the magnitude of the counterfactual. A $10^4$-firm network is not a harmless miniature of a $6.46\times 10^6$-firm economy. It is a coarser economy in which too much weight is concentrated on too few nodes. Hence an ESRI number computed on such a graph is biased upward. The level settles only when the graph is built at something close to the scale of the economy itself.

\begin{figure}[H]
  \centering
  \includegraphics[width=.66\linewidth]{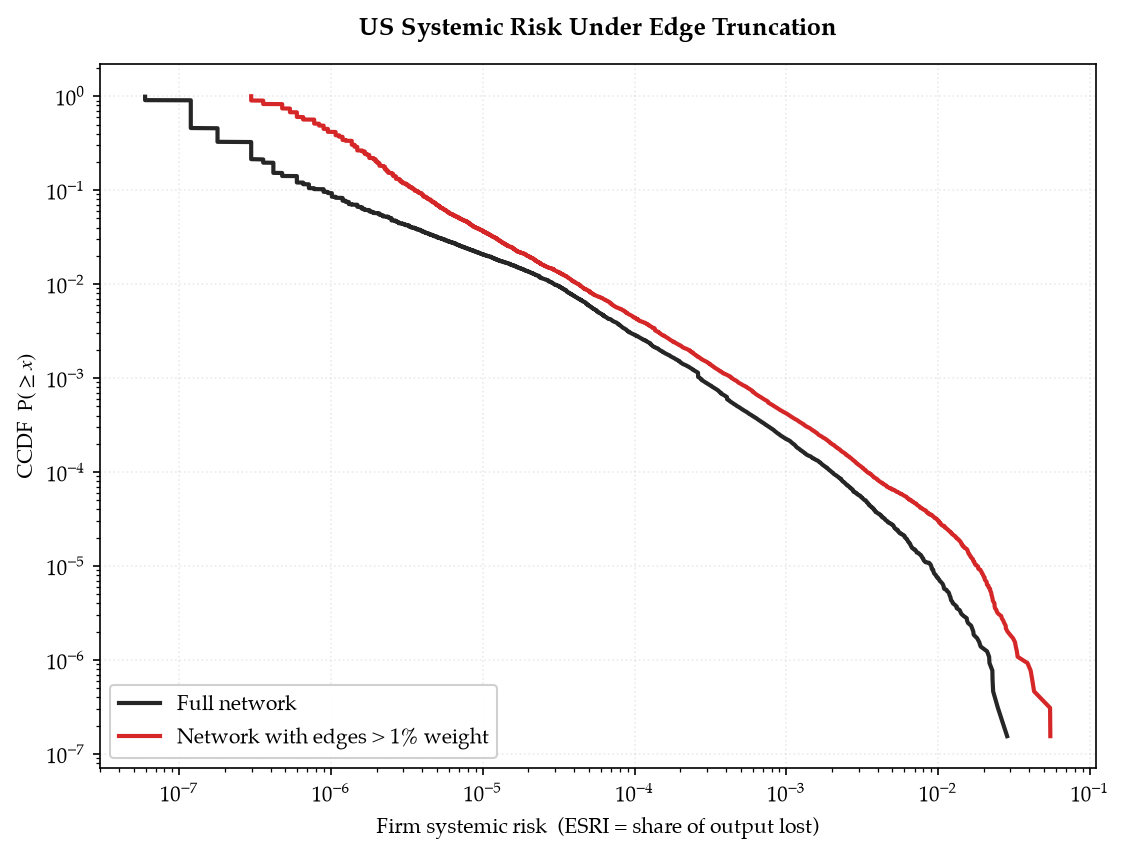}
  \caption{US firm-level systemic risk (CES) on the full network against the same network truncated to links above $1\%$ of the buyer's spend, the regime a thresholded real-world dataset occupies. Dropping the small suppliers removes substitution margins and shifts the whole distribution right: truncation overstates systemic risk by about $1.8\times$ in the tail (worst $0.1\%$: $1.7\times10^{-3}$ against $9.2\times10^{-4}$). Log axes.}
  \label{fig:esri_truncation}
\end{figure}

The second problem is thresholding. Figure~\ref{fig:esri_truncation} repeats the CES experiment on the full US network, but first deletes every link below $1\%$ of the buyer's spending. This is the kind of network one obtains when a dataset records only major suppliers. The effect is again in one direction. The ESRI distribution shifts right. In the worst $0.1\%$ of firms, systemic risk rises from $9.2\times 10^{-4}$ on the full graph to $1.7\times 10^{-3}$ on the thresholded graph, an increase of about $1.8$ times.

The reason is not mysterious. Small links are small as spending shares, but they are not irrelevant. They are substitution margins. When a supplier fails, these minor links give the buyer other routes through which production can continue. Deleting them makes the surviving major suppliers look more essential than they really are. It therefore turns a diversified buyer into a buyer with only a few apparent options. The same firm failure then produces a larger measured loss.

These two exercises give the same answer. If the graph is too small, systemic risk is too large. If the graph keeps only major suppliers, systemic risk is too large. And actual mapped production networks usually suffer from both problems at once. They miss firms, and they miss weak links among the firms they retain. The two biases therefore compound.

This is the first way a partial graph misleads: a smaller or thresholded network overstates systemic risk. A systemic-risk number read from a partial network should be interpreted as an upper bound, not as the risk of the fully resolved economy. The true economy is more diffuse, and therefore more resilient, than a small or thresholded network can show. Recovering that resilience requires putting back both the missing firms and the small links. This point matters even under CES, where substitution is present and the network should matter least. Under weaker substitution the case for the full graph only becomes stronger.

\subsection{Why No Sufficient Statistic Captures ESRI}
\label{subsec:no-sufficient-statistic}
Section~\ref{subsec:esri-scale} showed that the level of systemic risk depends on the whole graph. Does the ranking depend on it too? This is the harder question. A reader might grant that a small or thresholded network gets the level of risk wrong, but still think that a simple statistic identifies the risky firms. Perhaps the systemic firms are just the large firms. Or the firms with many customers. Or the firms in upstream sectors.

We test this directly. For each firm we compute its actual ESRI from the full removal experiment. We then ask how much of that number can be recovered from three cheaper objects: firm sales (weighted out-degree), out-degree, and a sector-level forward-linkage score. These are the natural candidates. Sales is the Hulten object. Out-degree is the simplest unweighted network object. The sectoral score is what remains when the firm network is collapsed back to an input--output table.

The comparison is deliberately generous to the reduced objects. For each predictor we do not impose a linear regression, nor a particular functional form. Instead, we fit the best monotone map from the predictor to ESRI, using Horvitz--Thompson-weighted isotonic regression. This gives the implied systemic risk $ \mathbb{E}[\mathrm{ESRI}\mid \text{predictor}] $, the most that any ranking of firms by that predictor can reproduce.

\begin{figure}[H]
  \centering
  \includegraphics[width=\linewidth]{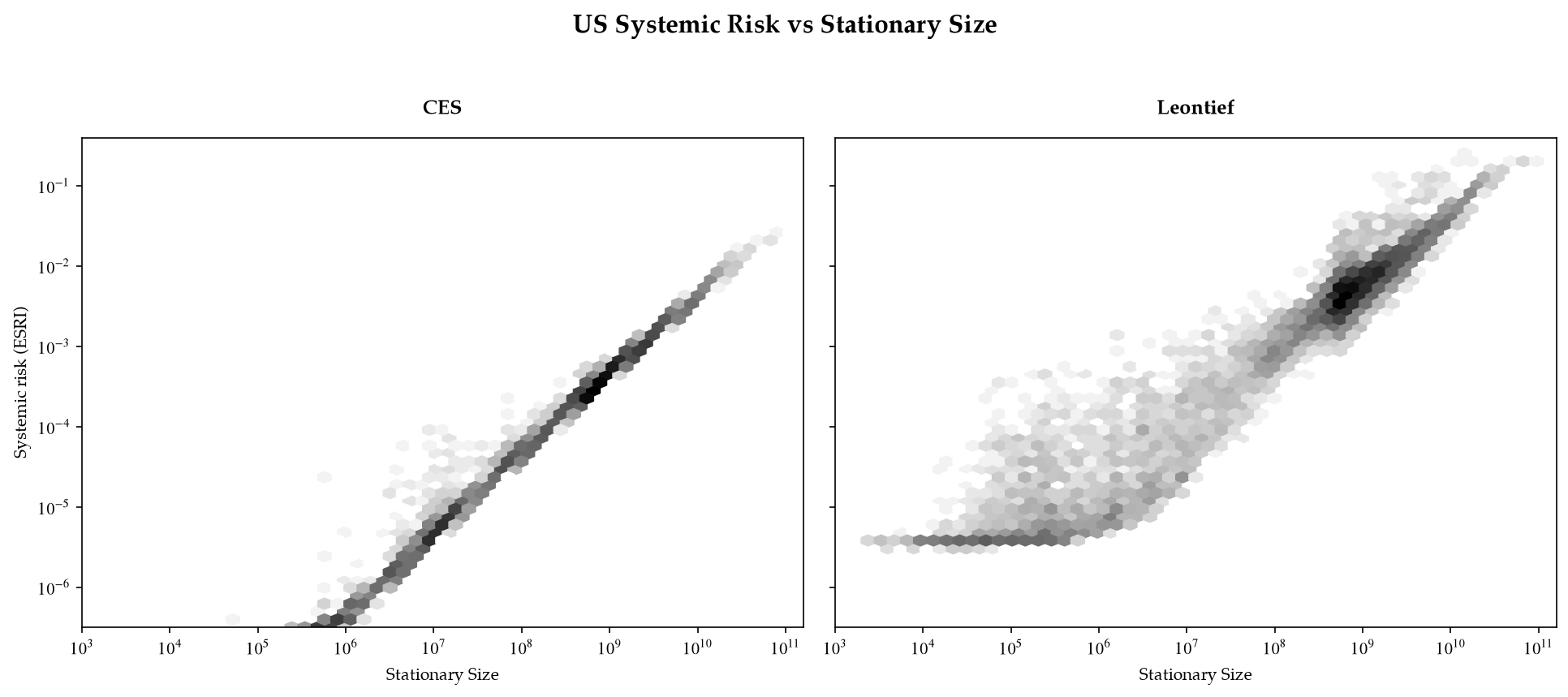}
  \caption{US firm-level systemic risk against stationary size, under CES and Leontief. Under CES,
  systemic risk rises with size but with wide scatter even above a floor, the Hulten tendency without a tight fit. Under Leontief,
  the relation loosens into a cloud: small but structurally pivotal firms can carry systemic risk that
  their size does not predict. Hexbin density, log axes.}
  \label{fig:esri_vs_size}
\end{figure}

CES is the most favorable case for the reduced objects. Inputs are substitutable, so a firm failure should behave close to a first-order perturbation. In that case firm sales does best among the three predictors. Firm sales is the Hulten object: to first order, a firm's aggregate importance is its Domar weight. Even in this most favorable case, however, firm sales is only a weak firm-level predictor. Because a firm's sales are its stationary size in the closed economy, this predictor is the size margin itself. Figure~\ref{fig:esri_vs_size} plots ESRI against it: the two are positively related but with wide scatter, not the tight monotone map a sufficient statistic would require.

\begin{figure}[H]
  \centering
  \includegraphics[width=\linewidth]{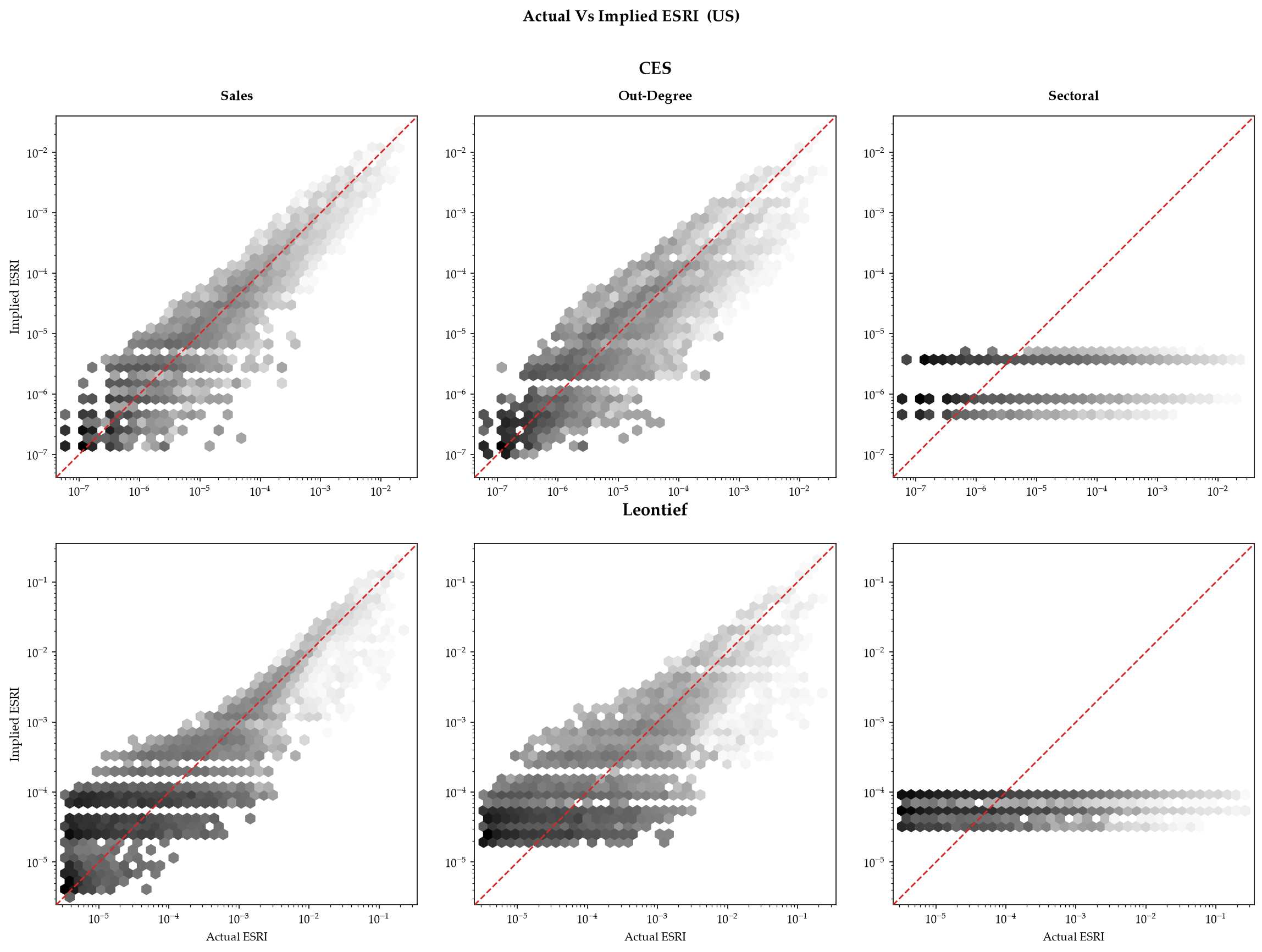}
  \caption{Actual firm-level ESRI against predictor-implied ESRI in the United States. The
  horizontal axis is the ESRI from the full removal experiment. The vertical axis is the best monotone
  prediction from the indicated reduced object. Hexagons are shaded by summed
  Horvitz--Thompson firm weight. The dashed line is the identity. A tight diagonal means good
  firm-level agreement. A horizontal band means the predictor cannot separate firms. Firm sales is
  informative under CES but deteriorates under Leontief. The sectoral score is nearly flat under both
  mechanisms because it is constant within sectors.}
  \label{fig:esri_predictor_implied}
\end{figure}

Figure~\ref{fig:esri_predictor_implied} plots actual against implied ESRI for all three predictors, and Table~\ref{tab:esri_proxy_errors} makes the gap quantitative. Even under CES, the best monotone map from sales leaves a typical per-firm error of about $0.46$ of the actual ESRI spread, a total-variation distance of $0.58$ from the true ESRI distribution, and a worst single-firm miss of about half the largest actual ESRI. On the table's $0$-to-$1$ scale---$0$ exact recovery, $1$ no better than assigning every firm the mean---sales sits near the middle. Out-degree performs worse. The sectoral score fails as a firm-level measure altogether. It assigns the same value to firms inside a sector and therefore cannot generate the firm-level tail that the full network produces.

\begin{table}[H]\centering
\caption{How well does each cheap predictor reproduce firm-level $\mathrm{ESRI}$? Bounded distances between the actual US $\mathrm{ESRI}$ distribution and the distribution implied by each predictor's best monotone (isotonic) map to $\mathrm{ESRI}$, HT-weighted, under CES and Leontief. Lower is closer. $0$ is exact recovery and $1$ is no better than the mean. Kuiper and total variation compare distribution shape. Normalized RMS is the paired per-firm error as a fraction of the actual spread. Normalized max is the worst single-firm miss as a fraction of the largest actual $\mathrm{ESRI}$.}
\label{tab:esri_proxy_errors}
\begin{tabular}{@{}lcccc@{}}
\toprule
Predictor & Kuiper & Total variation & Norm.\ RMS & Norm.\ max \\
\midrule
\multicolumn{5}{@{}l}{\emph{CES ($\rho=-1$)}}\\
Firm sales (weighted out-degree) & $0.46$ & $0.58$ & $0.46$ & $0.49$ \\
Out-degree (customer count)      & $0.46$ & $0.79$ & $0.70$ & $0.68$ \\
Sectoral forward-linkage score   & $0.89$ & $0.99$ & $1.00$ & $1.00$ \\
\addlinespace
\multicolumn{5}{@{}l}{\emph{Leontief (no substitution)}}\\
Firm sales (weighted out-degree) & $0.47$ & $0.83$ & $0.57$ & $0.87$ \\
Out-degree (customer count)      & $0.77$ & $0.95$ & $0.74$ & $0.90$ \\
Sectoral forward-linkage score   & $0.90$ & $0.99$ & $1.00$ & $1.00$ \\
\bottomrule
\end{tabular}
\end{table}

The Leontief case gives the sharper lesson. When inputs are complements, all three predictors deteriorate. The worst-firm miss of sales rises to $0.87$ of the largest actual ESRI. Figure~\ref{fig:esri_vs_size} shows why. Under CES, size and systemic risk lie close to a monotone curve. Under Leontief, that curve opens into a cloud. Some small firms become systemically important because they sit at chokepoints. They are not large. They need not have many customers. But they supply something essential to a part of the network that cannot easily re-source.

This is the sense in which the Leontief exercise is not merely a more dramatic version of the CES exercise. It changes what has to be measured. Under CES, firm size captures the direction of the first-order effect but still leaves large firm-level errors. Under Leontief, the relevant object is no longer just the firm's weight in aggregate sales. It is the set of higher-order paths that pass through the firm and the substitutability, or lack of substitutability, along those paths. That is the beyond-Hulten term: a firm removal is a large nonlinear shock, and the higher-order terms become more important as substitution falls.

The same pattern appears outside the United States. Appendix~\ref{app:esri} extends the sales-based comparison to all six economies. In every case, the sales-implied ESRI distribution moves farther from the actual distribution when we move from CES to Leontief. The failure of the reduced object is therefore not a feature of one reconstruction or one country. It is a feature of the mechanism. One qualification is important, however. The identity of the broad systemic set is not random. The systemic firms are largely upstream, input-supplying hubs, and this set is similar under CES and Leontief. In that sense the reconstruction produces a stable and interpretable ranking. But the point is exactly that this ranking is obtained from the weighted firm graph. It is not recovered from a sectoral table, from degree alone, or from sales alone. The conclusion is therefore not that simple statistics are useless. Sales carries the broad ordering of systemic firms, most of all under CES, but even there it is only a weak firm-level predictor. Out-degree contains less information. Sectoral position says something about broad upstreamness. But none is a sufficient statistic for firm-level systemic risk.

It is worth stating the principle behind these results. Hulten's theorem says that, to first order, the effect on aggregate output of a small productivity shock to firm $i$ is its Domar weight, the ratio of the firm's sales to GDP \citep{Hulten1978}. The result is an envelope argument: at an efficient equilibrium the induced reallocation of inputs across firms is second-order, so the network through which the shock travels drops out of the first-order term. A large amount of information collapses into a single number per firm, and that number is a sales margin, not a network object---it can be read from the size distribution without knowing which firms trade with which. In the first-order world the reduced statistics we tested are therefore close to sufficient by construction, and the reconstructed graph buys little. Our CES experiment sits near that world, with an elasticity close to the Cobb--Douglas case in which the theorem holds exactly, and this is why firm sales does best there. The lesson of the exercise is that even there it does so only loosely.

Two features of the systemic-risk question move it out of that world. First, a firm's failure is not a marginal shock: it forces the firm's output to zero, a large perturbation for which the first-order term is not a reliable approximation, so the second- and higher-order corrections of \citet{BaqaeeFarhi2019} enter. Second, those corrections are governed by the elasticities of substitution, and when inputs are complements they amplify rather than damp the loss. They also do not collapse to a scalar. They are sums over the higher-order paths that carry the shock---who supplies whom, how concentrated each buyer's sourcing is, and whether a lost input can be re-sourced further down the chain. These are properties of the weighted graph, not of the size margin. This is the precise sense in which granular networks matter when economic measurement moves beyond first-order Hulten logic. Under CES the beyond-Hulten term is small and sales, the first-order object, is a rough guide; under Leontief it dominates, systemic firms are set by chokepoints and reachability rather than by size, and no reduced statistic recovers them. The value of the reconstructed graph is thus concentrated exactly where the motivating questions lie: large, localized disruptions propagating through an economy in which inputs are hard to substitute. That is the regime in which sectoral aggregates and firm-size margins are least sufficient and the full weighted firm network is indispensable.

\section{Concluding Remarks} \label{sec:conclusion}
The motivation for this paper is simple. Production networks matter, but they are rarely observed. We showed that the missing network can be reconstructed from the two public objects it generates: the sectoral input--output table and the distribution of firm sizes by sector.  From these two objects alone, we reconstruct weighted, directed firm-to-firm networks at national scale for the United States, Japan, the United Kingdom, Australia, Finland, and Denmark.

Our reconstruction algorithm proceeds in two stages. First, a sector-aware gravity model assigns link probabilities to ordered firm pairs, and a Bernoulli draw realizes the binary backbone. We then regularize the drawn graph so that money can circulate through it as a primitive Markov chain. Second, a minimum-energy program puts weights on the realized links, subject to the observed firm-size and sector-flow restrictions. The result is not just a large graph. It is a weighted economy with a unique stationary distribution. The fitted one-step firm balances and sectoral flows satisfy the stated caps, and the stationary money vector remains close in aggregate when checked ex post. It also reproduces structure it was never shown: in every economy the customer (out-degree) tail comes out heavier than the supplier tail, even though a firm enters the model identically as buyer and as seller.

The algorithm is designed to preserve the objects central to the propagation of granular shocks: firm-size margins, degree-tail regimes, and stationary weights. It is not designed to reproduce every local motif of an observed production network. The independent-edge backbone cannot, by itself, generate the reciprocity, clustering, or degree correlation that may exist in a fully mapped economy. Where such local structure is the object of interest, this method is incomplete. Where the object is aggregate propagation through a weighted directed graph, it supplies a tractable national-scale stand-in for the missing network. Several choices in the paper are therefore best read as disciplined compromises. The mean degree is calibrated. The degree-tail shape is calibrated. The backbone is drawn with conditionally independent edges. The propagation exercise uses one particular closure of the inter-firm economy. These choices make the method scalable, but they also mark the places where it can be improved. Correlated edge draws would allow the reconstruction to match local clustering and reciprocity. And better data on firm degrees would discipline the degree-tail parameters.

Having developed the algorithm and reconstructed several national economies, we illustrated the networks' value through a canonical experiment: the aggregate output lost when an individual firm fails. For all six economies, the distribution of systemic risk is heavy-tailed. This is not new. The notable result is that a firm's size is not a good predictor of the systemic risk it poses. This illustrates a general difficulty of macroeconomic measurement. The micro realities that matter are often not properties of the individual units in isolation, but of how the units relate to one another through the web of relations we call the economy. It is precisely in the pursuit of such problems that our reconstructed networks can be a useful aid, perhaps even a routine laboratory for computational experiments.

National economies are not the largest scale at which the method could be used. A world-scale reconstruction with roughly \(10^7\)--\(10^8\) firms is conceivable, but the Bernoulli draw still requires quadratic pairwise work; only the sparse downstream stages are near-linear under bounded iteration counts and the sampling conditions above. Feasibility at that scale would therefore depend on parallel pair generation, memory, and the available hardware. The world input--output table is known. What is missing is the size distribution of firms linked to their sectoral code. If firms are further resolved into geographically located plants, the same graph could be used to study shocks that occur in space: floods, heatwaves, earthquakes, port closures, sanctions, or transport disruptions. Appendix~\ref{app:factory} explains how our algorithm can be extended to this geolocated factory setting. These are precisely the cases in which sectoral and national aggregates are least satisfactory, because the shock is local but the propagation is global. None of this is to suggest that our reconstructed networks are great substitutes for real-world production networks. They are good substitutes for having no network at all. Fortunately, the effort to assemble truly granular global production network data was set in motion some years ago \citep{INET2023Alliance}. If one were to gauge timelines from the effort that culminated in national and global input--output tables \citep{Leontief1953Studies}, the process is likely to span decades.  Until then, a reconstruction disciplined by the aggregates we observe is a placeholder worth having.

\clearpage
\bibliography{net_recon}
\bibliographystyle{aea}

\newpage
\appendix
\counterwithin{proposition}{section}
\counterwithin{lemma}{section}
\counterwithin{remark}{section}
\section{Notation} \label{app:notation}
Table~\ref{tab:notation} collects the notation used in the paper, grouped by the section in which each symbol is introduced. Symbols whose role is purely local (e.g.\ Lagrange multipliers inside a single proof) are flagged where they appear in the text.

\setstretch{1.05}
\begin{longtable}{@{} l p{0.76\linewidth} @{}}
\caption{Notation}\label{tab:notation}\\
\toprule
\textbf{Symbol} & \textbf{Description} \\
\midrule
\endfirsthead
\multicolumn{2}{@{}l}{\small\emph{Table~\ref{tab:notation} continued}}\\
\toprule
\textbf{Symbol} & \textbf{Description} \\
\midrule
\endhead
\midrule
\endfoot
\bottomrule
\endlastfoot
\multicolumn{2}{@{}l}{\emph{Indices and primitives}}\\
\addlinespace[2pt]
$i,j$ & firm indices, $i,j\in\{1,\dots,N_F\}$ \\
$k,\ell$ & sector indices, $k,\ell\in\{1,\dots,N_S\}$ \\
$N_F$, $N_S$ & number of firms; number of sectors \\
$\pi(\cdot)$ & map assigning each firm to its sector \\
$\Fset{\ell}$ & set of firms in sector $\ell$ \\
$u$ & index of a Bernoulli draw \\
$m_i$, $\mathbf m$ & size of firm $i$, normalized by the largest firm; vector of firm sizes \\
$s_\ell$ & empirical size of sector $\ell$, $s_\ell=\sum_{i\in\Fset{\ell}}m_i$ \\
\addlinespace
\multicolumn{2}{@{}l}{\emph{Sectoral data (Section~\ref{subsec:gravity})}}\\
\addlinespace[2pt]
$\mathbf{IO}$ & input--output table, buyer sectors in rows and seller sectors in columns \\
$\mathbf I$, $I_{k\ell}$ & row-stochastic input--output table: share of buyer-sector $k$'s expenditure allocated to seller-sector $\ell$ \\
$\sigma_{k\ell}$ & block flow-share, $IO_{k\ell}/\sum_{k'\ell'}IO_{k'\ell'}$ (the estimation target) \\
\addlinespace
\multicolumn{2}{@{}l}{\emph{Logistic-gravity model (Section~\ref{subsec:gravity})}}\\
\addlinespace[2pt]
$z$ & global density scalar, calibrated to the mean-degree target \\
$g(\cdot)$ & saturating role-symmetric firm fitness, $g(m)=m^{a}/(1+(m/m^\star)^{a})^{1-\eta}$ \\
$a$, $\eta$, $m^\star$ & fitness size elasticity, knee sharpness, and saturation knee (frozen per economy) \\
$\lambda_{k\ell}$, $\Delta$ & sector-pair multipliers on the unit simplex $\Delta$ \\
$x_{ij}$ & gravity intensity of the ordered firm pair $(i,j)$ \\
$p_{ij}$, $\mathbf P$ & link probability; matrix of link probabilities \\
$P_{k\ell}$, $\varphi_{k\ell}$ & expected links in block $(k,\ell)$; implied edge-share $P_{k\ell}/\sum P$ \\
$\mathcal B$, $\mathcal B_{\mathrm{tail}}(\lambda)$ & non-empty sector blocks; the $\lceil q_{\mathrm{tail}}|\mathcal B|\rceil$ blocks with the largest squared residuals $(\varphi_{k\ell}(\lambda)-\sigma_{k\ell})^2$, with ties broken by a fixed sector-pair ordering; $\mathcal B_{\mathrm{tail}}$ in \eqref{eq:lambda-program} denotes this set \\
$\bar d$ & target mean degree ($=50$) \\
$G_k$; $B_k$, $S_k$ & fitness mass of sector $k$; buyer-side and seller-side sector masses, with sparse limits $\widetilde B_k$, $\widetilde S_k$ (Proposition~\ref{prop:tail_inheritance}) \\
$\delta_{\mathrm{den}}$, $\tau_{\mathrm{tail}}$, $q_{\mathrm{tail}}$ & density band; worst-tail mismatch cap; tail fraction $q_{\mathrm{tail}}\in(0,1]$ (estimation) \\
\addlinespace
\multicolumn{2}{@{}l}{\emph{Bernoulli backbone (Section~\ref{subsec:bernoulli})}}\\
\addlinespace[2pt]
$\mathbf A^{(u)}$, $a^{(u)}_{ij}$ & adjacency matrix of draw $u$ and its entries; $\mathbf A=(a_{ij})$ denotes a generic draw \\
$\mathcal E$ & number of edges of a draw \\
$\mu_{\mathcal E}$, $\sigma_{\mathcal E}^2$ & mean and variance of $\mathcal E$ conditional on $\mathbf P$ \\
$d_i^{\mathrm O}$, $d_i^{\mathrm I}$ & out-degree and in-degree of firm $i$ in the stored money-flow adjacency $\mathbf A$; hence $d_i^{\mathrm O}$ is a supplier count and $d_i^{\mathrm I}$ is a customer count \\
$F^{\mathrm O}(h)$ & empirical out-degree probability mass function of a draw \\
$\vartheta_i^{\mathrm O}(h)$ & probability that firm $i$ has out-degree $h$ \\
$\mathcal H$, $\Sigma_{\mathrm O}(\mathcal H)$ & finite set of degree bins; asymptotic covariance matrix over $\mathcal H$ \\
$\xi$ & degree-floor spread (tilt); $\xi<0$ bunches repaired firms closer to the floor, $\xi>0$ spreads them farther above it (Section~\ref{subsec:degreefloor}) \\
\addlinespace
\multicolumn{2}{@{}l}{\emph{Markov closure (Section~\ref{subsec:markov})}}\\
\addlinespace[2pt]
$\mathcal C$, $V(c)$ & set of strongly connected components; firms in component $c$ \\
$\mathbf A_C$ & condensation directed acyclic graph \\
$\mathscr T$, $\mathscr U$ & source and sink components of the condensation \\
$\mathcal P$ & collection of ordered sink-source pairs $(c,c')$ receiving new arcs \\
$n_{cc'}$ & size of the smaller component in the pair $(c,c')$ \\
$f_\nu$, $\theta$, $\nu$ & saturating link-fraction function and its parameters \\
$k_{cc'}$ & number of links added from component $c$ to component $c'$ \\
$g_{\nu'}$, $\bar\gamma$, $n_0$, $\nu'$ & candidate-thinning function and its parameters ($\bar\gamma$ is distinct from the primitivity index $\gamma$) \\
$\Omega_{cc'}$, $\Omega$ & candidate sets; global candidate set \\
$y_{ij}$ & binary decision to activate candidate edge $(i,j)$ \\
$D_\ell$ & sectoral placement residual for sector $\ell$ \\
$d$, $\operatorname{CYC}(\cdot)$ & period of the irreducible closure; cyclic classes $\operatorname{CYC}(i)=h(i)\bmod d$ \\
$\Omega^{\mathrm{ap}}$, $(i^\star,j^\star)$ & aperiodicity candidate set; the single edge added to break periodicity \\
$\Ahat$, $\Aaug$ & irreducible closure; irreducible and aperiodic closure \\
\addlinespace
\multicolumn{2}{@{}l}{\emph{Weights and stationary money (Section~\ref{subsec:weights})}}\\
\addlinespace[2pt]
$\mathbf W$, $w_{ij}$ & row-stochastic weight matrix on the support of $\Aaug$ \\
$\widehat s_\ell(\mathbf W)$ & model-implied size of sector $\ell$ \\
$e_j$ & relative (percentage) one-step firm-size deviation at firm $j$ \\
$J$ & size-weighted root-mean-square relative deviation, $\big(\sum_j\mu_j e_j^2\big)^{1/2}$ \\
$\mathrm{RMS}_q$ & root-mean-square over the worst fraction $q$ of entries \\
$\delta^{\mathrm{firm}}$, $\delta^{\mathrm{tail}}$, $q$ & firm size-weighted RMS cap; firm worst-$q$ tail RMS cap; tail fraction \\
$\delta^{\mathrm{sec}}$, $\delta^{\mathrm{sec}}_{\mathrm{tail}}$, $\varepsilon_{\mathrm{floor}}$ & sector overall and worst-$q$ RMS caps; per-edge weight floor \\
$w_{\min}$ & smallest realized edge weight (Appendix~\ref{app:statmoney}) \\
$M$, $\boldsymbol\mu$ & total money $\mathbf 1^\top\mathbf m$; normalized firm-size vector $\mathbf m/M$ \\
$\mathbf r$ & one-step residual $\boldsymbol\mu-\mathbf W^\top\boldsymbol\mu$ \\
$\tau(\mathbf W)$ & Dobrushin ergodicity coefficient \\
$\Theta(\mathbf W)$ & ergodic sum $\sum_{t\ge0}\tau(\mathbf W^t)$ of the Dobrushin coefficients of the powers of $\mathbf W$ (Proposition~\ref{rem:stationary_vs_onestep}) \\
$\alpha_i$, $\gamma_j$ & buyer constant and seller balance price in the minimum-energy weights (Proposition~\ref{prop:kkt}) \\
$I^\star_{k\ell}$ & expenditure shares of the balanced inter-firm table $\mathbf F^\star$ (Section~\ref{subsec:interfirm}) \\
$\beta_{ij}$, $w^{\mathrm b}_{ij}$ & expected-balance weight factor and its row-stochastic version (Appendix~\ref{app:weights_structure}) \\
$N^{+}(S)$, $N^{-}(T)$ & sellers adjacent to a set $S$ of buyers; buyers adjacent to a set $T$ of sellers (Appendix~\ref{app:statmoney}) \\
$\gamma$ & primitivity index of $\mathbf W$, the least $t$ with $\mathbf W^{t}>0$ (Appendix~\ref{app:statmoney}) \\
\addlinespace
\multicolumn{2}{@{}l}{\emph{Computational complexity (Appendix~\ref{sec:compu})}}\\
\addlinespace[2pt]
$B$ & number of size bins per sector \\
$R_{\mathrm{bin}}$, $N_R$ & binning resolution (bins per unit log-size, $=32$); number of representative firms in the collapse \\
$E$ & number of realized edges \\
$I_{\mathrm{NLP}}$, $I_{\mathrm{FW}}$, $I_{\mathrm{QP}}$ & iterations of the gravity NLP; of the relaxed Markov-closure surrogate; of the weighting QP \\
$P_{\mathrm{gpu}}$ & concurrent GPU threads in the parallel Bernoulli draw (Appendix~\ref{sec:compu}) \\
\addlinespace
\multicolumn{2}{@{}l}{\emph{Census reconciliation and reconstruction fidelity (Sections~\ref{subsec:interfirm} and~\ref{subsec:fidelity})}}\\
\addlinespace[2pt]
$\kappa_\ell$ & inter-firm share of sector $\ell$ (fraction of output sold to other firms, not to final demand) \\
$r_i$, $R_\ell$ & firm $i$'s book receipts; sector $\ell$'s book-receipt total \\
$\tilde r_i$, $M_\ell$ & inter-firm receipts of firm $i$; inter-firm size of sector $\ell$ \\
$\mathbf F^\star$, $F^\star_{k\ell}$ & RAS-balanced inter-firm flow target and its entries \\
$\mathbf F_{\mathrm{eq}}$ & sectoral flows at the closed-market stationary state \\
$\widehat{\boldsymbol\pi}$, $\delta_j$ & stationary firm-size vector $\widehat{\boldsymbol\pi}=M\mathbf v$; per-firm drift $\delta_j=(\widehat\pi_j-m_j)/m_j$ (Appendix~\ref{app:robustness}) \\
$\zeta$ & tail exponent of the sector Domar-weight (sales-share) distribution, Gabaix--Ibragimov (Table~\ref{tab:recon_fidelity}) \\
\addlinespace
\multicolumn{2}{@{}l}{\emph{The production cascade and stationary vector (Section~\ref{sec:esri})}}\\
\addlinespace[2pt]
$\chi$ & degree-tail exponent (counter-cumulative CCDF index, estimated by the Hill estimator) \\
$\beta$ & tail exponent of the firm-size distribution (Hill top-$10\%$; Table~\ref{tab:eqsize}) \\
$\mathbf v$, $\|\mathbf v\|_2$ & stationary distribution of the money chain $\mathbf W$ (left Perron vector, $\mathbf v=\mathbf W^\top\mathbf v$, $\mathbf 1^\top\mathbf v=1$), equal to the Domar weights normalized to sum to one; its norm $\|\mathbf v\|_2=\sqrt{\mathrm{HHI}}$ \\
$h_j$, $h^{\star}_j(i)$ & normalized firm output (``health''), $h_j\in[0,1]$; its settled value with firm $i$ pinned to zero \\
$s$ & intermediate-input share of output, $=0.8$; $1-s$ is value added and $1/(1-s)=5$ the Domar multiplier \\
$Y$, $Y_0$, $Y_\infty(i)$ & aggregate output $Y=\sum_j m_j h_j/\sum_j m_j$; undisturbed baseline $Y_0=1$; settled value with firm $i$ pinned \\
$|\lambda_2|$ & subdominant-eigenvalue modulus of $\mathbf W$; spectral gap $=1-|\lambda_2|$ (Table~\ref{tab:lambda2}) \\
\addlinespace
\multicolumn{2}{@{}l}{\emph{Firm-level systemic risk (Section~\ref{sec:esri})}}\\
\addlinespace[2pt]
$\rho$, $p$, $\theta$ & CES order, per-edge essential probability, and Leontief essentiality threshold (Appendix~\ref{app:esri-impl}); headline $\rho=-1$, $p=0.01$; linear/Hulten $p\to0$, Leontief $\rho\to-\infty$ or $w_{jr}\ge\theta=0.05$ \\
$\mathrm{ESRI}_i$ & economic systemic risk of firm $i$: the settled size-weighted output loss when $i$'s output is pinned to zero, $1-Y_\infty(i)/Y_0$ \\
$h^{\mathrm{sup}}_j$, $h^{\mathrm{dem}}_j$ & supply- and demand-side output constraints on firm $j$; $\Phi(\mathbf h)_j=\operatorname{clip}(\min(h^{\mathrm{sup}}_j,h^{\mathrm{dem}}_j))$ \\
\addlinespace
\multicolumn{2}{@{}l}{\emph{Probability, asymptotics, and the appendix}}\\
\addlinespace[2pt]
$\Rightarrow$, $\xrightarrow{\mathbb P}$ & convergence in distribution and in probability, as $N_F\to\infty$ \\
$\mathbb E_{\mathbf P}$, $\mathbb P_{\mathbf P}$ & expectation and probability conditional on $\mathbf P$ \\
$\mathbf 1$, $\mathbf I_{N_F}$, $\mathbf 1_{\{\cdot\}}$ & vector of ones; identity matrix; indicator function \\
$\|\cdot\|_1$, $\|\cdot\|_F$ & $L^1$ norm; Frobenius norm \\
$\mu_i$, $\sigma_i^2$ & mean and variance of firm $i$'s degree (local to Appendix~\ref{appendix}; distinct from the size share $\mu_j$ of Section~\ref{subsec:weights}) \\
$q_v$, $L$ & degree thresholds and their number (Appendix~\ref{appendix}) \\
$\varrho$ & confidence level (Appendix~\ref{appendix}) \\
$\mathbf C$, $\mathbf d$, $\mathcal R_\eta$, $\eta$ & local balance operator, target, admissible residual set, and its Euclidean radius (Appendix~\ref{appendix}; this local $\eta$ is distinct from the fitness knee $\eta$ of Section~\ref{subsec:gravity}) \\
$\mathbf N_{\mathcal A}$, $\sigma_{\min}(\mathbf N_{\mathcal A})$ & active constraint matrix of the exact weighting problem and its smallest singular value, whose inverse is the sensitivity constant (Appendix~\ref{appendix}) \\
\end{longtable}
\setstretch{1.3}

\section{Mathematical Appendix} \label{appendix}

This appendix collects the proofs of the tail-inheritance proposition, the edge-count and degree central limit theorems, and the aperiodic-closure lemma of Section~\ref{sec:model}. It also collects three further results: finite-sample (Bernstein-type) bounds on a single Bernoulli draw, a local sensitivity bound on the weighted matrix, and the structure of the minimum-energy weights. The stationary-money bound (Proposition~\ref{rem:stationary_vs_onestep}) is a weaker guarantee whose real force is computational. We prove it in Appendix~\ref{app:robustness}, next to the evidence that backs it.

\subsection{Proofs of the Results in Section~\ref{sec:model}}

Throughout this appendix, degrees are taken with respect to the stored money-flow adjacency $\mathbf A$, where $i\to j$ means that buyer $i$ buys from seller $j$. Thus $d_i^{\mathrm O}$ is a supplier count and $d_i^{\mathrm I}$ is a customer count; the goods-flow convention used for reporting degrees reverses this orientation.

\begin{proposition}[Degree tails inherit the size tail]
\label{prop:tail_inheritance}
Fix the kernel \eqref{eq:kernel} with pure-power fitness $g(m)=m^{a}$ (equivalently, $\eta=1$),\footnote{With a saturation knee ($0<\eta<1$, $m^\star<1$), the same argument applied on the far-tail range $m\gg m^\star$, where $g(m)=m^{a\eta}(m^\star)^{a(1-\eta)}(1+o(1))$, yields the strict far-tail index $\beta/(a\eta)$. At $\eta=0$, fitness is asymptotically constant above the knee, so no finite power-law index is asserted there. Thus the pure-power relation $\chi=\beta/a$ is the calibration relation for the unsaturated, or empirically relevant, range where $g(m)$ behaves like $m^a$; the saturation deliberately steepens the extreme tail.} multipliers $\lambda$, and density $z$. Let $G_k:=\sum_{i\in\Fset{k}}g(m_i)$ be sector $k$'s fitness mass, and define the buyer-side and seller-side masses of sector $k$,
\[
B_k:=\sum_{\ell}\lambda_{k\ell}\,G_\ell,
\qquad
S_k:=\sum_{\ell}\lambda_{\ell k}\,G_\ell
\]
\begin{enumerate}[label=(\alph*), leftmargin=*]
\item Let $\bar x_i:=\max_{j\ne i}x_{ij}$ and $\bar x^{\mathrm I}_i:=\max_{j\ne i}x_{ji}$. Conditional on the sizes, the expected supplier count of a firm $i$ in sector $k$ satisfies
\[
\frac{z\,m_i^{a}\big(B_k-\lambda_{kk}m_i^{a}\big)}{1+\bar x_i}
\;\le\;
\mathbb E_{\mathbf P}\big[d_i^{\mathrm O}\big]
\;\le\;
z\,m_i^{a}\big(B_k-\lambda_{kk}m_i^{a}\big)
\]
and the expected customer count satisfies the same bounds with $S_k-\lambda_{kk}m_i^{a}$ in place of $B_k-\lambda_{kk}m_i^{a}$ and $\bar x^{\mathrm I}_i$ in place of $\bar x_i$.
\item Suppose in addition that raw firm sizes are independent draws whose upper tail is regularly varying with index $\beta>a$, $\Pr(\Xi>x)=L(x)\,x^{-\beta}$ with $L$ slowly varying. The reconstruction normalizes these to $m_i=\Xi_i/M_N$ with $M_N=\max_r\Xi_r$; under the pure-power kernel the common factor $M_N$ only rescales the link intensity and is absorbed into the density $z$, so we state the limit in the raw scale $\Xi$, which carries the cross-firm tail variation. Let $I_N$ be a firm drawn uniformly at random from sector $k$. Pass to the sparse triangular-array limit in which
\[
\frac{zB_k}{M_N^a}\xrightarrow{\mathbb P}\widetilde B_k\in(0,\infty),
\qquad
\frac{zS_k}{M_N^a}\xrightarrow{\mathbb P}\widetilde S_k\in(0,\infty),
\]
\[
\max_{j\neq I_N}p_{I_Nj}\xrightarrow{\mathbb P}0,
\qquad
\max_{j\neq I_N}p_{jI_N}\xrightarrow{\mathbb P}0,
\]
and, for every sector $\ell$, $\max_{r\in\Fset{\ell}}g(m_r)/G_\ell\xrightarrow{\mathbb P}0$. Then the supplier count of $I_N$ converges in distribution to a mixed Poisson law with mixing variable $\widetilde B_k\Xi^{a}$, where $\Xi$ is that firm's random size in this asymptotic scale. And its customer count converges to a mixed Poisson law with mixing $\widetilde S_k\Xi^{a}$.
\item Consequently both limiting degree distributions are regularly varying with the \emph{same} counter-cumulative tail index,
\[
\chi\;=\;\beta/a
\]
on the supplier and customer sides alike, in every sector and in any finite mixture of sectors. The masses $\widetilde B_k$ and $\widetilde S_k$ enter only the tail prefactors, as $\widetilde B_k^{\,\beta/a}$ and $\widetilde S_k^{\,\beta/a}$.\footnote{If within-sector size laws carry different indices $\beta_k$, the statement holds sector by sector and each side's aggregate index is $\min_k\beta_k/a$. The heaviest-tailed sector dominates. The hypothesis $\beta>a$ ensures that the mean degree is finite ($\mathbb E[\Xi^{a}]<\infty$), and it holds for every economy we reconstruct. Equivalently $\chi=\beta/a>1$ throughout Table~\ref{tab:country_powerlaw}.}
\end{enumerate}
\end{proposition}

\begin{proof}[Proof of Proposition~\ref{prop:tail_inheritance}]
(a) Since $p_{ij}=x_{ij}/(1+x_{ij})$ and $0\le x_{ij}\le\bar x_i$ for every $j\ne i$, every entry of row $i$ satisfies $x_{ij}/(1+\bar x_i)\le p_{ij}\le x_{ij}$. With $g(m)=m^{a}$, summing over sellers gives $\sum_{j\neq i}x_{ij}=z\,m_i^{a}\big(B_k-\lambda_{kk}m_i^{a}\big)$, the subtracted self-term being negligible under $\max_{r\in\Fset{k}}g(m_r)/G_k\to0$, and the display follows. The customer side is the same computation on the transpose, with $S_k-\lambda_{kk}m_i^{a}$ in place of $B_k-\lambda_{kk}m_i^{a}$. In the triangular-array regime of part (b) the self-pair correction is negligible and $\bar x_i\to0$, so both expected degrees are the same power of own size, scaled by a sector mass.

(b) Condition on the sampled firm's size $\Xi=m$ in the asymptotic scale just described. Conditioning on a single size leaves the three limit conditions of the statement in force, because the no-dominant-firm condition makes any one firm's contribution to $B_k$, $S_k$ and $G_\ell$ negligible. Its supplier indicators are independent Bernoulli variables. Let $\bar x^{\mathrm O}_{I_N,N}:=\max_{j\neq I_N}x_{I_Nj}$. Since $x_{ij}=p_{ij}/(1-p_{ij})$, the displayed incident-dyad condition gives $\bar x^{\mathrm O}_{I_N,N}\xrightarrow{\mathbb P}0$. Applying the sandwich in part (a) to the sampled row, its conditional mean satisfies $\mu_N(m)\xrightarrow{\mathbb P}\widetilde B_km^{a}$, with the self-pair correction vanishing by the no-dominant-firm condition. Le Cam's inequality bounds the total-variation distance between the supplier count and the Poisson law with the same mean by
\[
\sum_jp_{I_Nj}^2
\le
\max_{j\neq I_N}p_{I_Nj}\,\mu_N(m)
\xrightarrow{\mathbb P}0
\]
\citep{LeCam1960}. Hence the conditional law converges to $\mathrm{Poisson}(\widetilde B_km^{a})$. For every bounded test function $\varphi$, the corresponding conditional expectations are uniformly bounded by $\|\varphi\|_\infty$; convergence in probability therefore also gives convergence after averaging over the size law. This yields the mixed Poisson limit. The customer count is identical, using the sampled column and $\widetilde S_k$. Finiteness of $\mathbb E[\Xi^{a}]$, hence of the mean degree, is ensured by the hypothesis $\beta>a$.

(c) Let $D$ be mixed Poisson with mixing $\Lambda=\widetilde B_k\Xi^{a}$. By composition, $\Pr(\Lambda>t)=\Pr\big(\Xi>(t/\widetilde B_k)^{1/a}\big)$ is regularly varying with index $-\beta/a$ and prefactor scale $\widetilde B_k^{\,\beta/a}$. Fix $\varepsilon\in(0,1)$. Poisson concentration gives, for $\lambda\le(1-\varepsilon)t$, $\Pr\big(\mathrm{Po}(\lambda)>t\big)\le e^{-\delta(\varepsilon)t}$ with $\delta(\varepsilon)=-(\varepsilon+\log(1-\varepsilon))>0$ (Chernoff, the bound monotone in $\lambda$). And for $\lambda\ge(1+\varepsilon)t$, $\Pr\big(\mathrm{Po}(\lambda)\le t\big)\le e^{-c(\varepsilon)t}$ with $c(\varepsilon)=\varepsilon-\log(1+\varepsilon)>0$, the lower Chernoff bound, which is monotone in $\lambda$. Splitting on the mixing variable,
\[
\Pr\big(\Lambda>(1+\varepsilon)t\big)\big(1-e^{-c(\varepsilon)t}\big)\;\le\;\Pr(D>t)\;\le\;\Pr\big(\Lambda>(1-\varepsilon)t\big)+e^{-\delta(\varepsilon)t}
\]
and regular variation turns the outer probabilities into $(1\pm\varepsilon)^{-\beta/a}\,\Pr(\Lambda>t)\,(1+o(1))$ while the exponential terms vanish against any power. Letting $\varepsilon\downarrow0$ gives tail equivalence, $\Pr(D>t)\sim\Pr(\Lambda>t)$. The degree tail carries the index $\beta/a$ and the sector mass only the prefactor. A finite mixture across sectors of regularly varying tails with a common index retains that index, the prefactors adding.
\end{proof}

The three limit conditions in part (b) of Proposition~\ref{prop:tail_inheritance} are conditions on random quantities. We now show that they follow from a finite $a$-th moment of the size law together with the mean-degree calibration, with explicit constants. Regular variation with $\beta>a$, which implies the finite moment, is needed only for the tail index in part (c). Write $\pi_k$ for the limiting share of firms in sector $k$, $\bar\xi_a:=\mathbb E[\Xi^{a}]$, $b_k:=\sum_\ell\lambda_{k\ell}\pi_\ell$, $c_k:=\sum_\ell\lambda_{\ell k}\pi_\ell$, and $\Lambda:=\sum_{k,\ell}\lambda_{k\ell}\pi_k\pi_\ell$.

\begin{proposition}[The sparse triangular-array conditions under the calibration]
\label{prop:sparse_constants}
Let raw sizes be independent draws from a common law with $\mathbb E[\Xi^{a}]<\infty$, let $|\Fset k|/N_F\to\pi_k>0$ for every sector, let $g(m)=m^{a}$, fix $\lambda\in\Delta$ with $\Lambda>0$, and let $z=z_N$ be the root of \eqref{eq:zcalib}. Then the three limit conditions of Proposition~\ref{prop:tail_inheritance}(b) hold, with
\[
\widetilde B_k=\frac{\bar d\,b_k}{\bar\xi_a\Lambda}
\qquad\text{and}\qquad
\widetilde S_k=\frac{\bar d\,c_k}{\bar\xi_a\Lambda}.
\]
\end{proposition}

Since $\sum_k\pi_kb_k=\sum_k\pi_kc_k=\Lambda$, the two families of constants have the same $\pi$-mean $\bar d/\bar\xi_a$, so the limiting degree of a uniformly drawn firm has mean $\mathbb E[\widetilde B_{\pi(I)}\Xi^{a}]=\bar d$ and the limit reproduces the calibrated mean degree. The constants also quantify the prefactor asymmetry of part (c). The customer and supplier tails share the index $\beta/a$, and their prefactors are $C_S=\sum_k\pi_k\widetilde S_k^{\beta/a}$ and $C_B=\sum_k\pi_k\widetilde B_k^{\beta/a}$. Because $x\mapsto x^{\beta/a}$ is convex and the two families have the same mean, Jensen's inequality gives $C_S\ge C_B$ whenever the seller-side masses are a mean-preserving spread of the buyer-side masses. The customer survival function then lies above the supplier survival function by the factor $C_S/C_B$ at every large degree.

\begin{proof}[Proof of Proposition~\ref{prop:sparse_constants}]
Fix $\lambda$ and write $\kappa:=z/M_N^{2a}$, so that $x_{ij}=\kappa\lambda_{\pi(i)\pi(j)}\Xi_i^{a}\Xi_j^{a}$, and let $S_N(\kappa):=N_F^{-1}\sum_{i\ne j}p_{ij}$, which is continuous and strictly increasing in $\kappa$. Three facts about the sizes are used. First, the law of large numbers gives $N_F^{-1}\sum_{r\in\Fset\ell}\Xi_r^{a}\to\pi_\ell\bar\xi_a$ in probability. Second, since $\mathbb E[\Xi^{a}]<\infty$, the Borel--Cantelli lemma gives $\Xi_N^{a}/N\to0$ almost surely, hence $M_N^{a}/N_F\to0$. Third, for $Y:=\Xi_1^{a}\Xi_2^{a}$ and any constant $C$, dominated convergence gives $\mathbb E[Y\min\{1,CY/N_F\}]\to0$, because $Y\min\{1,CY/N_F\}\le Y$ and $\mathbb E[Y]=\bar\xi_a^{2}$.

We first locate the root. Fix $C>0$ and put $\kappa=C/N_F$. Then
\[
\frac{1}{N_F}\sum_{i\ne j}x_{ij}
=C\sum_{k,\ell}\lambda_{k\ell}\Big(\frac{1}{N_F}\sum_{i\in\Fset k}\Xi_i^{a}\Big)\Big(\frac{1}{N_F}\sum_{j\in\Fset\ell}\Xi_j^{a}\Big)
-\frac{C}{N_F^{2}}\sum_i\lambda_{\pi(i)\pi(i)}\Xi_i^{2a},
\]
and the first fact gives the double sum the limit $\bar\xi_a^{2}\Lambda$, while the second fact bounds the self-pair term by $C\lambda_{\max}(M_N^{a}/N_F)(N_F^{-1}\sum_i\Xi_i^{a})\to0$. The gap between $x_{ij}$ and $p_{ij}$ satisfies $0\le x_{ij}-p_{ij}=x_{ij}^{2}/(1+x_{ij})\le x_{ij}\min\{1,x_{ij}\}$, so with $Y_{ij}:=\Xi_i^{a}\Xi_j^{a}$,
\[
\frac{1}{N_F}\sum_{i\ne j}(x_{ij}-p_{ij})
\le\frac{C\lambda_{\max}}{N_F^{2}}\sum_{i\ne j}Y_{ij}\min\{1,C\lambda_{\max}Y_{ij}/N_F\},
\]
whose expectation is at most $C\lambda_{\max}\mathbb E[Y\min\{1,C\lambda_{\max}Y/N_F\}]\to0$ by the third fact, and Markov's inequality makes the left-hand side $o_p(1)$. Hence $S_N(C/N_F)\to C\bar\xi_a^{2}\Lambda$ in probability for every fixed $C$. Taking $C_\pm:=(1\pm\varepsilon)\bar d/(\bar\xi_a^{2}\Lambda)$, the monotonicity of $S_N$ gives $C_-/N_F<\kappa_N<C_+/N_F$ with probability tending to one, so $N_F\kappa_N\to\bar d/(\bar\xi_a^{2}\Lambda)$ in probability.

The constants follow. Since $G_\ell=M_N^{-a}\sum_{r\in\Fset\ell}\Xi_r^{a}$,
\[
\frac{zB_k}{M_N^{a}}=\kappa_N\sum_\ell\lambda_{k\ell}\sum_{r\in\Fset\ell}\Xi_r^{a}
=(N_F\kappa_N)\sum_\ell\lambda_{k\ell}\,\frac{1}{N_F}\sum_{r\in\Fset\ell}\Xi_r^{a}
\to\frac{\bar d\,b_k}{\bar\xi_a\Lambda},
\]
and likewise $zS_k/M_N^{a}\to\bar d\,c_k/(\bar\xi_a\Lambda)$. For the incident-dyad conditions,
\[
\max_{j\ne I_N}p_{I_Nj}\le\max_{j\ne I_N}x_{I_Nj}\le\kappa_N\lambda_{\max}\Xi_{I_N}^{a}M_N^{a}=(N_F\kappa_N)\,\lambda_{\max}\,\Xi_{I_N}^{a}\,\frac{M_N^{a}}{N_F}\to0,
\]
because $\Xi_{I_N}$ has the fixed size law, and the column condition is identical. Finally $\max_{r\in\Fset\ell}g(m_r)/G_\ell\le M_N^{a}/\sum_{r'\in\Fset\ell}\Xi_{r'}^{a}\to0$ by the first two facts.
\end{proof}

\begin{proposition}[CLT for the number of edges]
\label{lem:edges_mean_degree}
Fix the probability matrix $\mathbf P$ and let $\mathbf A$ be a single Bernoulli draw from $\mathbf P$. Define
\[
\mu_{\mathcal E} \;:=\; \sum_{i\neq j} p_{ij},
\qquad
\sigma_{\mathcal E}^2 \;:=\; \sum_{i\neq j} p_{ij}(1-p_{ij})
\]
so that, conditional on $\mathbf P$, the edge count $\mathcal E$ has mean $\mu_{\mathcal E}$ and variance $\sigma_{\mathcal E}^2$. If $\sigma_{\mathcal E}^2 \to \infty$ as $N_F\to\infty$, then
\[
\frac{\mathcal E - \mu_{\mathcal E}}{\sigma_{\mathcal E}} \;\Rightarrow\; \mathcal N(0,1)
\]
\end{proposition}

\begin{proof}[Proof of Proposition~\ref{lem:edges_mean_degree}]
Conditional on $\mathbf P$, the edge indicators $\{a_{ij}\}_{i\neq j}$ are independent (though not identically distributed) $\mathrm{Bernoulli}(p_{ij})$ variables. The centered variables $a_{ij} - p_{ij}$ are therefore independent, mean-zero, and uniformly bounded by one in absolute value, with total variance $\sigma_{\mathcal E}^2$. Because the summands are uniformly bounded while $\sigma_{\mathcal E}^2$ diverges, for any $\epsilon>0$ the sets $\{|a_{ij}-p_{ij}|>\epsilon\,\sigma_{\mathcal E}\}$ are empty for all sufficiently large $N_F$. So the Lindeberg condition for sums of independent, non-identically distributed variables is satisfied. The Lindeberg--Feller central limit theorem applied to $\mathcal E-\mu_{\mathcal E}=\sum_{i\neq j}\big(a_{ij}-p_{ij}\big)$ delivers the claim.
\end{proof}

\medskip
\noindent For the out-degree, write $d^{\mathrm O}_i := \sum_{j\neq i} a_{ij}$, the empirical out-degree probability mass function $F^{\mathrm O}(h) := \tfrac{1}{N_F}\sum_{i} \mathbf 1_{\{d_i^{\mathrm O}=h\}}$ ($h\in\mathbb N$), and $\vartheta_i^{\mathrm O}(h) := \mathbb P_{\mathbf P}\big(d^{\mathrm O}_i=h\big)$. The model-implied counterpart of $F^{\mathrm O}(h)$ is then $\mathbb E_{\mathbf P}\big[F^{\mathrm O}(h)\big] = \tfrac{1}{N_F}\sum_{i}\vartheta_i^{\mathrm O}(h)$.

\begin{proposition}[CLT for empirical degree distributions over finitely many bins]
\label{rem:degree_distr_CLT}
Fix a finite set of degrees $\mathcal H\subset\mathbb N$ and suppose that for each pair $(h,\ell)\in\mathcal H\times\mathcal H$ the limit defining $\big[\Sigma_{\mathrm O}(\mathcal H)\big]_{h\ell}$ exists. Then
\[
\sqrt{N_F}\,\Big(F^{\mathrm O}(h) - \mathbb E_{\mathbf P}\big[F^{\mathrm O}(h)\big]\Big)_{h\in\mathcal H}
\;\Rightarrow\;
\mathcal N\big(0,\Sigma_{\mathrm O}(\mathcal H)\big)
\]
where
\[
\big[\Sigma_{\mathrm O}(\mathcal H)\big]_{h\ell}
=
\lim_{N_F\to\infty}\frac{1}{N_F}\sum_{i=1}^{N_F}
\Big(
\mathbf 1_{\{h=\ell\}}\,\vartheta_i^{\mathrm O}(h)
-
\vartheta_i^{\mathrm O}(h)\,\vartheta_i^{\mathrm O}(\ell)
\Big)
\]
The same conclusion holds for the empirical in-degree distribution.
\end{proposition}

\begin{proof}[Proof of Proposition~\ref{rem:degree_distr_CLT}]
For each fixed $h$, the indicators $\mathbf 1_{\{d_i^{\mathrm O}=h\}}$ are independent across firms. Conditional on $\mathbf P$, each firm's degree depends only on the edges in its own row of $\mathbf A$, which are independent of the edges determining other firms' degrees.\footnote{The probability matrix $\mathbf P$ itself is obtained from an optimization problem that may impose a constraint on the mean degree. So the degree probabilities $\{\vartheta_i^{\mathrm O}(h)\}_{i\le N_F}$ are linked at the level of first moments. In our Bernoulli network model, however, once $\mathbf P$ is fixed we draw edges independently across $(i,j)$, so degrees are exactly independent across firms for any $N_F$. In alternative sampling schemes where one insists that the realized total number of edges or the realized mean degree match a target exactly (rather than in expectation), degrees can become weakly dependent, especially in very small networks. For large networks such dependence is typically negligible.} Hence $F^{\mathrm O}(h)$ is the average of $N_F$ independent random variables bounded in $[0,1]$, the $i^{\text{th}}$ of which has mean $\vartheta_i^{\mathrm O}(h)$. Because the centered summands $\mathbf 1_{\{d_i^{\mathrm O}=h\}} - \vartheta_i^{\mathrm O}(h)$ are independent and uniformly bounded by $1$, the Lindeberg condition for the triangular array normalized by $\sqrt{N_F}$ holds automatically. Each summand is eventually smaller than $\epsilon\sqrt{N_F}$, so the truncated second moments vanish. The hypothesis supplies the limiting covariance $\Sigma_{\mathrm O}(\mathcal H)$. Fix a linear combination of the coordinates $\big\{F^{\mathrm O}(h)\big\}_{h\in\mathcal H}$. If its limiting variance is positive, the Lindeberg--Feller central limit theorem applies to it. If its limiting variance is zero, the normalized combination converges to zero in probability by Chebyshev's inequality, which is convergence to the degenerate normal law. The Cram\'er--Wold device then yields the multivariate convergence, with a covariance matrix that may be singular. It is nonsingular in the direction of a degree $h^\star$ whenever $N_F^{-1}\sum_i\vartheta_i^{\mathrm O}(h^\star)\big(1-\vartheta_i^{\mathrm O}(h^\star)\big)$ stays bounded away from zero. The covariance entries follow from $\mathrm{Var}\big(\mathbf 1_{\{d_i^{\mathrm O}=h\}}\big)=\vartheta_i^{\mathrm O}(h)\big(1-\vartheta_i^{\mathrm O}(h)\big)$ and, for $h\neq\ell$, $\mathrm{Cov}\big(\mathbf 1_{\{d_i^{\mathrm O}=h\}},\mathbf 1_{\{d_i^{\mathrm O}=\ell\}}\big)=-\vartheta_i^{\mathrm O}(h)\,\vartheta_i^{\mathrm O}(\ell)$, the two events being disjoint, together with the hypothesis. The argument for in-degrees is identical after transposing $\mathbf A$.
\end{proof}

\begin{lemma}[Period after a single edge]\label{lem:aperiodic_edge}
Let $\widehat G$ be strongly connected with period $d$ and cyclic classes $\operatorname{CYC}(\cdot)$, and let $(i,j)$ be an absent ordered pair with $i\neq j$ (i.e.\ $\widehat a_{ij}=0$). Then $\widehat G+\{i\!\to\! j\}$ is strongly connected with period
\[
d'\;=\;\gcd\big(d,\ 1+\operatorname{CYC}(i)-\operatorname{CYC}(j)\big)\;=\;\gcd\big(d,\ 1+\operatorname{dist}_{\widehat G}(j,i)\big)
\]
In particular, if $i$ and $j$ lie in the same cyclic class ($\operatorname{CYC}(i)=\operatorname{CYC}(j)$), then $d'=1$.
\end{lemma}

\begin{proof}[Proof of Lemma~\ref{lem:aperiodic_edge}]
Adding an edge preserves strong connectivity. The period of a strongly connected digraph is the greatest common divisor of its closed-walk lengths, and every closed walk decomposes into simple cycles, so the period is the greatest common divisor of the simple-cycle lengths. A simple cycle of $\widehat G+\{i\!\to\! j\}$ either lies in $\widehat G$, where the lengths have greatest common divisor $d$, or uses the new edge exactly once, in which case it is the edge $i\!\to\! j$ followed by a directed $j\!\to\! i$ path in $\widehat G$. Because $\operatorname{CYC}(v)$ equals the length, modulo $d$, of any $r\!\to\! v$ path, every $j\!\to\! i$ path has length $\equiv \operatorname{CYC}(i)-\operatorname{CYC}(j)\pmod d$. So every new cycle has length $\equiv 1+\operatorname{CYC}(i)-\operatorname{CYC}(j)\pmod d$. Using $\gcd(d,L)=\gcd(d,L\bmod d)$ and that all new cycles share this residue,
\[
d'=\gcd\big(d,\ \{\text{new cycle lengths}\}\big)=\gcd\big(d,\ 1+\operatorname{CYC}(i)-\operatorname{CYC}(j)\big);
\]
the shortest new cycle has length $1+\operatorname{dist}_{\widehat G}(j,i)$, giving the second form. If $\operatorname{CYC}(i)=\operatorname{CYC}(j)$ the residue is $1$, whence $d'=\gcd(d,1)=1$.
\end{proof}

\subsection{Finite-Sample Error Bounds for a Randomly Drawn Adjacency Matrix}
In Proposition~\ref{lem:edges_mean_degree} we showed, via a CLT, that as $N_F\to\infty$ the total number of edges $\mathcal E$ of a Bernoulli draw concentrates around its model-implied counterpart $\mu_{\mathcal E}$. Here we provide finite-sample deviation bounds for the degrees of individual firms; the same Bernstein argument applied directly to all dyads gives the analogous bound for $\mathcal E$. As before, the entries $a_{ij}\sim\mathrm{Bernoulli}(p_{ij})$ of a draw are independent for $i\neq j$ with $a_{ii}=0$. Let $\mathbb E$ and $\mathbb P$ denote expectation and probability under this Bernoulli product measure. For each firm $i$, define
\[
\mu_i:=\sum_{j\neq i} p_{ij},
\qquad
\sigma_i^2:=\sum_{j\neq i} p_{ij}(1-p_{ij})
\]
the model-implied mean and variance of its (out-)degree,\footnote{For concreteness we state all results for out-degrees. The corresponding statements for in-degrees follow by transposing $\mathbf{P}$ and are entirely analogous.} and write $d_i:=\sum_{j\neq i} a_{ij}$ for the degree of firm $i$ in the draw. By construction $\mathbb{E}[d_i]=\mu_i$ and $\mathrm{Var}(d_i)=\sigma_i^2$. The centered summands $\{a_{ij}-p_{ij}\}_{j\neq i}$ that produce them are independent, mean-zero, and bounded by one in absolute value, with total variance $\sigma_i^2$---the standing array to which the concentration bounds below apply. (The pair $(\mu_i,\sigma_i^2)$ is local to this appendix. Elsewhere $\mu_j$ denotes the normalized firm-size share of Section~\ref{subsec:weights}.)

\begin{proposition}[Bernstein bounds for firm degrees]
\label{prop:bernstein_degrees}
For every firm $i$ and every $t>0$,
\[
\mathbb{P}\!\left(\big|d_i-\mu_i\big|\ge t\right)
\le 2\exp\!\left(-\frac{t^2}{\,2\sigma_i^2+\tfrac{2}{3}t\,}\right)
\]
Equivalently, for any $\varrho\in(0,1)$, with probability at least $1-\varrho$,
\[
\big|d_i-\mu_i\big|
\le \sqrt{2\sigma_i^2\log\tfrac{2}{\varrho}}
   +\tfrac{2}{3}\log\tfrac{2}{\varrho}
\]
Moreover, the latter bound holds simultaneously for all firms $i=1,\dots,N_F$ with probability at least $1-\varrho$ after replacing $\varrho$ by $\varrho/N_F$ on the right-hand side.
\end{proposition}

\begin{proof}[Proof of Proposition~\ref{prop:bernstein_degrees}]
Fix a firm $i$ and write
\[
d_i-\mu_i
= \sum_{j\neq i} \big(a_{ij}-p_{ij}\big)
\]
The summands form the standing centered array of the previous paragraph---independent, mean-zero, bounded by one, with total variance $\sigma_i^2$. These are precisely the conditions under which Bernstein's inequality for sums of independent, bounded random variables applies (with variance proxy $\sigma_i^2$ and bound $1$), yielding the first display. The second display is the standard inversion of the first, and the simultaneous statement follows from a union bound over $i=1,\dots,N_F$.
\end{proof}

(We write $\varrho$ for the confidence level to avoid collision with the tolerances $\delta^{\mathrm{firm}},\delta^{\mathrm{tail}}$ of the weighting program in Section~\ref{subsec:weights}.) The bounds are most informative when no single dyad carries a large probability mass and a firm's expected degree is moderate or large. They quantify the draw-to-draw dispersion around the degree encoded in $\mathbf P$; they do not imply that the relative error of every bounded-degree firm vanishes as $N_F$ grows. The aggregate degree distribution does concentrate, in relative terms across the tail, as Proposition~\ref{prop:unif_degree_dist} establishes.

The same concentration machinery yields finite-sample control for the empirical degree distribution, and control in relative terms, which is what the upper tail requires. Fix thresholds $q_1<\dots<q_L$. For each $v$ define the model-implied and empirical survival fractions
\[
\bar F_v
:=\frac{1}{N_F}\sum_{i=1}^{N_F}\mathbb P\big(d_i>q_v\big),
\qquad
\hat{\bar F}_v
:=\frac{1}{N_F}\sum_{i=1}^{N_F}\mathbf 1_{\{d_i>q_v\}},
\qquad
n_v:=N_F\bar F_v ,
\]
where $n_v$ is the expected number of firms with degree above $q_v$.

\begin{proposition}[Uniform finite-sample control of the degree tail]
\label{prop:unif_degree_dist}
For any $\varrho\in(0,1)$, with probability at least $1-\varrho$, simultaneously for all $v=1,\dots,L$,
\[
\big|\hat{\bar F}_v-\bar F_v\big|
\le
\sqrt{\frac{2\bar F_v\log(2L/\varrho)}{N_F}}+\frac{2\log(2L/\varrho)}{3N_F}.
\]
\end{proposition}

Dividing by $\bar F_v$, the relative error at threshold $q_v$ is at most $\sqrt{2\log(2L/\varrho)/n_v}+2\log(2L/\varrho)/(3n_v)$, which is small wherever the expected number of firms above the threshold is large.

\begin{proof}[Proof of Proposition~\ref{prop:unif_degree_dist}]
Fix $v$. The indicators $\mathbf 1_{\{d_i>q_v\}}$ depend on disjoint rows of the Bernoulli product array, so they are independent across firms, bounded by one, with variances $\bar F_{v,i}(1-\bar F_{v,i})\le\bar F_{v,i}$ that sum to at most $N_F\bar F_v$. Bernstein's inequality with deviation $N_Fu$ gives
\[
\mathbb P\left(\big|\hat{\bar F}_v-\bar F_v\big|\ge u\right)
\le 2\exp\left\{-\frac{N_Fu^{2}}{2\bar F_v+2u/3}\right\}.
\]
Inverting at level $\varrho/L$ and applying a union bound over $v=1,\ldots,L$ gives the stated bound.
\end{proof}

Thus a single draw tracks the model-implied survival function of the degree distribution in relative terms wherever the expected number of firms above the threshold is large. That range is the one over which the degree-tail exponents are estimated.\footnote{The analogous statement for in-degree thresholds follows by applying the same argument to columns.}

All statements in this subsection are conditional on the fixed probability matrix \(\mathbf P\) and concern the independent Bernoulli draw, before the degree-floor and Markov-closure additions, not the repaired graph.

\subsection{Error Bound on the Weighted Matrix}
We close with a \emph{local} sensitivity statement. If the support admits an exactly money-preserving matrix, then in a neighborhood made precise below our minimum-energy weights lie close to it. Relaxing the exact firm and sector balances to the small slack the weighting caps permit moves the minimum-energy point by a controlled amount, provided the relaxation does not change which constraints bind.

Assume the fixed support $\Aaug$ admits a row-stochastic matrix satisfying the per-edge bounds and preserving firm balances exactly, $\widetilde{\mathbf{W}}^\top\mathbf{m}=\mathbf{m}$ (so $\boldsymbol{\mu}=\mathbf{m}/M$ is stationary for it). Since the support is primitive and every realized edge remains positive, such a matrix is irreducible and aperiodic. Among all such matrices let $\widetilde{\mathbf{W}}$ be the one of least energy, the matrix our weights would equal if the balances held exactly. Vectorize the weights on the fixed support as $w=\mathrm{vec}(\mathbf{W})\in\mathbb{R}^{E}$. Let $\mathcal{X}_0$ collect the fixed polyhedral constraints that are not balance-error caps: row sums and the per-edge bounds on the fixed support. Stack the scaled firm- and sector-balance residuals as $\mathbf Cw-\mathbf d$, removing redundant equations so that $\mathbf C$ has full row rank. Exact firm balance already implies exact sector balance, but the two sets of residuals may carry different scalings. Let $\mathcal R_\eta$ denote the closed convex set of residual vectors admitted by the balance caps, and write $\eta$ for a Euclidean radius such that
\[
        \mathbf 0\in\mathcal R_\eta
        \subseteq
        \{r:\|r\|_2\le\eta\}.
\]
(The slack $\eta$ is local to the present subsection, distinct from the fitness knee sharpness of Section~\ref{subsec:gravity}.) Consider the two convex quadratic programs
\[
\begin{aligned}
\text{(Exact)}\quad
&\widetilde w
=\arg\min_{w}\tfrac12\|w\|_2^2
\quad\text{s.t.}\quad w\in\mathcal{X}_0,\ \ \mathbf{C}w=\mathbf{d},
\\[0.25em]
\text{(Approximate)}\quad
&w^\star
=\arg\min_{w}\tfrac12\|w\|_2^2
\quad\text{s.t.}\quad w\in\mathcal{X}_0,\ \ \mathbf{C}w-\mathbf d\in\mathcal R_\eta.
\end{aligned}
\]
Both problems are strongly convex (Hessian $=\mathbf{I}_{E}$), hence each has a unique minimizer.

\begin{proposition}[Local sensitivity of the weighted matrix]
\label{rem:error_bound_W}
Suppose the exact problem is feasible with solution $\widetilde w=\mathrm{vec}(\widetilde{\mathbf{W}})$, and that at $\widetilde w$ the linear independence constraint qualification and strict complementarity hold. Let $\mathbf N_{\mathcal A}$ be the matrix whose rows are the row-sum constraints, the unit vectors of the per-edge bounds active at $\widetilde w$, and the rows of $\mathbf C$. Then there is $\bar\eta>0$ such that, for every $\eta\in[0,\bar\eta]$, the approximate solution $w^\star=\mathrm{vec}(\mathbf{W})$ has the same active bounds as $\widetilde w$, and
\[
\|\mathbf{W}-\widetilde{\mathbf{W}}\|_F
=
\|w^\star-\widetilde w\|_2
\le
\frac{\eta}{\sigma_{\min}(\mathbf N_{\mathcal A})}.
\]
\end{proposition}

The constant is the norm of the pseudo-inverse of the active constraint matrix. The radius $\bar\eta$ is the smallest $\|r\|_2$ at which the affine branch $w(r)=\widetilde w+\mathbf N_{\mathcal A}^\top(\mathbf N_{\mathcal A}\mathbf N_{\mathcal A}^\top)^{-1}(\mathbf 0;r)$ reaches an inactive bound or an active bound multiplier vanishes. Both are computable at the solved point. No uniform-in-$N_F$ conclusion follows unless these quantities and the residual scaling are themselves controlled uniformly.

\begin{proof}[Proof of Proposition~\ref{rem:error_bound_W}]
Fix $r$ with $\|r\|_2\le\bar\eta$ and consider the equality-perturbed problem with $\mathbf{C}w=\mathbf{d}+r$. On the active set of $\widetilde w$ the problem is $\min\tfrac12\|w\|_2^{2}$ subject to $\mathbf N_{\mathcal A}w=b_{\mathcal A}+(\mathbf 0;r)$, whose unique solution is the minimum-norm point $w(r)=\mathbf N_{\mathcal A}^\top(\mathbf N_{\mathcal A}\mathbf N_{\mathcal A}^\top)^{-1}(b_{\mathcal A}+(\mathbf 0;r))$, since $\mathbf N_{\mathcal A}$ has full row rank by the constraint qualification. The multipliers $(\mathbf N_{\mathcal A}\mathbf N_{\mathcal A}^\top)^{-1}(b_{\mathcal A}+(\mathbf 0;r))$ are affine in $r$, so strict complementarity keeps the signs of the bound multipliers, and the inactive bounds remain slack, for $\|r\|_2\le\bar\eta$. The point $w(r)$ therefore satisfies the Karush--Kuhn--Tucker conditions of the perturbed problem and is its minimizer by convexity. The operator $\mathbf N_{\mathcal A}^\top(\mathbf N_{\mathcal A}\mathbf N_{\mathcal A}^\top)^{-1}$ is the pseudo-inverse of $\mathbf N_{\mathcal A}$, whose spectral norm is $1/\sigma_{\min}(\mathbf N_{\mathcal A})$, which gives $\|w(r)-\widetilde w\|_2\le\|r\|_2/\sigma_{\min}(\mathbf N_{\mathcal A})$. Let $r^\star:=\mathbf{C}w^\star-\mathbf{d}$. Since $r^\star\in\mathcal R_\eta$, we have $\|r^\star\|_2\le\eta\le\bar\eta$. Moreover, $w^\star$ minimizes the equality-perturbed problem at $r^\star$, because a lower-energy point in $\mathcal X_0$ with the same residual would improve the approximate problem. Hence $w^\star=w(r^\star)$ and the display follows. Finally $\|w^\star-\widetilde w\|_2=\|\mathbf{W}-\widetilde{\mathbf{W}}\|_F$.
\end{proof}

Two qualifications travel with the statement. It is genuinely \emph{local}. The bound holds only while no edge enters or leaves its active bound and the equality-perturbed solution remains on the same regular branch. And the constant $1/\sigma_{\min}(\mathbf N_{\mathcal A})$ is large if the active constraint matrix is ill-conditioned. Their economic counterpart is how the one-step weighting residual actually propagates to the stationary firm-size gap, measured directly in Appendix~\ref{app:robustness}. Across the full US reconstruction, a one-step residual of root-mean-square $5\%$ (the firm-size cap $\delta^{\mathrm{firm}}$) propagates to a small stationary drift. The median stationary-to-census ratio is $1.04$, the total-variation money relocation about $3\%$, and every firm stays within a factor of two of its reconstructed size. So the census sits nearly at a fixed point of the network's own money dynamics. We make no claim beyond this neighborhood, and we do not chain the bound into a statement about the stationary distribution. That closeness is established separately, in the size-weighted norm it requires, by Proposition~\ref{rem:stationary_vs_onestep}.

\subsection{Structure of the Minimum-Energy Weights}
\label{app:weights_structure}

This subsection proves Proposition~\ref{prop:kkt} and records an explicit weighting that balances every firm and every sector block in expectation over the Bernoulli draw. Together they say which part of the balance adjustment in program~\eqref{eq:weight-program} is systematic and which part the program must supply on the realized graph.

\begin{proof}[Proof of Proposition~\ref{prop:kkt}]
Fix a solution $\mathbf W$ and write $\onestep:=\mathbf W^\top\mathbf m$ for the one-step money vector, so that $\onestep_j=\sum_im_iw_{ij}$ and $\widehat s_\ell(\mathbf W)=\sum_{j\in\Fset\ell}\onestep_j$. Each of the four caps in program~\eqref{eq:weight-program}, written in squared form, is a convex function of $\onestep$ alone, so its derivative with respect to $w_{ij}$ is $m_i$ times its derivative with respect to $\onestep_j$. The program is convex with affine equalities, box constraints, and convex inequality caps, so under the strict-feasibility hypothesis the Karush--Kuhn--Tucker conditions are necessary and sufficient. Let $\alpha'_i$ be the multiplier of buyer $i$'s row sum, $\beta_{ij}\ge0$ that of the floor $w_{ij}\ge\varepsilon_{\mathrm{floor}}$, and $\nu_c\ge0$ that of cap $c$, with $\nu_c=0$ when cap $c$ is slack. The two worst-fraction caps are maxima of finitely many smooth convex functions, and we take the gradient of an active piece, that is, of a worst set $T$ of firms or $T^{\mathrm{sec}}$ of sectors attained at $\mathbf W$. Stationarity in $w_{ij}$ reads
\[
2w_{ij}=\alpha'_i-m_i\sum_c\nu_c\,\frac{\partial c}{\partial\onestep_j}+\beta_{ij},
\]
and complementary slackness gives $\beta_{ij}>0$ only when $w_{ij}=\varepsilon_{\mathrm{floor}}$. The upper bounds $w_{ij}\le1$ never bind, because every buyer has at least two realized edges each carrying at least $\varepsilon_{\mathrm{floor}}$. Putting $\alpha_i:=\alpha'_i/2$ and $\gamma_j:=-\tfrac12\sum_c\nu_c\,\partial c/\partial\onestep_j$ gives $w_{ij}=\max\{\varepsilon_{\mathrm{floor}},\alpha_i+m_i\gamma_j\}$. Differentiating the four caps gives, with $\ell=\pi(j)$,
\[
\gamma_j=\frac{\nu^{\mathrm{firm}}}{M}\,\frac{m_j-\onestep_j}{m_j}
+\frac{\nu^{\mathrm{tail}}}{\lceil qN_F\rceil}\,\frac{m_j-\onestep_j}{m_j^{2}}\,\mathbf 1\{j\in T\}
+\Big(\frac{\nu^{\mathrm{sec}}}{N_S}+\frac{\nu^{\mathrm{sec}}_{\mathrm{tail}}\,\mathbf 1\{\ell\in T^{\mathrm{sec}}\}}{\lceil qN_S\rceil}\Big)\frac{s_\ell-\widehat s_\ell(\mathbf W)}{s_\ell^{2}},
\]
which is the combination of shortfalls described in Section~\ref{subsec:weights}. The projection form follows because the Karush--Kuhn--Tucker conditions of $\min_{\mathbf w}\tfrac12\|\mathbf w-m_i\boldsymbol\gamma\|_2^{2}$ over the rows that sum to one and respect the floor are $w_j=\max\{\varepsilon_{\mathrm{floor}},m_i\gamma_j+\alpha\}$ with $\alpha$ fixed by the row sum, the same system.
\end{proof}

We next compute what flat weights do to a seller's inflow, which is the computation behind the sign of $\gamma_j$. Under $w_{ij}=a_{ij}/d_i^{\mathrm O}$, with each out-degree replaced by its mean $z\,g(m_i)B_{\pi(i)}$ and in the sparse regime $p_{ij}=x_{ij}$, the expected inflow of seller $j$ in sector $\ell$ is
\[
\sum_{i}m_i\,\frac{x_{ij}}{z\,g(m_i)B_{\pi(i)}}
=g(m_j)\sum_k\lambda_{k\ell}\frac{s_k}{B_k},
\]
proportional to the seller's fitness rather than to its size. Since the target inflow is $m_j$, the ratio of inflow to target varies across sellers as $m_j^{a-1}$ below the saturation knee, so the sellers short of inflow under flat weights are the large ones.

We now record a weighting that removes this systematic shortfall. Let $I^\star_{k\ell}$ be the expenditure shares of the balanced target of Section~\ref{subsec:interfirm}, so that $\sum_ks_kI^\star_{k\ell}=s_\ell$ by its margin condition. For a realized edge $i\to j$ with $k=\pi(i)$ and $\ell=\pi(j)$ define
\[
\beta_{ij}:=\frac{I^\star_{k\ell}}{\lambda_{k\ell}\,s_\ell}\cdot\frac{m_j}{g(m_j)},
\qquad
\bar w_{ij}:=\frac{a_{ij}\beta_{ij}}{z\,g(m_i)},
\qquad
w^{\mathrm b}_{ij}:=\frac{a_{ij}\beta_{ij}}{\sum_{j'}a_{ij'}\beta_{ij'}} .
\]

\begin{proposition}[Expected-balance weights]
\label{prop:balance}
Suppose $x_{ij}\le1$ for every ordered pair and let edges be drawn independently with $p_{ij}=x_{ij}$. Then, for every buyer $i$ in sector $k$, every seller $j$ in sector $\ell$, and every block $(k,\ell)$,
\begin{gather*}
\mathbb E\Big[\sum_{j'}a_{ij'}\beta_{ij'}\Big]=z\,g(m_i)\Big(1-\frac{I^\star_{kk}m_i}{s_k}\Big),
\qquad
\mathbb E\Big[\sum_im_i\bar w_{ij}\Big]=m_j\Big(1-\frac{I^\star_{\ell\ell}m_j}{s_\ell}\Big),\\
\mathbb E\Big[\sum_{i\in\Fset k}\sum_{j\in\Fset\ell}m_i\bar w_{ij}\Big]=s_kI^\star_{k\ell}-\mathbf 1\{k=\ell\}\,I^\star_{kk}\,\frac{\sum_{i\in\Fset k}m_i^{2}}{s_k}.
\end{gather*}
\end{proposition}

Up to the self-pair terms, whose relative size is at most $\max_im_i/s_{\pi(i)}$, the weights $\bar w$ preserve every firm's money and every block flow in expectation, and $w^{\mathrm b}$ is their row-stochastic version, obtained by replacing the expected row normalizer by the realized one. Within a buyer's row the weight rises with seller size as $m_j/g(m_j)$, which is $m_j^{1-a}$ below the saturation knee and $m_j^{1-a\eta}$ above it, times the block factor $I^\star_{k\ell}/(\lambda_{k\ell}s_\ell)$, the ratio of the block's expenditure share to its link intensity. This is the systematic part of the balance adjustment, and it is why the firm cap $\delta^{\mathrm{firm}}$ binds. It is also consistent with Proposition~\ref{prop:kkt}: solving the exact balance $\sum_im_ia_{ij}(\alpha_i+m_i\gamma_j)=m_j$ in expectation gives $\gamma_j=(zQ_\ell)^{-1}\big(m_j/g(m_j)-zR_\ell\big)$ with $Q_\ell:=\sum_k\lambda_{k\ell}\sum_{i\in\Fset k}m_i^{2}g(m_i)$ and $R_\ell:=\sum_k\lambda_{k\ell}\sum_{i\in\Fset k}m_ig(m_i)\alpha_i$, so the balance price is an increasing affine function of $m_j/g(m_j)$ with a sector-specific threshold below which it is negative. What the closed form does not do is balance the realized graph. The residual it leaves comes from the logistic saturation of the largest pairs and from the sampling noise in the customer counts of sellers with few customers, and that residual is what program~\eqref{eq:weight-program} corrects. The weights $w^{\mathrm b}$ cost one pass over the edge list and serve as a warm start for the weighting solve.

\begin{proof}[Proof of Proposition~\ref{prop:balance}]
Fix a buyer $i$ in sector $k$. Since $p_{ij'}=z\lambda_{k\ell'}g(m_i)g(m_{j'})$ for $j'$ in sector $\ell'$,
\[
\mathbb E\sum_{j'\ne i}a_{ij'}\beta_{ij'}
=\sum_{\ell'}\frac{I^\star_{k\ell'}}{\lambda_{k\ell'}s_{\ell'}}\sum_{\substack{j'\in\Fset{\ell'}\\ j'\ne i}}z\lambda_{k\ell'}g(m_i)g(m_{j'})\frac{m_{j'}}{g(m_{j'})}
=z\,g(m_i)\sum_{\ell'}I^\star_{k\ell'}\,\frac{s_{\ell'}-m_i\mathbf 1\{\ell'=k\}}{s_{\ell'}},
\]
and $\sum_{\ell'}I^\star_{k\ell'}=1$. This proves the first identity. Now fix a seller $j$ in sector $\ell$. The expected inflow is
\[
\mathbb E\sum_{i\ne j}m_i\bar w_{ij}
=\sum_k\sum_{\substack{i\in\Fset k\\ i\ne j}}m_i\,\frac{z\lambda_{k\ell}g(m_i)g(m_j)}{z\,g(m_i)}\cdot\frac{I^\star_{k\ell}}{\lambda_{k\ell}s_\ell}\cdot\frac{m_j}{g(m_j)}
=\frac{m_j}{s_\ell}\sum_kI^\star_{k\ell}\big(s_k-m_j\mathbf 1\{k=\ell\}\big),
\]
and $\sum_ks_kI^\star_{k\ell}=s_\ell$ by the margin condition. This proves the second identity. The third is the same sum restricted to buyers in sector $k$, which gives $\sum_{i\in\Fset k}m_iI^\star_{k\ell}(s_\ell-m_i\mathbf 1\{k=\ell\})/s_\ell$.
\end{proof}

\section{Algorithmic Repairs and Markov Closure}\label{app:repairs}
This appendix records the two repair steps summarized in Sections~\ref{subsec:degreefloor} and~\ref{subsec:markov}: the tilted conditional law that imposes the minimum-degree floor, and the component-pairing and aperiodization construction that makes the binary backbone primitive.

\subsection{Degree-Floor Repair}\label{app:degfloor}
The minimum-degree floor of Section~\ref{subsec:degreefloor} requires \(d_i^{\mathrm O}\ge2\) and \(d_i^{\mathrm I}\ge2\) for every firm. Deficient rows are repaired while preserving the gravity ranking of counterparties, as follows.

\subsubsection{Out-degree floor}

Consider first outgoing links. If firm \(i\) already has
\[
        d_i^{\mathrm O}=\sum_{j\neq i}a_{ij}\ge2
\]
its row is left unchanged. If it has fewer than two outgoing links, its row is repaired.

The repair preserves the gravity ranking of potential suppliers. Let \(r_i:=2-d_i^{\mathrm O}\in\{1,2\}\) be the number of missing outgoing links, and let \(\mathrm{PB}_i^{+}\) be the Poisson-binomial law generated by the probabilities \(\{p_{ij}:j\neq i,\ a_{ij}=0\}\) on currently absent entries. Whenever \(\Pr(\mathrm{PB}_i^{+}\ge r_i)>0\), the number \(K_i\) of added outgoing links is drawn from the tilted conditional law
\begin{equation}
\label{eq:tilt}
        \Pr(K_i=k)
        \propto
        \mathrm{PB}_i^{+}(k)e^{\xi k},
        \qquad k\ge r_i
\end{equation}
where \(\xi\in\mathbb R\) is a spread parameter. Equivalently, on the absent entries the tilt rescales the edge odds by
\[
        \frac{p_{ij}}{1-p_{ij}}
        \mapsto
        e^\xi\frac{p_{ij}}{1-p_{ij}}
\]
and then the added-edge set is drawn from this tilted Bernoulli product law conditional on $K_i\ge r_i$. If the conditioning event has probability zero, we instead add the $r_i$ absent non-self entries with the largest $p_{ij}$, breaking ties uniformly. For $N_F\ge3$, a deficient row always contains at least $r_i$ such entries. Thus the repair changes how many additional suppliers a deficient firm receives, but not the gravity logic that ranks those suppliers.

The repaired row is coupled monotonically with the original deficient row. The repair keeps any
edge already drawn and only adds edges. Thus the lower tail of the degree distribution is made
denser without changing the rest of the draw. The parameter \(\xi\) controls how far the repaired
firms are moved above the floor. When \(\xi=0\), the repair uses the gravity product law on
the absent entries, conditioned to add at least \(r_i\) links. When \(\xi<0\), repaired firms bunch
closer to the floor. When \(\xi>0\), the repair assigns more mass to degrees above the floor.

\subsubsection{In-degree floor}

The buyer-side floor is imposed in the same way, but on the transpose. Rows of \(\mathbf A^\top\)
collect the buyers of a firm. Applying the same tilted conditional law to those rows enforces
\[
        d_j^{\mathrm I}=\sum_{i\neq j}a_{ij}\ge2
\]
for every firm \(j\).

Because each repair only adds edges, the two passes do not undo each other. The out-degree repair
raises out-degrees. The in-degree repair can only add additional outgoing edges for some firms, so
the out-degree floor remains satisfied. Symmetrically, after the second pass every firm has both at
least two suppliers and at least two customers.\footnote{Because the repair only adds edges, the repaired row dominates the deficient row coordinatewise and the coupling is monotone; only deficient rows in the orientation being repaired are selected, and no already-satisfied floor is undone. If the aggregate expected repair count is \(O(N_F)\)---not implied by monotonicity alone, since a rare deficient high-mean row need not have a bounded conditional repair---the edge count stays linear, the network remains sparse, and the calibrated \(z\) is unchanged. The concentration results in Propositions~\ref{lem:edges_mean_degree} and~\ref{rem:degree_distr_CLT} are stated for the independent pre-repair draw; the transpose pass can couple rows, so after repair we rely only on these deterministic facts.}

\subsubsection{Lower-tail under-connection}

The floor corrects a mechanical problem created by independent Bernoulli drawing. A probability
model does not force small firms to have trading partners. For every firm,
\[
        \Pr(d_i^{\mathrm O}=0)
        =
        \prod_{j\neq i}(1-p_{ij})>0
\]
This probability is largest for small firms, because their link probabilities are small. Hence the
deficient firms are not just sampling accidents. They are the visible part of a systematic
under-connection of the lower tail.

For this reason we use a positive tilt, \(\xi>0\). The positive tilt moves repaired firms above the
bare floor and gives the lower tail a more plausible level of connectivity. The operation remains
disciplined in two ways. First, firm sizes are unchanged. The repair acts only on the realized graph,
not on the data. Second, the hard floor is delivered by the conditioning rule, with the deterministic feasibility fallback above, not by the value of \(\xi\). The parameter \(\xi\) only determines how much mass lies above the floor when the conditional law is used.

\subsection{Markov Closure}\label{app:closure}
To make the repaired backbone primitive, Section~\ref{subsec:markov} adds a minimal component-level set of ordinary inter-firm links, using more links where the smaller component is larger. The construction, and the sectoral placement objective that places them, is as follows.

\subsubsection{Strongly connected components}

Let \(\mathbf A\in\{0,1\}^{N_F\times N_F}\) be the repaired Bernoulli adjacency matrix, with
\(a_{ii}=0\). Let
\[
        \mathcal C=\{1,\ldots,C\}
\]
be the set of strongly connected components of \(\mathbf A\). For each component \(c\), let \(V(c)\)
be its set of firms. Collapsing each component to a node gives the condensation graph
\(\mathbf A_C\), which is a directed acyclic graph.

Let
\[
\mathscr T
=
\{c\in\mathcal C:\mathrm{indeg}_{\mathbf A_C}(c)=0\},
\qquad
\mathscr U
=
\{c\in\mathcal C:\mathrm{outdeg}_{\mathbf A_C}(c)=0\}
\]
be the source and sink components of the condensation graph. If \(C=1\), the graph is already
strongly connected and no irreducibility repair is needed. Otherwise, source and sink components
must be connected.

Choose a set
\[
        \mathcal P\subseteq \mathscr U\times\mathscr T
\]
of ordered sink-source component pairs such that adding one component-level arc \(c\to c'\) for
each \((c,c')\in\mathcal P\) would make the condensation graph strongly connected. Such a set can
always be chosen with
\[
        |\mathcal P|= \max\{|\mathscr T|,|\mathscr U|\}.
\]
The Eswaran--Tarjan construction \citep{eswaran-tarjan1976} produces such a set in linear time, and it is this pairing that we use, since an arbitrary pairing of sinks with sources of the same size need not strongly connect the condensation. In a minimum augmentation, every component on the longer of the source and sink lists appears exactly once as the relevant endpoint; components on the shorter list may be repeated, and the pairs can be chosen without self-arcs.
For each \((c,c')\in\mathcal P\), define
\[
        n_{cc'}
        =
        \min\{|V(c)|,|V(c')|\}
\]
This is the size of the smaller of the two components. We add
\[
        k_{cc'}
        =
        \left\lceil f_\nu(n_{cc'})\,n_{cc'}\right\rceil
\]
firm-level edges from component \(c\) to component \(c'\), where
\[
        f_\nu(n)
        =
        \theta(1-e^{-\nu n}),
        \qquad
        0<\theta<1,\quad \nu>0
\]
Thus the added fraction is increasing in the size of the smaller component and saturates at $\theta$, while the number of added links grows at most linearly in $n_{cc'}$.
The purpose is to avoid connecting large components by a single fragile edge while keeping the
total number of added edges small.

\subsubsection{Placing the new firm-level edges}

It remains to decide which firm pairs receive the new edges. We choose these edges jointly, using a
sectoral placement objective.

Let \(s_\ell\) be the empirical size of sector \(\ell\):
\[
        s_\ell=\sum_{j\in\Fset{\ell}}m_j
\]
Given the current graph, define the baseline sectoral placement residual for sector \(\ell\) as
\[
D_\ell
=
\sum_{k=1}^{N_S}
\sum_{i\in\Fset{k}}
\sum_{j\in\Fset{\ell}}
m_i a_{ij} I_{k\ell}
-
s_\ell
\]
Here \(I_{k\ell}\) is the expenditure share from Section~\ref{subsec:gravity}. It is used as a proxy
for the weight that an edge from sector \(k\) to sector \(\ell\) will later carry. Since each realized edge contributes separately, \(D_\ell\) is a placement residual, not an accounting measure of realized sectoral money inflow.

For each component pair \((c,c')\in\mathcal P\), let
\[
        \Omega_{cc'}^{\mathrm{raw}}
        =
        \{(i,j):i\in V(c),\ j\in V(c'),\ a_{ij}=0\}
\]
be the notional set of absent firm-level edges from \(c\) to \(c'\). We neither optimize over nor materialize the full raw set.
Instead, we sample a candidate set
\[
        \Omega_{cc'}\subseteq\Omega_{cc'}^{\mathrm{raw}}
\]
of size \(O(n_{cc'})\), with \(|\Omega_{cc'}|\ge k_{cc'}\). Appendix~\ref{sec:compu} gives the
saturating sampling rule. In the cyclic pairing above, one endpoint list is used once, while the
other is repeated. Since \(n_{cc'}\) is no larger than the size of the component on the unrepeated
list,
\[
        \sum_{(c,c')\in\mathcal P} n_{cc'}\le N_F.
\]
Thus the global candidate set
\[
        \Omega=\bigcup_{(c,c')\in\mathcal P}\Omega_{cc'}
\]
has size \(O(N_F)\).

For each candidate edge \((i,j)\in\Omega\), let \(y_{ij}\in\{0,1\}\) indicate whether the edge is
added. The placement problem is
\begin{equation}
\label{eq:closure-program}
\min_{y}
\sum_{\ell=1}^{N_S}
\frac{1}{s_\ell^2}
\left(
D_\ell
+
\sum_{(i,j)\in\Omega}
m_i I_{\pi(i)\ell}\mathbf 1\{\pi(j)=\ell\}y_{ij}
\right)^2
\end{equation}
subject to
\[
        y_{ij}\in\{0,1\},
        \qquad (i,j)\in\Omega
\]
and
\[
        \sum_{(i,j)\in\Omega_{cc'}}y_{ij}=k_{cc'},
        \qquad (c,c')\in\mathcal P
\]
The term multiplying \(y_{ij}\) is the contribution of edge \((i,j)\) to the placement score of the sector
containing \(j\). The objective penalizes proportional deviations in these scores, so large and small sectors
are treated symmetrically in relative terms.

Although the problem is written with binary variables, it is not solved as a hard integer program.
The candidate sets \(\Omega_{cc'}\) form a partition of \(\Omega\), and the constraints require a fixed
number of edges to be selected from each group. After relaxing \(y_{ij}\in\{0,1\}\) to \(0\le y_{ij}\le1\), the feasible set is therefore the base polytope of a
partition matroid. The relaxed quadratic minimizer need not itself be integral; we solve the relaxed convex problem as a tractable stand-in for the integer program. Let \(\Phi(y)\) denote the objective in \eqref{eq:closure-program}. At a relaxed point \(\bar y\), the partition-matroid linear oracle uses the first-order scores
\[
        R_{ij}(\bar y)
        =
        -\left.\frac{\partial\Phi(y)}{\partial y_{ij}}\right|_{y=\bar y}.
\]
Within each group \(\Omega_{cc'}\), the oracle selects the \(k_{cc'}\) candidates with the largest \(R_{ij}(\bar y)\), equivalently the smallest gradient entries. The final added support is obtained by this group-wise selection using the gradient at the final relaxed point, rather than by solving the binary program to global optimality.\footnote{Appendix~\ref{sec:compu}
gives the partition-matroid separation argument, the \(O(I_{\mathrm{FW}}|\Omega|)\) relaxed-placement cost, and the \(O(|\Omega|)\) final group-wise oracle cost. The single
aperiodizing edge below is the \(k=1\) version of the same construction.}

The selection is nevertheless within a computable distance of the global binary optimum. Write $\Delta_\ell(y):=\sum_{(i,j)\in\Omega}m_iI_{\pi(i)\ell}\mathbf 1\{\pi(j)=\ell\}y_{ij}$ for the change in the placement score of sector $\ell$ induced by $y$, so that $\Phi(y)=\Phi(0)+L(y)+Q(y)$ with $L(y):=2\sum_\ell s_\ell^{-2}D_\ell\Delta_\ell(y)$ linear in $y$ and $Q(y):=\sum_\ell s_\ell^{-2}\Delta_\ell(y)^{2}\ge0$. The group-wise oracle at $y=0$ selects, within each group, the $k_{cc'}$ candidates with the smallest linear scores $2D_{\pi(j)}m_iI_{\pi(i)\pi(j)}/s_{\pi(j)}^{2}$, and this selection $y^{\mathrm g}$ minimizes $L$ over the base polytope and hence over its integer points. For any integral selection $y$ and any global binary minimizer $y^{\star}$, since $L(y^{\mathrm g})\le L(y^{\star})$ and $Q(y^{\star})\ge0$,
\[
\Phi(y)=\Phi(0)+L(y)+Q(y)
\le\Phi(y^{\star})+Q(y)+\big[L(y)-L(y^{\mathrm g})\big].
\]
The final selection $\bar y$ is therefore within $Q(\bar y)+L(\bar y)-L(y^{\mathrm g})$ of the global optimum. Both terms are computed in one pass over the added edges, and when they are negligible against $\Phi(0)$ the relaxation can be replaced by the selection $y^{\mathrm g}$ itself.

\subsubsection{Aperiodicity without self-loops}
\label{subsubsec:aperiodic}

Let \(\widehat{\mathbf A}\) be the strongly connected adjacency matrix obtained after the
irreducibility repair. It may still be periodic. We make it aperiodic by adding one ordinary directed
edge between distinct firms. We do not add self-loops, because a self-loop would mean that a firm
buys from itself. That is a different economic object from an inter-firm trade link.

For a strongly connected digraph \(\widehat G=(V,\widehat E)\), the period is
\[
        d(\widehat G)
        =
        \gcd\{\ell:\widehat G\text{ contains a directed cycle of length }\ell\}
\]
The graph is aperiodic if and only if \(d(\widehat G)=1\). The period can be computed without
enumerating cycles. Root a breadth-first search at an arbitrary vertex \(r\), set \(h(r)=0\), and let
\(h(i)\) be the search depth of vertex \(i\). Then
\[
        d(\widehat G)
        =
        \gcd_{(i,j)\in\widehat E}
        |h(i)+1-h(j)|
\]
where the gcd is taken over nonzero terms \citep{Denardo1977}. The same computation gives the
cyclic classes
\[
        \operatorname{CYC}(i)=h(i)\bmod d
\]
If \(d=1\), no edge is added and \(\Aaug=\widehat{\mathbf A}\).

If \(d>1\), adding an absent edge \(i\to j\) changes the period to
\[
        \gcd\bigl(d,\,1+\operatorname{CYC}(i)-\operatorname{CYC}(j)\bigr)
\]
Thus an edge satisfying
\[
        \gcd\bigl(d,\,1+\operatorname{CYC}(i)-\operatorname{CYC}(j)\bigr)=1
\]
makes the graph aperiodic. Such an edge always exists after the degree floor, and Lemma~\ref{lem:noexception} says where.

\begin{lemma}[An aperiodizing edge between distinct firms always exists]
\label{lem:noexception}
Let $\widehat G$ be strongly connected without self-loops, with period $d\ge2$ and cyclic classes $C_0,\dots,C_{d-1}$, and let every vertex have out-degree at least two. Then every cyclic class contains at least two vertices, and for any two distinct vertices $i,j$ in a common class the ordered pair $(i,j)$ is absent and $\widehat G+\{i\to j\}$ has period one.
\end{lemma}

\begin{proof}[Proof of Lemma~\ref{lem:noexception}]
Fix $\widehat G$ as in the statement. Every edge of a strongly connected digraph of period $d$ runs from some class $C_r$ to the class $C_{r+1\bmod d}$. A vertex of $C_r$ has two distinct out-neighbours, both in $C_{r+1\bmod d}$, so that class contains at least two vertices. Every class is nonempty, so the conclusion holds for every class. If $i$ and $j$ lie in a common class $C_r$, an edge $i\to j$ would run from $C_r$ to $C_r$, which is impossible for $d\ge2$, so the pair $(i,j)$ is absent. Lemma~\ref{lem:aperiodic_edge} then gives the new period $\gcd(d,1+\operatorname{CYC}(i)-\operatorname{CYC}(j))=\gcd(d,1)=1$.
\end{proof}

The degree floor of Section~\ref{subsec:degreefloor} gives every firm at least two suppliers, and the irreducibility closure only adds edges, so Lemma~\ref{lem:noexception} applies to $\widehat{\mathbf A}$.

Among admissible edges, we choose the one that least changes the sectoral placement score, with $D_\ell$ updated after the irreducibility edges have been inserted. If adding edge
\((i,j)\) changes the score for sector \(\ell\) by
\[
        \Delta_\ell(i,j)
        =
        m_i I_{\pi(i)\ell}\mathbf 1\{\pi(j)=\ell\}
\]
then we choose
\[
(i^\star,j^\star)
\in
\arg\min_{(i,j)\in\Omega^{\mathrm{ap}}}
\sum_{\ell=1}^{N_S}
\frac{1}{s_\ell^2}
\bigl(D_\ell+\Delta_\ell(i,j)\bigr)^2
\]
subject to
\[
        \gcd\bigl(d,\,1+\operatorname{CYC}(i)-\operatorname{CYC}(j)\bigr)=1
\]
Here \(\Omega^{\mathrm{ap}}\) is an \(O(N_F)\) thinned sample of absent non-self edges, drawn by the same
gravity-weighted rule as above. If it holds no admissible edge, we enlarge it with absent pairs drawn inside a cyclic class that contains at least two firms, every one of which is admissible by Lemma~\ref{lem:noexception}, so correctness does not depend on the thinning. The linear bound applies when an admissible edge is found after
\(O(N_F)\) generated candidates. The final primitive backbone \(\Aaug\) is obtained by activating
\(i^\star\to j^\star\).

\section{Computational Complexity of the Algorithm}
\label{sec:compu}

The algorithm is useful only if it can be run at the scale of an economy. For the United States the
object has millions of firms and hundreds of millions of links. A reconstruction method whose cost
is quadratic at every stage would therefore be of little practical value.

The main point of this appendix is that the algorithm is not quadratic at every stage. Three steps
are essentially linear in the number of firms: the gravity estimation after binning, the Markov
closure per relaxed-placement iteration and final oracle, and the weighting step per iteration. They
are linear because they either work with a small sector-bin representation or touch only realized
edges. The Bernoulli draw is the exception. It makes
one decision for each ordered pair of firms, and so its total work is quadratic. But those decisions
are independent. They can therefore be sent to the GPU, where the relevant wall-clock cost is closer
to \(O(N_F^2/P_{\mathrm{gpu}})\) than to \(O(N_F^2)\), with \(P_{\mathrm{gpu}}\) concurrent threads.

In practice the slowest part is not the quadratic draw. It is the weighting solve. The weighting
problem is linear per iteration on the sparse edge list, but the solver uses sequential sweeps that
parallelize poorly. Thus the theoretical work count and the practical wall-clock bottleneck are
different. In work, the draw is the large step. In wall-clock time, the weight assignment is the large
step. Throughout this appendix we assume that the reconstructed network is sparse,
\(E=\Theta(N_F)\), and that the number of sectors is much smaller than the number of firms,
\(N_S\ll N_F\).

\subsection{Logistic Gravity}

The gravity step estimates the parameters that determine the probability of a link between two
firms. If this were done directly on firms, each evaluation of the objective would visit all ordered
pairs \((i,j)\), and the cost would be \(O(N_F^2)\) per solver iteration. We avoid this by using
sector-size bins.

The idea is simple. In a one-time preprocessing pass, each firm is assigned to a sector-bin cell. We
then store the relevant bin-level quantities, such as counts and powers of firm size. This pass costs
\(O(N_F)\). After that, the optimization no longer loops over firms. It loops over pairs of
sector-bin cells. If there are \(N_S\) sectors and \(B\) size bins per sector, one evaluation of the
objective, constraints, and gradients costs \(O(N_S^2B^2)\). This does not grow with \(N_F\) when
\(N_S\) and \(B\) are fixed.

Thus the cost of the gravity step is
\[
        O(N_F)+O(I_{\mathrm{NLP}}N_S^2B^2)
\]
where \(I_{\mathrm{NLP}}\) is the number of nonlinear-programming iterations. For fixed sector count,
fixed bin count, and moderate iteration count, this is essentially linear in the number of firms.\footnote{\emph{Binning error.} Let \(\phi(x_i,x_j)\) be a twice continuously differentiable pairwise score with bounded second derivatives on the binned range, and represent each cell by its weighted centroid. A pointwise replacement can have a first-order \(O(\Delta)\) error, but in the cell average the linear Taylor terms cancel. The remaining error is controlled by the weighted within-cell second moments, \(O(\sum_b\omega_b\operatorname{Var}_b)\). If the outer normalized objective or gradient has bounded derivatives on the relevant range, its discrepancy has the same order. Under equal-width binning on a bounded range, this is \(O(\Delta^2)=O(B^{-2})\).}

\subsection{Bernoulli Draw}

The draw is the only genuinely quadratic step. For each admissible ordered pair of firms we decide
whether the buyer-seller link is present. The total number of pair decisions is therefore
\(\Theta(N_F^2)\). At the full US scale this is an enormous number of potential pairs.

This step is feasible because the decisions are independent. There is no shared state across pairs.
Our implementation assigns one GPU thread to each ordered pair. The thread computes the link
probability and decides the Bernoulli outcome using a stateless counter-based hash of the pair
index. No \(N_F\times N_F\) probability matrix is ever stored. The output is only the realized edge
list. Since the calibrated mean degree is fixed, the expected number of realized edges is
\[
        E=O(N_F)
\]
Thus the draw has total work \(\Theta(N_F^2)\), but wall-clock cost
\[
        O(N_F^2/P_{\mathrm{gpu}})
\]
where \(P_{\mathrm{gpu}}\) is the number of concurrent GPU threads. The degree-floor repair that follows touches only deficient rows and their candidate entries. Under a sparse non-materialized sampler with bounded expected work per added edge, its expected cost is \(O(N_F+E)\) when the aggregate expected repair count is \(O(N_F)\); otherwise the sampling work must be counted separately.

There is a serial alternative. Once firms are binned, all pairs in a sector-bin cell share the same
probability. One could draw the number of links in each cell from a binomial distribution and then
place those links by geometric skipping, in expected time \(O(E)\) plus the number of cells
\citep{batagelj-brandes2005}. We do not use that route. The GPU draw is exact at the firm-pair
level, requires no binning approximation in the realized support, is simple to reproduce, and is not
the practical bottleneck of the pipeline.

\subsection{Markov Regularity}

After the draw, the graph must be made suitable for money circulation. This means it must be
strongly connected and aperiodic. Since the realized support is sparse, \(E=O(N_F)\), the strongly
connected components can be found in linear time using standard algorithms \citep{tarjan1972}.
The condensation graph, its sources and sinks, and the candidate component pairs can also be
computed in \(O(N_F)\). If the aim were only to add the minimum number of arcs to make the
condensation strongly connected, the Eswaran--Tarjan construction would do this in linear time
\citep{eswaran-tarjan1976}.

Our closure is slightly more disciplined. We do not merely add arbitrary component-level arcs. We
choose firm-level links using the sectoral placement objective above. For each ordered
sink-source component pair \((c,c')\in\mathcal P\), let \(\Omega_{cc'}^{\mathrm{raw}}\) be the set of absent
firm-level edges from component \(c\) to component \(c'\). We do not optimize over all of this set.
We thin it to a candidate set \(\Omega_{cc'}\) of size
\[
        L_{cc'} =
        \min\left\{
        |\Omega_{cc'}^{\mathrm{raw}}|,\,
        \max\left\{
        k_{cc'},\,
        \left\lceil g_{\nu'}(n_{cc'})\,n_{cc'} \right\rceil
        \right\}
        \right\}
\]
where \(k_{cc'}\) is the number of new links prescribed between the two components,
\(n_{cc'}\) is the size of the smaller component, and
\[
        g_{\nu'}(n)
        :=
        \bar{\gamma}
        \left(\frac{n}{n_0+n}\right)^{\nu'},
        \qquad n\in\mathbb N,\quad \bar\gamma>0,\quad n_0>0,\quad \nu'>0
\]
The function \(g_{\nu'}\) is increasing and saturates at \(\bar\gamma\). The parameter \(\nu'\) governs
the speed of saturation and is distinct from the parameter \(\nu\) used elsewhere. Because \(c\) is
a sink and \(c'\) a distinct source of the original condensation graph, no edge runs from \(c\) to
\(c'\). Hence \(|\Omega_{cc'}^{\mathrm{raw}}|=|V(c)|\,|V(c')|\). Since
\(k_{cc'}\le n_{cc'}\le|\Omega_{cc'}^{\mathrm{raw}}|\), the sampling step is well defined, and
\(L_{cc'}=O(n_{cc'})\). The cardinality of the raw set is computed from component sizes, not by
listing its pairs.
The \(L_{cc'}\) candidates are sampled sequentially without replacement, with probabilities proportional to the pre-repair gravity probabilities \(p_{ij}\), but \(\Omega_{cc'}^{\mathrm{raw}}\) is never materialized. Since \(p_{ij}=x_{ij}/(1+x_{ij})\), we propose pairs from \(V(c)\times V(c')\) with probability proportional to \(x_{ij}\), using the sector-pair mixture in \(\lambda\) and the factorized firm fitness within each block. Each proposal is accepted with probability \((1+x_{ij})^{-1}\), rejecting repeats, so conditional on acceptance the next distinct pair is again proportional to \(p_{ij}\). If fewer than \(L_{cc'}\) pairs carry positive intensity, all such pairs are kept and the rest are filled uniformly from the unchosen pairs. The linear candidate-generation bound assumes \(O(L_{cc'})\) expected proposals; any component pair for which this fails is enumerated directly, preserving correctness but not the linear bound for that pair.
By this pairing,
\[
        |\Omega|
        =
        \sum_{(c,c')\in\mathcal P}|\Omega_{cc'}|
        =
        O\!\left(\sum_{(c,c')\in\mathcal P} n_{cc'}\right)
        =
        O(N_F).
\]

The optimization over \(\Omega\) has binary variables \(y_{ij}\), but we do not solve it as a hard combinatorial
problem. The candidate sets \(\{\Omega_{cc'}\}\) partition \(\Omega\), and the only constraints are
one equality per group,
\[
        \sum_{(i,j)\in\Omega_{cc'}}y_{ij}=k_{cc'}
\]
together with \(0\le y_{ij}\le 1\). This is the base polytope of a partition matroid. Its constraint
matrix is totally unimodular, so the relaxation has integral vertices and the Frank--Wolfe linear
minimization step separates across component pairs. After the relaxed solve terminates, let
\(\bar y\) denote the final relaxed point: the converged relaxed minimizer if the relaxation is solved
to tolerance, or the last Frank--Wolfe iterate if a fixed iteration cap \(I_{\mathrm{FW}}\) is used. Using the gradient at
\(\bar y\), the final group-wise oracle selects the \(k_{cc'}\) candidates with the smallest gradient entries inside each group. As in Appendix~\ref{app:closure}, integrality is a property of the vertices and the linear oracle, not of the relaxed quadratic minimizer, so this is a placement rule rather than a global-optimality certificate. If \(I_{\mathrm{FW}}\) relaxed Frank--Wolfe iterations are used, the relaxed placement costs \(O(I_{\mathrm{FW}}|\Omega|)\) and the final group-wise top-\(k\) oracle \(O(|\Omega|)\); since \(|\Omega|=O(N_F)\), these become \(O(I_{\mathrm{FW}}N_F)\) and \(O(N_F)\).

Aperiodicity is then handled by a linear-time period test and, if needed, one additional edge chosen
from the thinned aperiodicity candidate set to minimize the induced change in the placement score. This
final pass costs \(O(N_F+|\widehat E|+|\Omega^{\mathrm{ap}}|)\), which is \(O(N_F)\) under sparsity and
the same linear candidate-generation condition. An admissible edge always exists by Lemma~\ref{lem:noexception}, so no full absent-set screen is needed.

\medskip\noindent\emph{Weight assignment.}\ 
The last step puts weights on the realized support. Since weights live only on realized edges, the
number of variables is the number of edges,
\[
        E=\sum_i \deg(i)=\Theta(N_F)
\]
Each iteration of the weighting solver consists of sparse operations. First, we compute firm-level
and sector-level quantities such as \((\mathbf W^\top \mathbf m)_j\) and
\(\widehat s_\ell(\mathbf W)\). Second, we project each buyer row of \((w_{ij})\) onto a capped
probability simplex, enforcing row-stochasticity and the relevant bounds. Both operations cost
\(O(E)\) per iteration.

If \(I_{\mathrm{QP}}\) denotes the number of iterations needed to reach the desired tolerance, the total
cost of the weighting step is therefore
\[
        O(I_{\mathrm{QP}}E)=O(I_{\mathrm{QP}}N_F)
\]
This is linear per iteration. But linear per iteration does not mean fast. The consensus-ADMM
implementation uses Gauss--Seidel sweeps. Buyer rows are updated in sequence, and each update
uses the running seller-column sums left by the previous updates. This makes the step hard to
parallelize. As a result, the weighting solve is the binding wall-clock cost at full scale, even though
its arithmetic cost per iteration is linear.

The pipeline has two different complexity stories. In total work, the Bernoulli draw dominates:
it performs \(\Theta(N_F^2)\) independent pair decisions. The other stages are essentially linear
under sparsity. Gravity estimation costs \(O(N_F)\) preprocessing plus
\(O(I_{\mathrm{NLP}}N_S^2B^2)\) optimization work on the sector-bin collapse. Markov closure costs
\(O(I_{\mathrm{FW}}N_F)\) for the relaxed placement plus \(O(N_F)\) for the final oracle and period pass when candidate generation is linear, while weighting costs \(O(I_{\mathrm{QP}}N_F)\). In wall-clock time, the ordering is different. The quadratic draw is embarrassingly parallel and
runs on the GPU, with wall-clock cost \(O(N_F^2/P_{\mathrm{gpu}})\). The weighting step is only
linear per iteration, but its Gauss--Seidel sweeps parallelize poorly and must be repeated
\(I_{\mathrm{QP}}\) times. At full scale this makes the weighting solve the practical bottleneck of the
pipeline. These are practical scaling laws under sparsity and calibrated mean degree, not worst-case
complexity guarantees.

Table~\ref{tab:complexity_summary} summarizes the cost of each step. The bounds use
\(N_F\) for firms, \(N_S\) for sectors, \(B\) for size bins, \(E=\Theta(N_F)\) for realized edges,
\(P_{\mathrm{gpu}}\) for concurrent GPU threads, and \(I_{\mathrm{NLP}}\), \(I_{\mathrm{FW}}\), and \(I_{\mathrm{QP}}\) for the relevant
iteration counts.

\begin{table}[H]
\centering
\caption{Computational complexity of each step of the network reconstruction algorithm}
\label{tab:complexity_summary}
\setlength{\tabcolsep}{4pt}
\begin{tabular}{
  >{\raggedright\arraybackslash}p{0.14\linewidth}
  >{\raggedright\arraybackslash}p{0.50\linewidth}
  >{\raggedright\arraybackslash}p{0.28\linewidth}
}
\toprule
\textbf{Step} & \textbf{Description} & \textbf{Complexity} \\
\midrule
1: Logistic gravity
&
Estimate link probabilities using a sector-size-bin representation rather than all firm pairs.
&
\(O(N_F)+O(I_{\mathrm{NLP}}N_S^2B^2)\)
\\
\midrule
2: Bernoulli draw
&
Make one independent Bernoulli decision per ordered pair on the GPU; store only the realized edge
list; repair deficient firms with sparse non-materialized sampling.
&
Work \(\Theta(N_F^2)\); wall-clock \(O(N_F^2/P_{\mathrm{gpu}})\); expected repair \(O(N_F+E)\) if the repair count is \(O(N_F)\) and expected work per added edge is bounded
\\
\midrule
3: Markov regularity
&
Find SCCs, add cross-component links using the sector-flow partition-matroid surrogate, and add
one aperiodizing edge if needed.
&
\(O(I_{\mathrm{FW}}N_F)\) for relaxed placement; \(O(N_F)\) final pass under linear candidate generation
\\
\midrule
4: Weight assignment
&
Solve the minimum-energy weighting problem on the realized support. The arithmetic cost is linear
per iteration, but the Gauss--Seidel sweeps are the practical bottleneck.
&
Per iteration \(O(N_F)\); total \(O(I_{\mathrm{QP}}N_F)\)
\\
\bottomrule
\end{tabular}
\end{table}

\section{Robustness: Firm-Size Drift, Sectoral Fidelity, and Mixing}
\label{app:robustness}

This appendix collects three checks on the reconstructed money-flow matrix. The first asks whether
the empirical firm-size vector remains close to the stationary distribution of the reconstructed
economy. The second asks whether the sectoral flows implied by the stationary economy remain
close to the input--output table. The third asks how fast the money chain mixes, and how much
draw-to-draw dispersion remains in the small-scale ensembles.

The issue arises because the weighting step does not directly impose stationarity. It makes the
observed firm-size vector nearly a one-step fixed point of the money chain. That is, it makes
\(\mathbf W^\top \mathbf m\) close to \(\mathbf m\). But the object used in the propagation exercises is
the stationary distribution of \(\mathbf W\). We therefore need to know whether a small one-step error
can accumulate into a large stationary error.

We first prove a bound. If \(J\) is the size-weighted one-step error and \(\Theta(\mathbf W)\) is the sum of the Dobrushin
coefficients of the powers of \(\mathbf W\), then the stationary money vector is within \(\Theta(\mathbf W)J\) of the empirical
money vector in \(L^1\), and the gap is an exact linear response to the one-step residual. The bound is useful because it gives an analytic guarantee. It is also weak
in large networks, because \(\Theta(\mathbf W)\) cannot be bounded a priori for a sparse chain. We therefore check the actual drift in all six reconstructions and report the detailed firm-level
diagnostics for the full US network below. In every economy the stationary money vector relocates
only a few percent of total money; in the largest case it also remains close firm by firm. The
proposition is therefore conservative.

\subsection{The Stationary-Money Bound}
\label{app:statmoney}

The weighting step makes the empirical firm-size vector almost invariant after one application of
the money-flow matrix. This subsection shows what that implies for the stationary distribution.

Let \(\mathbf W\) be the row-stochastic weight matrix. Let \(\mathbf m\) be the fitted firm-size vector,
\(M=\mathbf 1^\top \mathbf m\), and
\[
        \boldsymbol\mu=\mathbf m/M
\]
be the normalized empirical money vector. Let \(\mathbf v\) be the stationary distribution of
\(\mathbf W\), so that
\[
        \mathbf v^\top=\mathbf v^\top\mathbf W,
        \qquad
        \mathbf 1^\top\mathbf v=1
\]
The question is how far \(\mathbf v\) can be from \(\boldsymbol\mu\).

Eigenvalues alone are not enough for this purpose. The matrix \(\mathbf W\) is generally non-normal,
so the modulus of the second eigenvalue does not control the norm of the resolvent
\((\mathbf I_{N_F}-\mathbf W^\top)^{-1}\). We therefore use Dobrushin's ergodicity coefficient,
\[
\tau(\mathbf W)
:=
\tfrac{1}{2}
\max_{i,i'}
\sum_{j=1}^{N_F}
\big|w_{ij}-w_{i'j}\big|
\in[0,1]
\]
This is the largest total-variation distance between two rows of \(\mathbf W\). It contracts
zero-sum vectors:
\[
        \|\mathbf W^\top \mathbf u\|_1
        \le
        \tau(\mathbf W)\|\mathbf u\|_1
\]
for every \(\mathbf u\) such that \(\mathbf 1^\top\mathbf u=0\)
\citep[Ch.~4.3]{seneta2006nonnegative}.

Let \(e_j=|(\mathbf W^\top\mathbf m)_j-m_j|/m_j\) be the relative one-step deviation of firm \(j\) and \(J=(\sum_j\mu_je_j^{2})^{1/2}\) its size-weighted root mean square, as in Section~\ref{subsec:weights}. For every $t\ge0$ the power $\mathbf W^t$ is row-stochastic, and we write
\[
        \Theta(\mathbf W):=\sum_{t\ge0}\tau(\mathbf W^t)
\]
for the ergodic sum of $\mathbf W$. Let $\gamma$ be the primitivity index of $\mathbf W$, the least $t$ with $\mathbf W^t>0$.

\begin{proposition}[Stationary money near the empirical firm-size vector]
\label{rem:stationary_vs_onestep}
Let \(\mathbf W\) be row-stochastic and primitive with stationary distribution \(\mathbf v\), and let \(\mathbf r:=\boldsymbol\mu-\mathbf W^\top\boldsymbol\mu\) be the one-step residual of the normalized firm-size vector. Then
\begin{gather*}
\boldsymbol\mu-\mathbf v=\sum_{t\ge0}(\mathbf W^\top)^t\mathbf r,
\qquad
\|\boldsymbol\mu-\mathbf v\|_1\le\Theta(\mathbf W)\,\|\mathbf r\|_1\le\Theta(\mathbf W)\,J,\\
\Theta(\mathbf W)\le\min\Big\{\frac{1}{1-\tau(\mathbf W)},\ \frac{\gamma}{1-\tau(\mathbf W^{\gamma})}\Big\}<\infty .
\end{gather*}
\end{proposition}

The residual norm is the size-weighted mean relative error, \(\|\mathbf r\|_1=\sum_j\mu_je_j\). The squared error \(J^{2}=\sum_j(\mathbf W^\top\mathbf m-\mathbf m)_j^{2}/(Mm_j)\) is the \(\chi^{2}\)-divergence of the one-step money vector \(\mathbf W^\top\boldsymbol\mu\) from \(\boldsymbol\mu\), so the firm-size cap \(\delta^{\mathrm{firm}}\) is a cap on that divergence. The first display says that the stationary gap is the one-step residual pushed through the powers of the chain, and nothing else. The ergodic sum is finite for every primitive \(\mathbf W\), and it is infinite for every reducible or periodic \(\mathbf W\), since two rows of \(\mathbf W^t\) then have disjoint supports for every \(t\).

\begin{proof}[Proof of Proposition~\ref{rem:stationary_vs_onestep}]
Fix \(\mathbf W\) as in the statement. Both \(\boldsymbol\mu\) and \(\mathbf W^\top\boldsymbol\mu\) are probability vectors, so the residual \(\mathbf r\) has zero sum. Telescoping gives \(\boldsymbol\mu-(\mathbf W^\top)^n\boldsymbol\mu=\sum_{t<n}(\mathbf W^\top)^t\mathbf r\) for every \(n\), and primitivity gives \((\mathbf W^\top)^n\boldsymbol\mu\to\mathbf v\). This proves the identity.

For the bound, the Dobrushin contraction applied to the stochastic matrix \(\mathbf W^t\) and the zero-sum vector \(\mathbf r\) gives \(\|(\mathbf W^\top)^t\mathbf r\|_1\le\tau(\mathbf W^t)\|\mathbf r\|_1\) for every \(t\). Summing over \(t\) in the identity gives \(\|\boldsymbol\mu-\mathbf v\|_1\le\Theta(\mathbf W)\|\mathbf r\|_1\). The ergodic coefficient is submultiplicative, \(\tau(\mathbf W^{s+t})\le\tau(\mathbf W^{s})\tau(\mathbf W^{t})\) \citep[Ch.~4.3]{seneta2006nonnegative}, and \(\tau(\mathbf W^{\gamma})<1\) because \(\mathbf W^{\gamma}\) is entrywise positive. Writing \(t=b\gamma+u\) with \(0\le u<\gamma\) therefore gives \(\tau(\mathbf W^{t})\le\tau(\mathbf W^{\gamma})^{b}\), and summing the geometric series gives \(\Theta(\mathbf W)\le\gamma/(1-\tau(\mathbf W^{\gamma}))\). The same argument with \(\gamma\) replaced by one gives \(\Theta(\mathbf W)\le1/(1-\tau(\mathbf W))\) whenever \(\tau(\mathbf W)<1\).

Finally, \(|r_j|=|m_j-(\mathbf W^\top\mathbf m)_j|/M=\mu_je_j\), so \(\|\mathbf r\|_1=\sum_j\mu_je_j\), and the Cauchy--Schwarz inequality gives \(\sum_j\mu_je_j\le(\sum_j\mu_j)^{1/2}(\sum_j\mu_je_j^{2})^{1/2}=J\).
\end{proof}

The proposition needs only primitivity. The minimum-degree floor is not needed for the existence of a unique stationary distribution, and it is not needed for the algebra above. Its role is quantitative: it removes the most degenerate rows of the money-flow matrix, and it makes the last inequality of the proposition explicit. Let
\[
        w_{\min}
        =
        \min_{(i,j):a_{ij}=1}w_{ij}>0
\]
be the smallest realized edge weight. The per-edge floor gives \(w_{\min}\ge
\varepsilon_{\mathrm{floor}}\), and the minimum-energy objective keeps every backbone edge active.
Since every firm has out-degree at least two and no self-loop, no row of \(\mathbf W\) is a point mass.
Every firm sends money to at least two distinct successors, each with weight at least \(w_{\min}\).
This rules out deterministic rows and near-absorbing degree-one bottlenecks. If two firms share a common supplier \(k\), then \(\min(w_{ik},w_{i'k})\ge w_{\min}\), so their one-step row overlap is bounded below and their pairwise Dobrushin distance is at most \(1-w_{\min}\). The floor creates local overlap wherever the graph structure permits it.

The floor also gives a crude explicit constant. Since every realized edge has weight at least \(w_{\min}\) and \(\mathbf W^{\gamma}>0\), every entry of \(\mathbf W^{\gamma}\) is at least \(w_{\min}^{\gamma}\), hence \(\tau(\mathbf W^{\gamma})\le1-N_Fw_{\min}^{\gamma}\) and
\[
        \|\boldsymbol\mu-\mathbf v\|_1
        \le
        \Theta(\mathbf W)\,J
        \le
        \frac{\gamma J}{1-\tau(\mathbf W^{\gamma})}
        \le
        \frac{\gamma J}{N_F w_{\min}^{\gamma}} .
\]
This constant is not meant to be sharp. Its purpose is to show where the floor enters. Adding floor edges cannot lengthen any existing reachability path, but the primitive index $\gamma$ may still be large, and the factor $w_{\min}^{-\gamma}$ may make the display numerically weak. A useful global constant would require stronger path-overlap or scrambling assumptions than the degree floor alone provides. The in-degree floor plays the same role for the reversed chain: no firm receives all of its inflow through a single source.

It is worth being explicit about what these results do and do not deliver, since neither certifies in advance that a given reconstruction is close to its own stationary state. What we prove is a local, conditional statement---that on a neighborhood of an exactly balanced weighting the fitted weights stay within a controlled ball---together with a Dobrushin inequality whose constant we cannot bound a priori. Neither is a general guarantee, and it is useful to see precisely why.

Two obstructions block an unconditional bound of the form \(\|\boldsymbol\mu-\mathbf v\|_1\le\varepsilon\). The first is spectral. The stationary vector \(\mathbf v\) is a global function of \(\mathbf W\), and for a non-normal \(\mathbf W\) the resolvent \((\mathbf I_{N_F}-\mathbf W^\top)^{-1}\) on the zero-sum subspace is not controlled by the spectral gap \(1-|\lambda_2|\). Pseudospectral transients can be large even when \(|\lambda_2|\) is small \citep{TrefethenEmbree2005}. This is why the argument runs through the Dobrushin coefficient rather than the spectrum. But the Dobrushin route has its own limitation. \(\tau(\mathbf W)=1\) as soon as two firms have disjoint supplier sets, which is generic at mean degree \(50\ll N_F\), so the first term of the ergodic sum \(\Theta(\mathbf W)\) in Proposition~\ref{rem:stationary_vs_onestep} is one and only some power \(\mathbf W^\gamma\) is scrambling. The explicit constant \(\gamma/(N_F w_{\min}^{\gamma})\) is then exponentially weak in the reachability index unless the backbone expands. There is no closed-form global constant that is both valid and useful for a sparse non-normal chain.

The second obstruction is combinatorial. The map from the data to the weights runs through the minimum-energy program, whose solution is only piecewise smooth. As firm sizes or the support move, the active set of the program---which realized edges sit at the floor \(\varepsilon_{\mathrm{floor}}\), which firms bind the worst-tail cap---changes, and at each such change the map \(\text{data}\mapsto\mathbf W\), and hence \(\text{data}\mapsto\mathbf v\), is non-differentiable. A single smooth bound cannot survive these switches. Compounding both obstructions, a fixed sparse support need not admit any exactly money-preserving row-stochastic matrix \(\widetilde{\mathbf W}\) with \(\widetilde{\mathbf W}^\top\mathbf m=\mathbf m\), and without such a matrix there is no center around which to expand. Whether it does is a question about the binary backbone alone, and we now record the answer. Write \(N^{+}(S)\) for the sellers adjacent to a set \(S\) of buyers and \(N^{-}(T)\) for the buyers adjacent to a set \(T\) of sellers. Put \(m'_i:=m_i(1-\varepsilon_{\mathrm{floor}}d_i^{\mathrm O})\) and \(m''_j:=m_j-\varepsilon_{\mathrm{floor}}\sum_i\widetilde a_{ij}m_i\), the masses left after every support edge carries its floor flow.

\begin{proposition}[Existence of an exactly money-preserving weighting]
\label{prop:gale}
A row-stochastic \(\widetilde{\mathbf W}\) supported on \(\Aaug\), with \(\widetilde w_{ij}\ge\varepsilon_{\mathrm{floor}}\) on every support edge and \(\widetilde{\mathbf W}^\top\mathbf m=\mathbf m\), exists if and only if \(m'_i\ge0\) and \(m''_j\ge0\) for all firms and
\[
\sum_{i\in S}m'_i\le\sum_{j\in N^{+}(S)}m''_j\qquad\text{for every set }S\text{ of buyers}.
\]
\end{proposition}

The condition is equivalent to its seller form, \(\sum_{j\in T}m''_j\le\sum_{i\in N^{-}(T)}m'_i\) for every set \(T\) of sellers, and it is checked by one maximum-flow computation on a network with \(E\) arcs. Two families of necessary conditions cost \(O(E)\): no buyer is larger than its suppliers combined, \(m'_i\le\sum_{j\in N^{+}(i)}m''_j\), and no seller is larger than its customers combined, \(m''_j\le\sum_{i\in N^{-}(j)}m'_i\). A firm that fails either check is a firm for which no weighting on the given support can be balanced, and the cap \(\delta^{\mathrm{firm}}\) has to absorb it.

\begin{proof}[Proof of Proposition~\ref{prop:gale}]
Fix the support \(\Aaug\) and write \(f_{ij}=m_iw_{ij}\) for dollar flows. A matrix \(\widetilde{\mathbf W}\) as in the statement exists if and only if a nonnegative matrix \(\mathbf F\) supported on \(\Aaug\) has row sums \(\mathbf m\), column sums \(\mathbf m\), and \(f_{ij}\ge\varepsilon_{\mathrm{floor}}m_i\) on the support. Subtracting the floor flows, \(f_{ij}=\varepsilon_{\mathrm{floor}}m_i\widetilde a_{ij}+f'_{ij}\), reduces this to a nonnegative \(\mathbf F'\) supported on \(\Aaug\) with row sums \(m'\) and column sums \(m''\), which have equal totals. Build the network with a source joined to each buyer \(i\) by an arc of capacity \(m'_i\), each seller \(j\) joined to a sink by an arc of capacity \(m''_j\), and an uncapacitated arc for each support edge. Such an \(\mathbf F'\) exists if and only if the maximum flow equals \(\sum_im'_i\), and by the max-flow min-cut theorem this holds if and only if every cut has capacity at least \(\sum_im'_i\), which reduces to the displayed condition \citep{Gale1957}. The seller form is the same condition on the transposed network.
\end{proof}

What remains provable is therefore local, and Proposition~\ref{rem:error_bound_W} is exactly that: a local sensitivity bound valid until the first combinatorial change in the binding constraints. Three features of the problem make even this much possible. First, the objective \(\tfrac12\|w\|_2^2\) is strongly convex with Hessian \(\mathbf I_E\), so the minimizer is unique and the solution map is well defined. Second, after fixing a residual vector \(r\), the row sums, the support and floor bounds, and the balance equalities \(\mathbf Cw=\mathbf d+r\) form a polyhedral quadratic program, so on a fixed active set the perturbed Karush--Kuhn--Tucker system gives a local Lipschitz dependence of the solution on \(r\). Third, at an exact-preserving center meeting the linear independence constraint qualification and strict complementarity, the active Karush--Kuhn--Tucker system is a nonsingular square system whose active set is locally constant. The radius \(\bar\eta\) is chosen before the first perturbation at which an active multiplier or an inactive slack reaches zero, that is, before the first active-set change. Past it a constraint switches, the system defining the branch changes, and the local Lipschitz bound lapses. This is the sense in which a ``there exists a ball'' statement is available but a global one is not: the ball is the region on which the binding constraints, and therefore the active Karush--Kuhn--Tucker system governing the weights, do not change.

Neither result certifies that our fitted \(\mathbf W\) lands in the favorable regime: that \(\Theta(\mathbf W)\) is moderate in Proposition~\ref{rem:stationary_vs_onestep}, or that the achieved residual satisfies \(\eta\le\bar\eta\) around an existing center in Proposition~\ref{rem:error_bound_W}. The closeness of the fitted and stationary firm-size vectors is therefore ultimately an empirical matter. The aggregate check relocates only a few percent of total money in each of the six economies. The self-consistency check below (Appendix~\ref{app:selfconsistency}) reports the detailed firm-level drift for the full US reconstruction, the largest case; the remaining cross-country evidence in this appendix concerns sectoral fidelity, mixing, and the ESRI comparisons.

\subsection{Firm-Size Self-Consistency and Sectoral Fidelity}
\label{app:selfconsistency}

What happens in the actual reconstructed economy? This matters because the reconstruction is
only as trustworthy as the closeness of its stationary firm sizes to the data, and the analytic
bound above, though useful, can be loose. The empirical check is more direct: start from the reconstructed firm sizes, run
the money chain to stationarity, and compare the stationary firm sizes with the original ones.

Let \(\widehat{\boldsymbol\pi}=M\mathbf v\) denote the stationary firm-size vector of \(\mathbf W\), scaled to total money \(M\). For each firm define
\[
        \delta_j
        =
        \frac{\widehat\pi_j-m_j}{m_j}
\]
Table~\ref{tab:drift} reports the drift for the full US network. The movement is small. The median
stationary-to-census ratio is \(1.04\). About \(92\%\) of firms remain within \(10\%\) of their
reconstructed size, and every firm remains within a factor of two. In aggregate, only \(2.8\%\) of
money relocates in total variation.

\begin{table}[H]\centering
\caption{Firm-size drift from the reconstructed census sizes to the stationary state of the money
chain, full US network.}
\label{tab:drift}
\begin{tabular}{lc}
\toprule
 & US \(6.46\times10^6\) \\
\midrule
Median per-firm ratio (stationary/census) & \(1.04\) \\
Money relocated (total variation) & \(2.8\%\) \\
Firms within \(5\%\) of census size & \(63\%\) \\
Firms within \(10\%\) of census size & \(92\%\) \\
Firms within a factor of \(2\) & \(100\%\) \\
Firm-size tail exponent: census \(\to\) stationary & \(0.72\to0.73\) \\
\bottomrule
\end{tabular}
\end{table}

\begin{figure}[H]
  \centering
  \includegraphics[width=\linewidth]{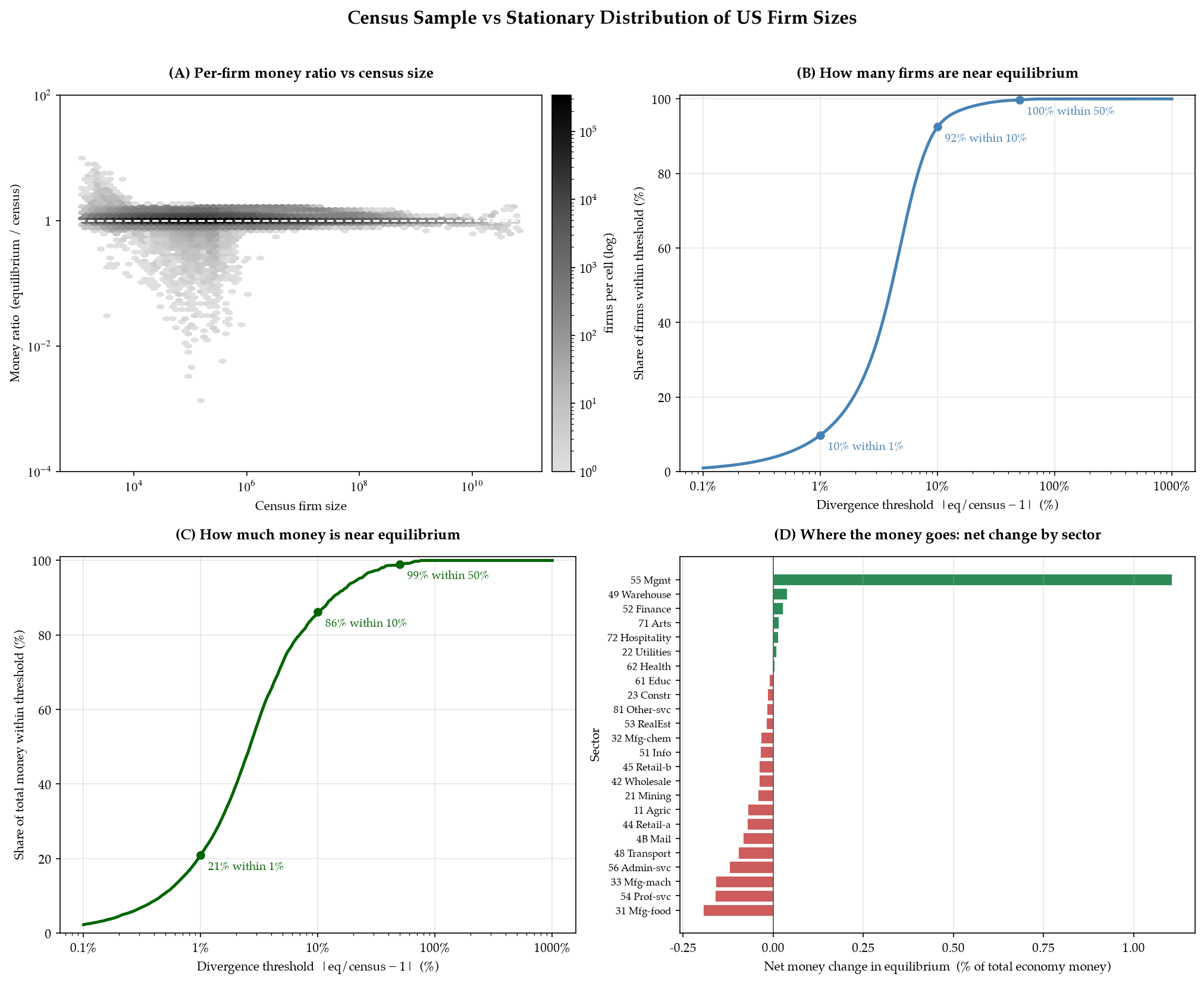}
  \caption{Census versus stationary US firm sizes over the same \(6.46\) million firms. Panel A
  plots the stationary-to-census ratio against census size, both on logarithmic axes and shaded by
  firm density. Panels B and C report the share of firms and the share of money within a given
  tolerance of equilibrium. Panel D reports net sectoral relocation as a share of total money.}
  \label{fig:money_divergence}
\end{figure}

\begin{figure}[H]
  \centering
  \includegraphics[width=.85\linewidth]{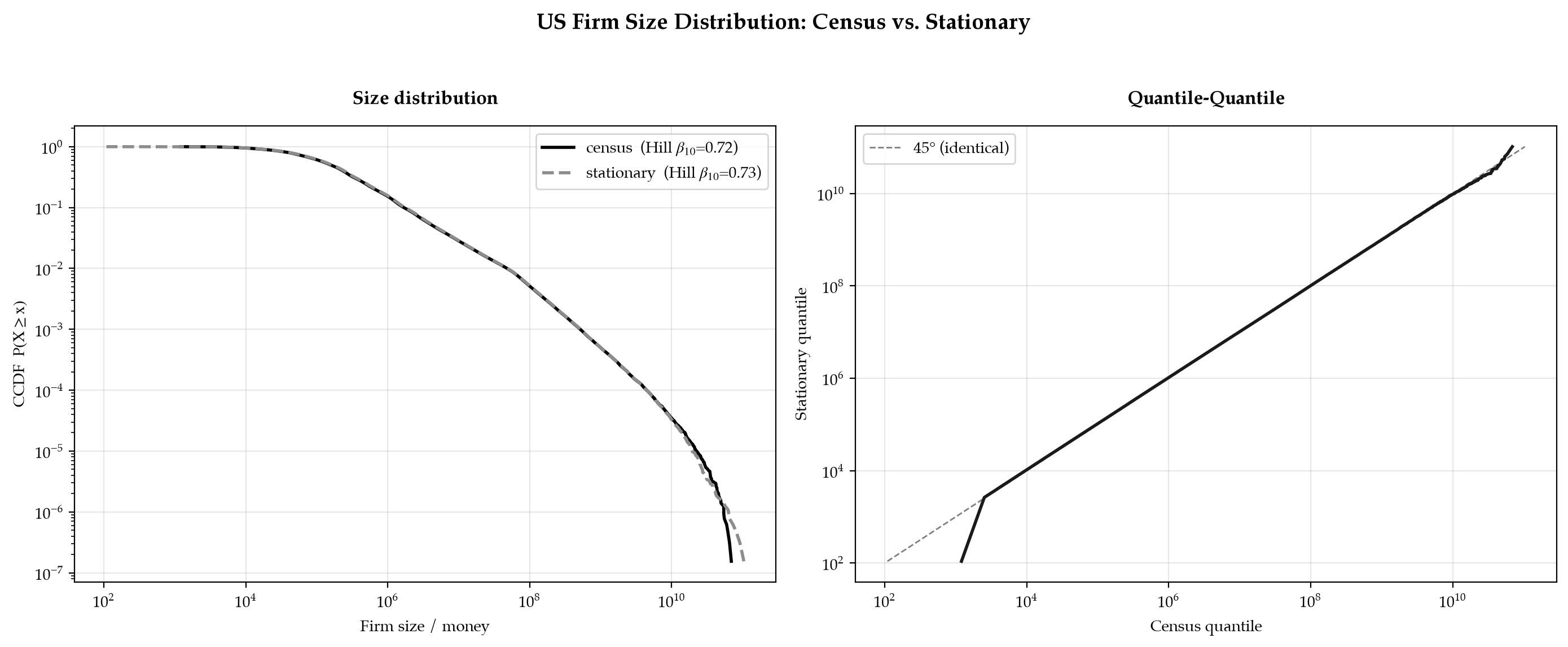}
  \caption{Aggregate size-distribution change, census versus stationary, full US network. The left
  panel overlays the two firm-size CCDFs on log-log axes and reports the Hill top-\(10\%\) tail
  exponent. The right panel gives the quantile-quantile comparison.}
  \label{fig:size_dist_compare}
\end{figure}

Figure~\ref{fig:money_divergence} shows where the drift occurs. The largest firms are almost fixed.
Most of the dispersion is among small and mid-sized firms. At the sector level, the net relocation is
also small, and it mainly pools into the dominant management-of-companies hub. The shape of the
firm-size distribution is essentially unchanged. The census and stationary CCDFs overlap through
the extreme tail, and the Hill top-\(10\%\) exponent changes only from \(0.72\) to \(0.73\)
(Figures~\ref{fig:money_divergence} and~\ref{fig:size_dist_compare}). Thus the propagation objects in
Section~\ref{sec:esri}, which are computed from the stationary economy, are not sensitive to whether
one starts from the census or the stationary money vector.

\begin{figure}[H]
  \centering
  \includegraphics[width=\linewidth]{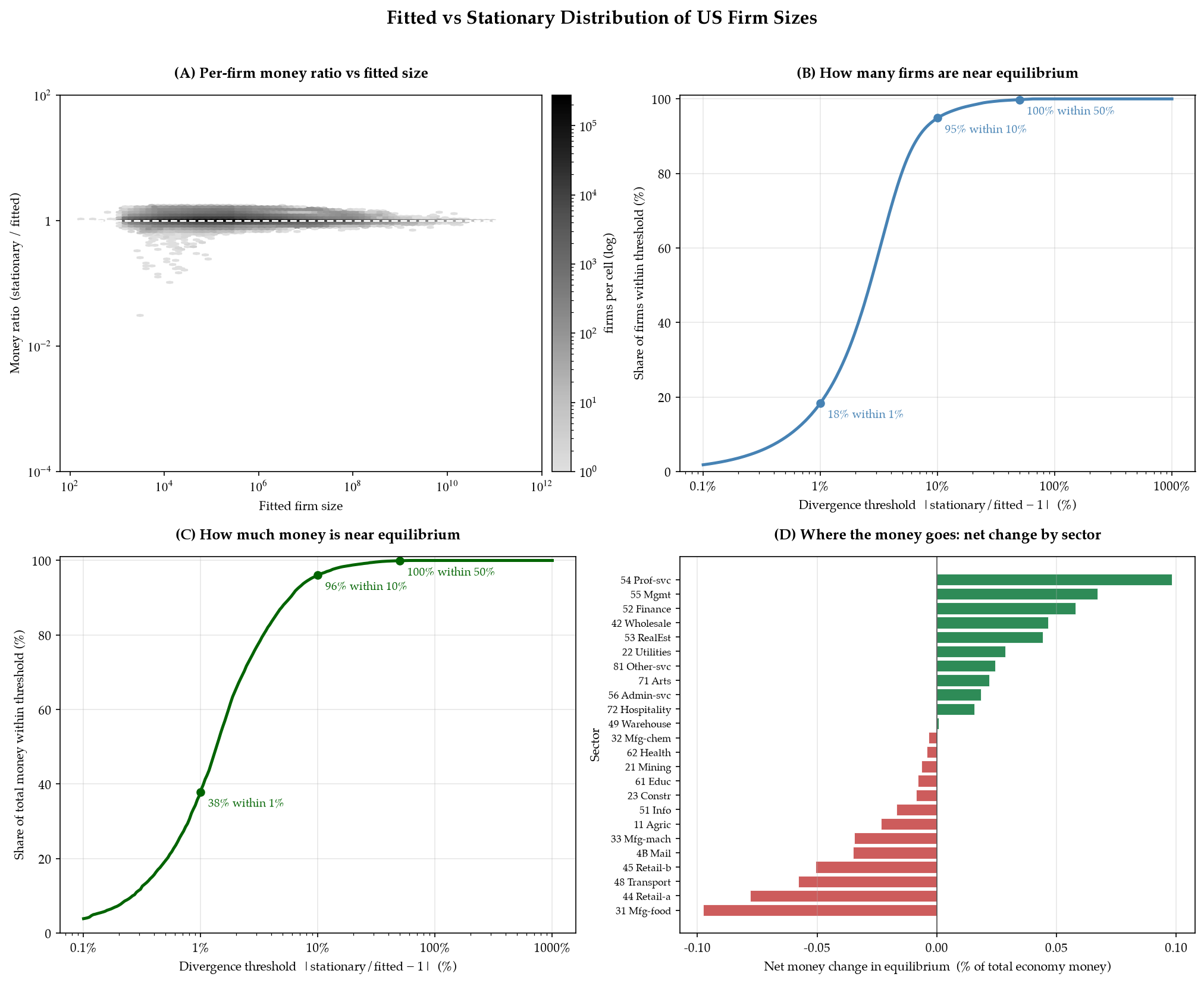}
  \caption{Fitted-money versus stationary firm sizes, full US network, in the same four panels and
  on the same logarithmic axes as Figure~\ref{fig:money_divergence}.}
  \label{fig:fitted_divergence}
\end{figure}

The fitted money vector is even closer to the stationary state. Let
\[
        \mathbf m_1=\mathbf W^\top\mathbf m
\]
be the one-step fitted money vector produced by the optimized weights. Figure~\ref{fig:fitted_divergence}
compares \(\mathbf m_1\) with the stationary vector. The agreement tightens. About \(18\%\) of firms
and \(38\%\) of money lie within \(1\%\) of equilibrium, compared with \(10\%\) and \(21\%\) when
the comparison is made from the census vector. Net sectoral relocation is at most \(0.1\%\). Thus
the model's own fitted output is itself a near-fixed point of the money dynamics.

\medskip\noindent\emph{Sectoral flow residuals.}
\label{app:sector_resid}

The same check can be made at the sector level. The main text compares three sectoral-flow objects:
the RAS-balanced input--output target \(\mathbf F^\star\), the fitted flows, and the stationary
closed-market flows. We record here the signed residuals behind that comparison.

\begin{figure}[H]
  \centering
  \includegraphics[width=\linewidth]{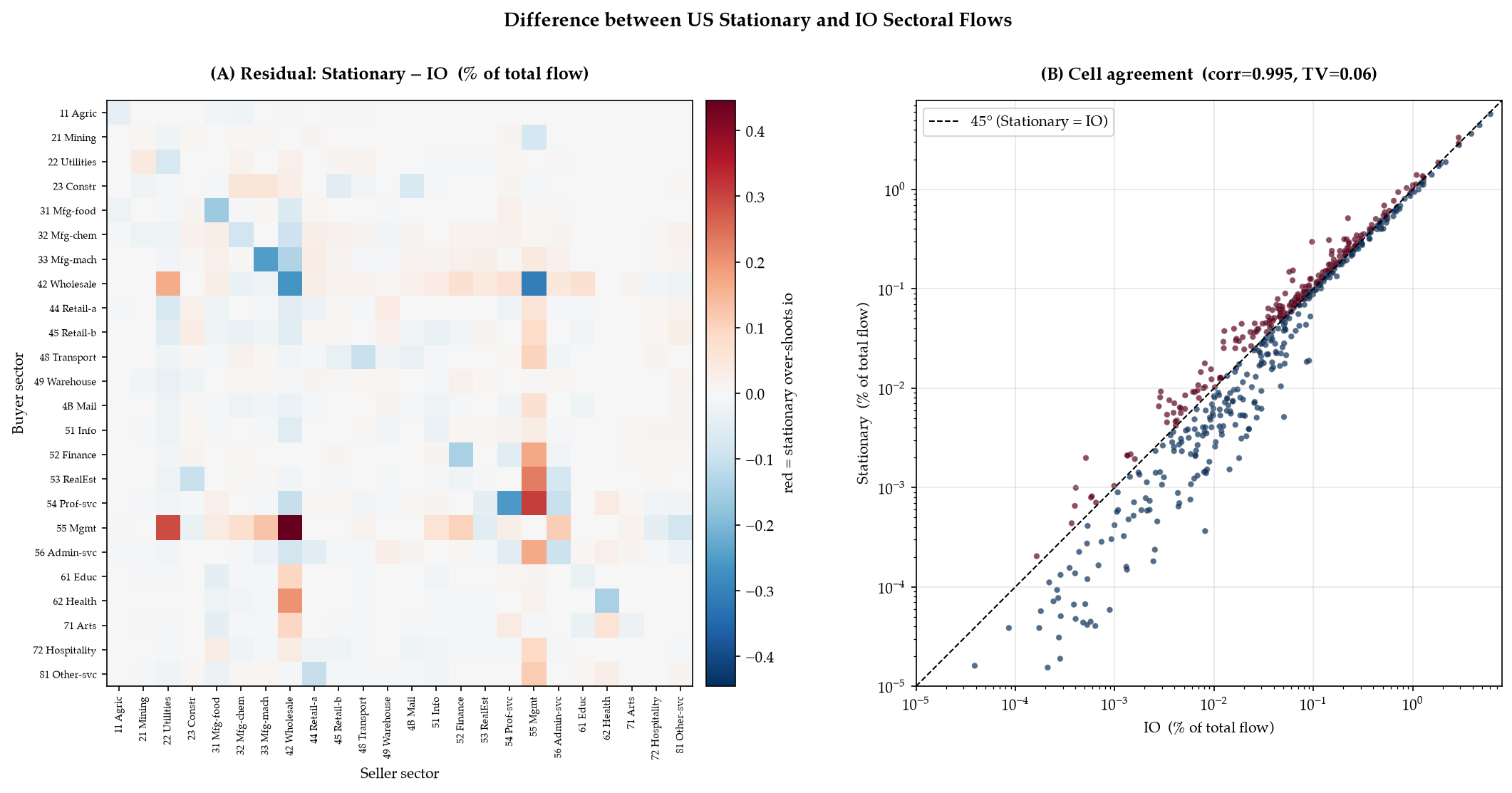}
  \caption{Sectoral-flow residual, stationary minus input--output target, full US network. The left
  panel is a signed heat map over the \(24\times24\) buyer-to-seller sector cells, measured as a
  percent of total flow. The right panel is a cell-by-cell scatter of the stationary share against the
  target share, with correlation and total variation reported in the title.}
  \label{fig:sector_flow_diff}
\end{figure}

Figure~\ref{fig:sector_flow_diff} plots stationary flows minus the input--output target. The residual
is small and structured. The cell-level correlation is \(0.995\), total variation is \(0.057\), and the
main deviation is the over-routing of flow into the management hub. Between the fitted and the
stationary flows, by contrast, the residual is essentially zero: the correlation is
\(1.000\), and the largest cell differs by only \(0.04\) percentage points. The closed-market
stationary dynamics add almost nothing beyond the fitted-flow residual.

\subsection{Mixing of the Money Chain}
\label{app:spectral}

The last check concerns mixing. A propagation experiment is easier to interpret and easier to
compute when the money chain reaches its stationary state quickly. We summarize mixing with the
second-eigenvalue modulus \(|\lambda_2|\) of \(\mathbf W\). The spectral gap is
\(1-|\lambda_2|\), so smaller \(|\lambda_2|\) means faster asymptotic mixing.

There is one qualification. Since \(\mathbf W\) is non-normal, \(|\lambda_2|\) controls only the
asymptotic rate. It does not rule out finite-horizon transients \citep{TrefethenEmbree2005}. This is
why the stationary-money bound above used the Dobrushin coefficient rather than the spectrum.
The eigenvalue table should therefore be read as an ordering of asymptotic mixing speeds, not as a
finite-time error bound.

Table~\ref{tab:lambda2} reports \(|\lambda_2|\) across scales and economies, together with the
corresponding value for the row-normalized sectoral input--output matrix. Three facts are useful. First, the leading eigenvalue is exactly one because the matrices are row-stochastic. Irreducibility and aperiodicity then give a unique stationary distribution and convergence. Second, at full scale the firm-level chains mix quickly. The full-scale values are
\(0.66\) for the United States, \(0.34\) for Japan, and \(0.38\) for the United Kingdom. These values
make forward iteration practical at national scale. Third, the firm networks often mix faster than
the sectoral input--output tables from which they are built. For example, the US firm network has
\(|\lambda_2|=0.66\), while the sector table has \(|\lambda_2|=0.75\). Japan gives \(0.34\) versus
\(0.44\), and the United Kingdom \(0.38\) versus \(0.47\). The reconstructed firm network is
therefore not merely a mechanical disaggregation of the sector table. It is a better-connected
circulation object.

The small-scale ensembles also show why we use the full-scale reconstruction for the propagation
exercises. At \(10^4\) and \(10^5\) firms, the draws have more dispersion, and the \(10^5\) ensembles
are sometimes bimodal. This is the spectral trace of the fit-convergence instability addressed by
the gravity solver in Section~\ref{subsec:gravity}. At national scale the estimates are
much more stable.

\begin{table}[H]\centering
\caption{Second-eigenvalue modulus \(|\lambda_2|\) of the row-stochastic money-flow matrix across
scales, and of the row-normalized sector input--output table. The \(10^4\) and \(10^5\) columns are
mean \(\pm\) standard deviation over \(100\) networks. ``--'' marks scales not built.}
\label{tab:lambda2}
\setlength{\tabcolsep}{5pt}\small
\begin{tabular}{lccccc}
\toprule
Economy & \(10^4\) (mean\(\pm\)sd) & \(10^5\) (mean\(\pm\)sd) & \(10^6\) & full & sector IO (\(S\)) \\
\midrule
United States  & \(0.78\pm0.11\) & \(0.54\pm0.06\) & \(0.63\) & \(0.66\) & \(0.75\) (\(24\)) \\
Japan          & \(0.67\pm0.15\) & \(0.51\pm0.22\) & \(0.32\) & \(0.34\) & \(0.44\) (\(18\)) \\
United Kingdom & \(0.66\pm0.17\) & \(0.55\pm0.22\) & \(0.34\) & \(0.38\) & \(0.47\) (\(17\)) \\
Australia      & \(0.70\pm0.12\) & \(0.66\pm0.18\) & \(0.77\) & \(0.59\) & \(0.44\) (\(19\)) \\
Finland        & \(0.70\pm0.11\) & \(0.60\pm0.13\) & --     & \(0.56\) & \(0.51\) (\(18\)) \\
Denmark        & \(0.70\pm0.15\) & \(0.39\pm0.10\) & --     & \(0.79\) & \(0.44\) (\(19\)) \\
\bottomrule
\end{tabular}
\end{table}

\section{Implementation of the Systemic-Risk Cascade}
\label{app:esri-impl}

This appendix describes the cascade used to compute firm-level systemic risk. The object is simple.
For each firm \(i\), we set its production health to zero, let the shock propagate through suppliers
and customers, and measure the resulting aggregate output loss.

Throughout this appendix, \(\mathbf W\) is the reconstructed row-stochastic input-share matrix.
The entry \(w_{jr}\) is the share of buyer \(j\)'s spending that goes to seller \(r\), so
\[
        \sum_r w_{jr}=1
\]
Firm \(j\)'s size is \(m_j\). The dollar flow from buyer \(b\) to seller \(r\) is
\[
        f_{br}=w_{br}m_b
\]
Let \(\mathbf F=(f_{br})_{b,r}\) denote the corresponding firm-level flow matrix. Seller \(j\)'s total sales are
\[
        \mathrm{sales}_j=\sum_b f_{bj}
\]
The health vector is \(h\in[0,1]^N\). A firm with \(h_j=1\) operates at baseline. A firm with
\(h_j=0\) is shut down. The intermediate-input share is \(s=0.8\), as in the main text. It enters the supply channels as the weight of intermediate inputs in output. In the demand channel it acts as a damping coefficient, which we set equal to the intermediate share for parsimony.
The essentiality threshold \(\theta\), the per-edge essential probability \(p\), and the CES order \(\rho\) are local to this
appendix.

For each knocked-out firm \(i\), we compute
\[
        \mathrm{ESRI}_i
        =
        \frac{\sum_j m_j(1-h_j)}{\sum_j m_j},
        \qquad
        h_i\equiv0
\]
This is the size-weighted aggregate output loss after the cascade settles.

The cascade has two channels. A firm can be hurt because its suppliers are impaired. It can also be
hurt because its customers are impaired. We compute both channels and take the smaller health:
\[
        h_j
        \leftarrow
        \min\!\left(h_j^{\mathrm{sup}},h_j^{\mathrm{dem}}\right),
        \qquad
        h_i:=0,
        \qquad
        h_j\in[0,1]
\]
The demand channel is common to all variants:
\[
        h_j^{\mathrm{dem}}
        =
        1
        -
        s\left(
        1-
        \frac{\sum_b f_{bj}h_b}{\mathrm{sales}_j}
        \right),
        \qquad
        h_j^{\mathrm{dem}}=1
        \quad\text{if } \mathrm{sales}_j=0
\]
Thus a firm loses demand when its buyers are impaired. The effect is damped through \(s\).

The variants differ only in the supply channel. The linear reference case is
\[
        h_j^{\mathrm{sup}}
        =
        1
        -
        s\sum_r w_{jr}(1-h_r)
\]
The Leontief and CES variants replace this linear rule with different notions of input essentiality.

\medskip
\noindent\emph{Well-posedness and settling.}

For the permanent protocol, the knocked-out firm \(i\) is held at zero at every step. Let
\(\Phi^{(i)}\) be the resulting constrained update map: it applies the two-channel update to all firms
except \(i\), and sets \(h_i=0\). The map
\[
        \Phi^{(i)}:[0,1]^N\to[0,1]^N
\]
is order-preserving. The supply channels, the demand channel, the minimum operator, and the
clipping operation are all nondecreasing in \(h\). With the conventions for empty input sets and,
in the CES case, the \(\varepsilon\)-floor before negative powers of Appendix~\ref{app:ces-esri}, this finite-dimensional
map is also continuous on \([0,1]^N\).

By the Knaster--Tarski theorem, \(\Phi^{(i)}\) has a greatest fixed point. Starting
from the healthy economy, with firm \(i\) set to zero, the iteration
\[
        h^{(k+1)}=\Phi^{(i)}(h^{(k)})
\]
is monotonically nonincreasing. By continuity, its limit satisfies \(h^\infty=\Phi^{(i)}(h^\infty)\). It is the greatest fixed point: any fixed point \(h^{\star}\) lies below the top element \(h^{(0)}\) of the constrained lattice, so monotonicity gives \(h^{\star}=(\Phi^{(i)})^{k}(h^{\star})\le(\Phi^{(i)})^{k}(h^{(0)})=h^{(k)}\) for every \(k\), and hence \(h^{\star}\le h^{\infty}\). We stop when
\[
        \|h^{(k+1)}-h^{(k)}\|_\infty<10^{-6}
\]
with a cap of about \(200\) iterations. The cap is a numerical safeguard rather than a convergence guarantee; a run that reaches it before meeting the tolerance is treated as unconverged. The settled state gives the permanent loss
\[
        \mathrm{ESRI}_i
        =
        1-\frac{Y_\infty(i)}{Y_0}
\]
Both the CES and Leontief mechanisms use the same knock, the same iteration, and the same output
measure.

\subsection{Leontief ESRI: Perfect Complements}
\label{app:leontief-esri}

The Leontief variant asks what happens when some inputs are essential. An input from supplier \(r\) to buyer \(j\) is classified as essential if its spending share exceeds a threshold:
\[
        w_{jr}\ge\theta,
        \qquad
        \theta=0.05
\]
Firm \(j\)'s supply health is then the minimum health among its essential suppliers:
\[
        h_j^{\mathrm{sup}}
        =
        \min_{r:\,w_{jr}\ge\theta} h_r,
        \qquad
        h_j^{\mathrm{sup}}=1
        \quad\text{if \(j\) has no essential input.}
\]
This is the all-or-nothing mechanism. If one essential supplier is dead, the buyer is pulled down to
that supplier's health. The supply channel itself is undamped. Boundedness comes from the
demand channel and from the restriction \(h\in[0,1]\).

This mechanism is useful as an extreme case. It shows what the cascade looks like when inputs are
perfect complements and substitution margins are absent. In that case a single firm failure can
generate an avalanche that is limited only by the structure and size of the network.

A discrete-failure cousin of this mechanism uses binary states \(x_j\in\{0,1\}\), with failures
absorbing. Define the input-loss and output-loss exposures by
\[
        L_j^{\mathrm{in}}
        =
        \sum_r w_{jr}(1-x_r)
        =
        [\mathbf W(1-x)]_j
\]
and
\[
        L_j^{\mathrm{out}}
        =
        \frac{[\mathbf F^\top(1-x)]_j}{\mathrm{sales}_j}
\]
Firm \(j\) fails if either exposure exceeds the threshold:
\[
        x_j\to0
        \quad\text{if}\quad
        L_j^{\mathrm{in}}>\theta
        \ \text{or}\
        L_j^{\mathrm{out}}>\theta
\]
This is the Watts/bootstrap-percolation version of the cascade. The reported object in that case is
the avalanche size,
\[
        \mathrm{av}_i
        =
        \#\{\text{failed firms}\}
\]
and \(\theta\) can be swept to locate the critical threshold \(\theta_c\).

\subsection{CES ESRI: Finite Substitution}
\label{app:ces-esri}

The CES variant allows some inputs to be substitutable. Each edge receives a fixed draw
\[
        u_e\sim U(0,1)
\]
The draws are seeded and held fixed, so a sweep over \(p\) is nested. At substitutability level \(p\),
the input from supplier \(r\) to buyer \(j\) is essential if
\[
        u_e<p
\]
Otherwise it is substitutable. This partitions buyer \(j\)'s inputs into two sets,
\[
        \mathrm{Ess}(j)
        \quad\text{and}\quad
        \mathrm{Sub}(j)
\]
with essential-input weight
\[
        W_j^{\mathrm{ess}}
        =
        \sum_{r\in\mathrm{Ess}(j)}w_{jr}
\]

The supply channel combines linear aggregation over substitutable inputs with a CES aggregate over
essential inputs:
\[
h_j^{\mathrm{sup}}
=
(1-s)
+
s\left[
\sum_{r\in\mathrm{Sub}(j)}w_{jr}h_r
+
W_j^{\mathrm{ess}}
\left(
\frac{1}{W_j^{\mathrm{ess}}}
\sum_{r\in\mathrm{Ess}(j)}
w_{jr}h_r^\rho
\right)^{1/\rho}
\right]
\]
If \(W_j^{\mathrm{ess}}=0\), the essential-input term is set to zero.

The limiting cases are useful to keep in mind:
\[
\begin{cases}
p=0
&\Rightarrow\ \text{linear DRS ESRI},\\
\rho\to-\infty
&\Rightarrow\ \text{Leontief aggregation over essential inputs.}
\end{cases}
\]
The canonical CES run uses
\[
        p=0.01,
        \qquad
        \rho=-1
\]
so only a small share of edges are essential and those essential inputs are gross complements.

This CES supply channel is damped by the envelope \((1-s)+s[\cdot]\). It is therefore not identical
to the Leontief rule, even at \(p=1\) and \(\rho\to-\infty\). The two mechanisms differ by this
damping term. The Leontief rule is a pure minimum over essential suppliers. The CES rule is a
decreasing-returns version of a CES input aggregator.

There is one numerical detail. Since \(\rho<0\), the computation raises health values to negative
powers. We therefore replace
\[
        h_r
        \quad\text{by}\quad
        \max(h_r,\varepsilon)
\]
before exponentiation. For negative integer \(\rho\), we compute \(h_r^{-|\rho|}\) as
\(1/h_r^{|\rho|}\), using exponentiation by squaring. We also clamp the inner ratio to avoid
overflow.

\subsection{Dynamics of the CES Variant}
\label{app:ces-dynamics}

The CES variant is also used for a temporary-outage experiment. Its finite-horizon summaries are
reported separately from the settled ESRI defined above. Let \(\mathbf h^0\) denote the no-shock fixed
point. We set \(h_{j,0}=h_j^0\) for \(j\ne i\) and \(h_{i,0}=0\), and apply the CES map synchronously
for \(t=0,\ldots,T_H-1\), pinning firm \(i\) at zero through date \(T_{\mathrm{out}}-1\). At date
\(T_{\mathrm{out}}\), we set \(h_{i,T_{\mathrm{out}}}=1\) once; from the following update onward it
evolves under the ordinary CES map. The seeded essential-input partition is held fixed throughout.
The map is run for a fixed horizon \(T_H\), not to convergence. The default values are
\[
        T_{\mathrm{out}}=12,
        \qquad
        T_H=30
\]

Losses are measured relative to the no-shock settled baseline \(Y_0\). The baseline is obtained by
iterating the unshocked map to a fixed point, with \(Y_0=1\) when rows sum to one. The state
\(\mathbf h_t\), after any outage constraint or release at date \(t\), is used to compute \(Y_t\).
Thus,
\[
        Y_t
        =
        \frac{\sum_j m_j h_{j,t}}{\sum_j m_j},
        \qquad
        \ell_t
        =
        1-\frac{Y_t}{Y_0}
\]
For each cascade we report three summaries. The first is peak output loss:
\[
        \mathrm{peakloss}_i
        =
        \max_{0\le t\le T_H} \ell_t
\]
We also report the integrated loss,
\[
        \mathrm{cumloss}_i
        =
        \sum_{t=0}^{T_H} \ell_t
\]
and the post-release recovery half-life \(T_{1/2,i}\), defined as the smallest nonnegative integer such that
\(\ell_{T_{\mathrm{out}}+T_{1/2,i}}\le \ell_{T_{\mathrm{out}}}/2\). If this does not occur by \(T_H\),
the half-life is right-censored at \(T_H-T_{\mathrm{out}}\).

\subsection{Estimation}
\label{app:esri-estimation}

Computing exact ESRI for every firm requires one cascade per firm. This is \(O(N^2)\) at national
scale and is not feasible for the full network. We therefore estimate the ESRI distribution from a
stratified knockout sample.

The strata are log-spaced by the proxy equilibrium firm size, namely the stationary (Perron) vector
\(\mathbf v\), where
\[
        \mathbf v=\mathbf W^\top\mathbf v
\]
The top \(0.1\%\) of firms are sampled exhaustively. The remaining non-empty strata are sampled
using a two-phase Neyman allocation. Let \(\mathcal H_{\mathrm{nonex}}\) denote these non-exhaustive strata, let \(n_{\mathrm{ESRI}}\) be their final-sample budget, and let \(n_{\min}\) be the stratum floor, with
\[
        \sum_{h\in\mathcal H_{\mathrm{nonex}}}\min\{N_h,n_{\min}\}\le n_{\mathrm{ESRI}}\le\sum_{h\in\mathcal H_{\mathrm{nonex}}}N_h.
\]
A separate pilot pass draws \(\min\{N_h,n_{\min}\}\) firms from each
non-empty stratum and estimates \(\widehat{\mathrm{sd}}_h\); these pilot draws are not used in the
plotted estimates. The final allocation first assigns the floor and then distributes the remaining
budget in proportion to
\[
        N_h\,\widehat{\mathrm{sd}}_h,
\]
subject to the cap \(n_h\le N_h\), redistribution of any capped allocation among the remaining strata, and largest-remainder integer rounding. If all remaining values $N_h\widehat{\mathrm{sd}}_h$ are zero, the residual budget is allocated in proportion to $N_h$. Firms are then drawn uniformly without replacement
within each stratum. Empty strata are ignored. All plotted counter-cumulative distributions are
Horvitz--Thompson weighted, with weights
\[
        \omega_i=\frac{N_h}{n_h}
\]
for non-exhaustive strata and \(\omega_i=1\) in the exhaustive top stratum. Uncertainty is computed
by a within-stratum bootstrap, resampling within each non-exhaustive stratum and holding
exhaustive strata fixed.

For validation, both cascade variants are compared with the brute-force \(O(N^2)\) reference on
small networks. The maximum absolute error is below \(10^{-5}\).

\section{Firm-Level Systemic Risk: Supplementary Figures}
\label{app:esri}

\begin{figure}[H]
  \centering
  \includegraphics[width=\linewidth]{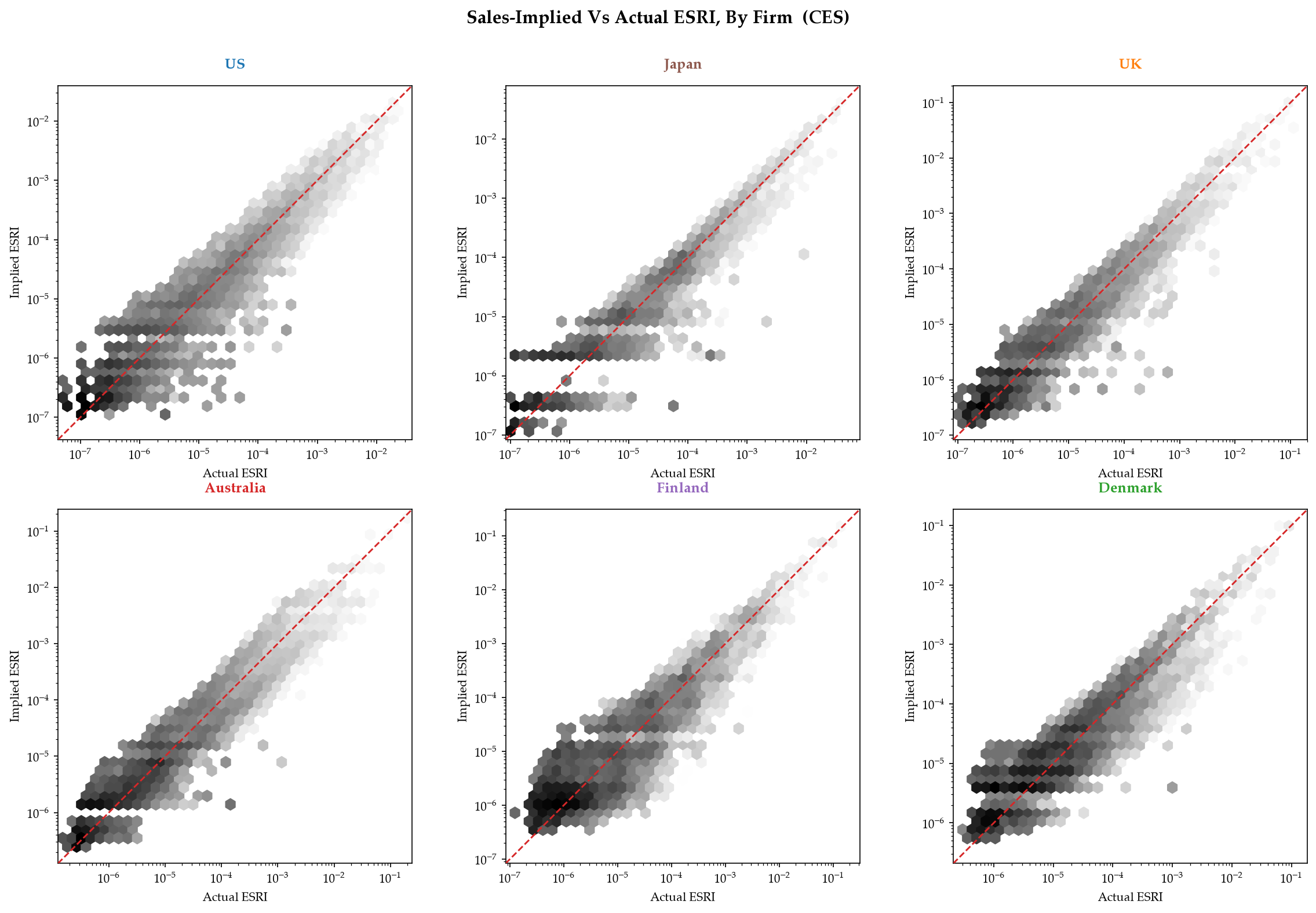}
  \caption{Actual firm-level \(\mathrm{ESRI}\) against sales-implied \(\mathrm{ESRI}\) under CES, one
  panel per economy. Hexagons are shaded by summed Horvitz--Thompson firm weight. The dashed
  line is the identity. In every economy the cloud follows the diagonal over several decades. With
  substitution, firm sales recovers much of the ordering of systemic risk.}
  \label{fig:esri_sales_scatter_ces}
\end{figure}

\begin{figure}[H]
  \centering
  \includegraphics[width=\linewidth]{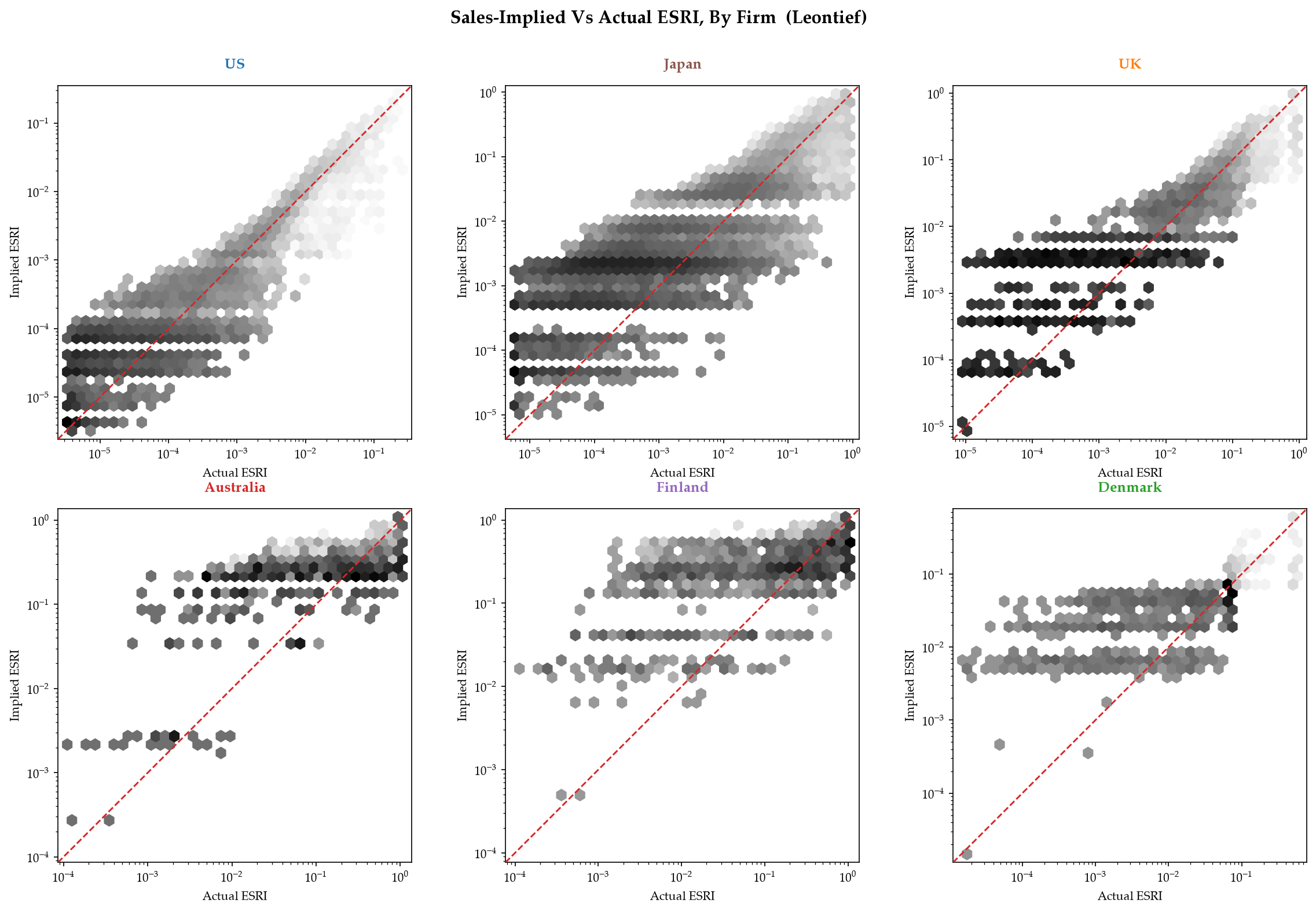}
  \caption{Actual firm-level \(\mathrm{ESRI}\) against sales-implied \(\mathrm{ESRI}\) under Leontief,
  one panel per economy. The fit deteriorates in every economy. The panels show horizontal bands
  and systematic over-prediction above the identity line, especially in the smaller economies. With
  no substitution, firm sales no longer pins down firm-level systemic risk. The Leontief samples are
  smaller for the small economies, which makes those panels sparser.}
  \label{fig:esri_sales_scatter_leontief}
\end{figure}

\begin{figure}[H]
  \centering
  \includegraphics[width=\linewidth]{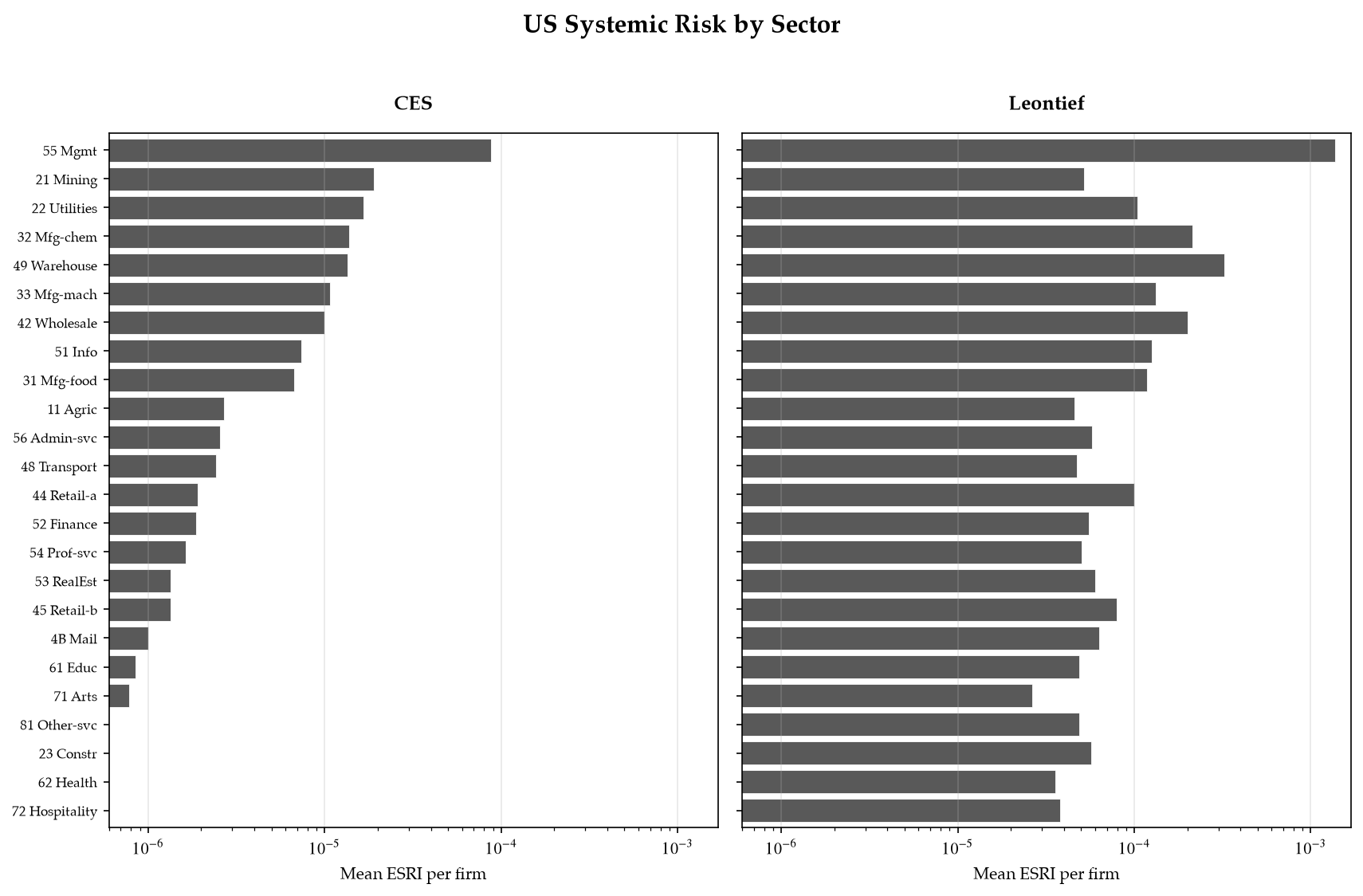}
  \caption{US sectors ranked by mean firm-level systemic risk, under CES and Leontief.
  Management of companies, mining, utilities, wholesale, warehousing, and chemical and machinery
  manufacturing are among the most systemic sectors. Arts, accommodation, and health are among
  the least systemic. The broad ranking is similar across the two production functions.}
  \label{fig:esri_sectors}
\end{figure}

\begin{figure}[H]
  \centering
  \includegraphics[width=.6\linewidth]{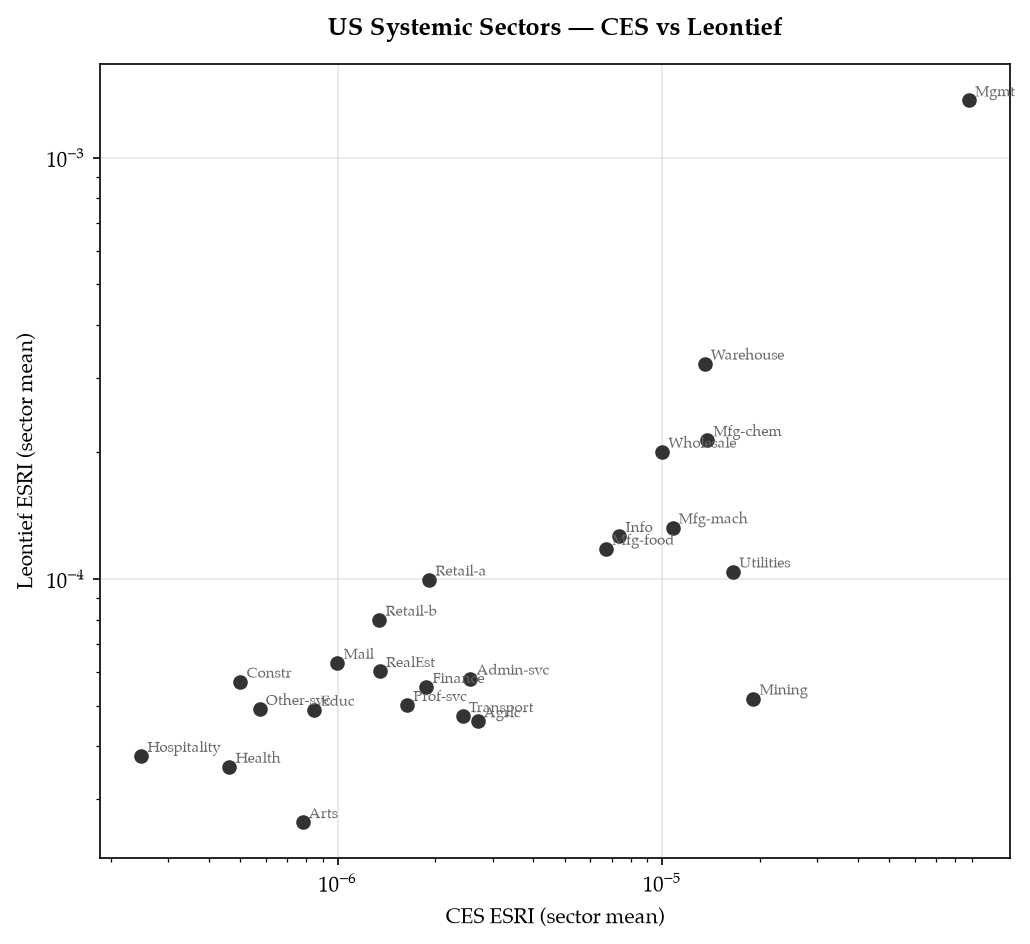}
  \caption{US sector-mean systemic risk under CES against Leontief, one point per NAICS sector.
  The sectors lie close to an upward diagonal, with Spearman correlation approximately one. The
  same upstream hubs are systemic under both mechanisms. The off-diagonal sectors show the
  channel of transmission. Warehousing and logistics move up under Leontief because they are
  bottlenecks. Utilities and mining sit lower because substitution under CES attenuates their
  impact. Log axes.}
  \label{fig:esri_sectors_scatter}
\end{figure}

\begin{figure}[H]
  \centering
  \includegraphics[width=\linewidth]{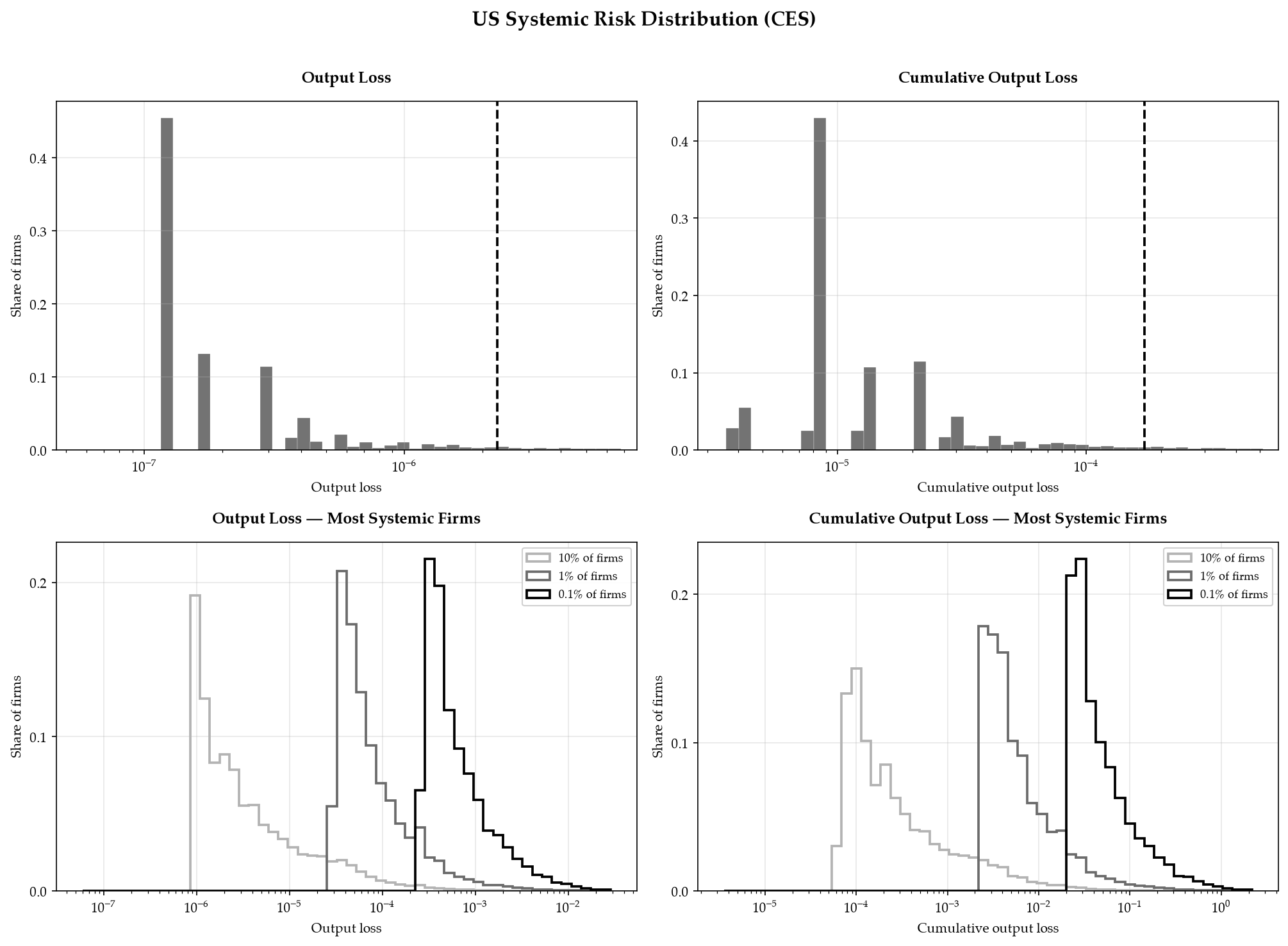}
  \caption{Distribution of per-firm impact in the United States under the temporary CES outage.
  The columns compare peak output loss with cumulative output loss over the outage horizon. The top
  row shows all firms, with the size-weighted mean marked. The bottom
  row shows the most systemic \(10\%\), \(1\%\), and \(0.1\%\) of firms, ranked by their own impact.}
  \label{fig:esri_dist}
\end{figure}

\begin{figure}[H]
  \centering
  \includegraphics[width=\linewidth]{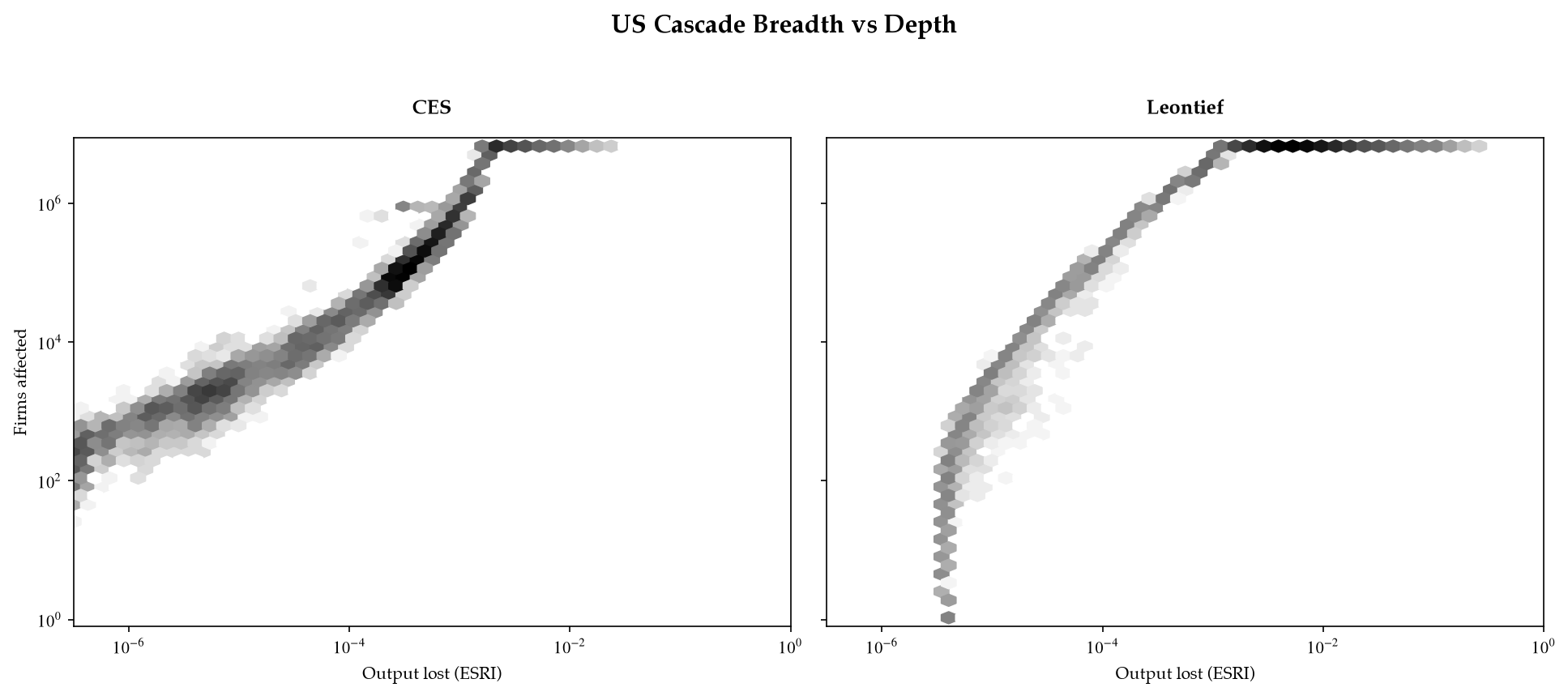}
  \caption{Cascade breadth and depth in the United States. For each knocked-out firm, the vertical
  axis gives the number of firms affected and the horizontal axis gives the output loss,
  \(\mathrm{ESRI}_i\). The left panel is CES. The right panel is Leontief. Systemic cost rises with the
  breadth of propagation and then saturates as the largest cascades reach almost the whole economy.
  Hexbin density, log axes.}
  \label{fig:esri_breadth}
\end{figure}

\begin{figure}[H]
  \centering
  \includegraphics[width=\linewidth]{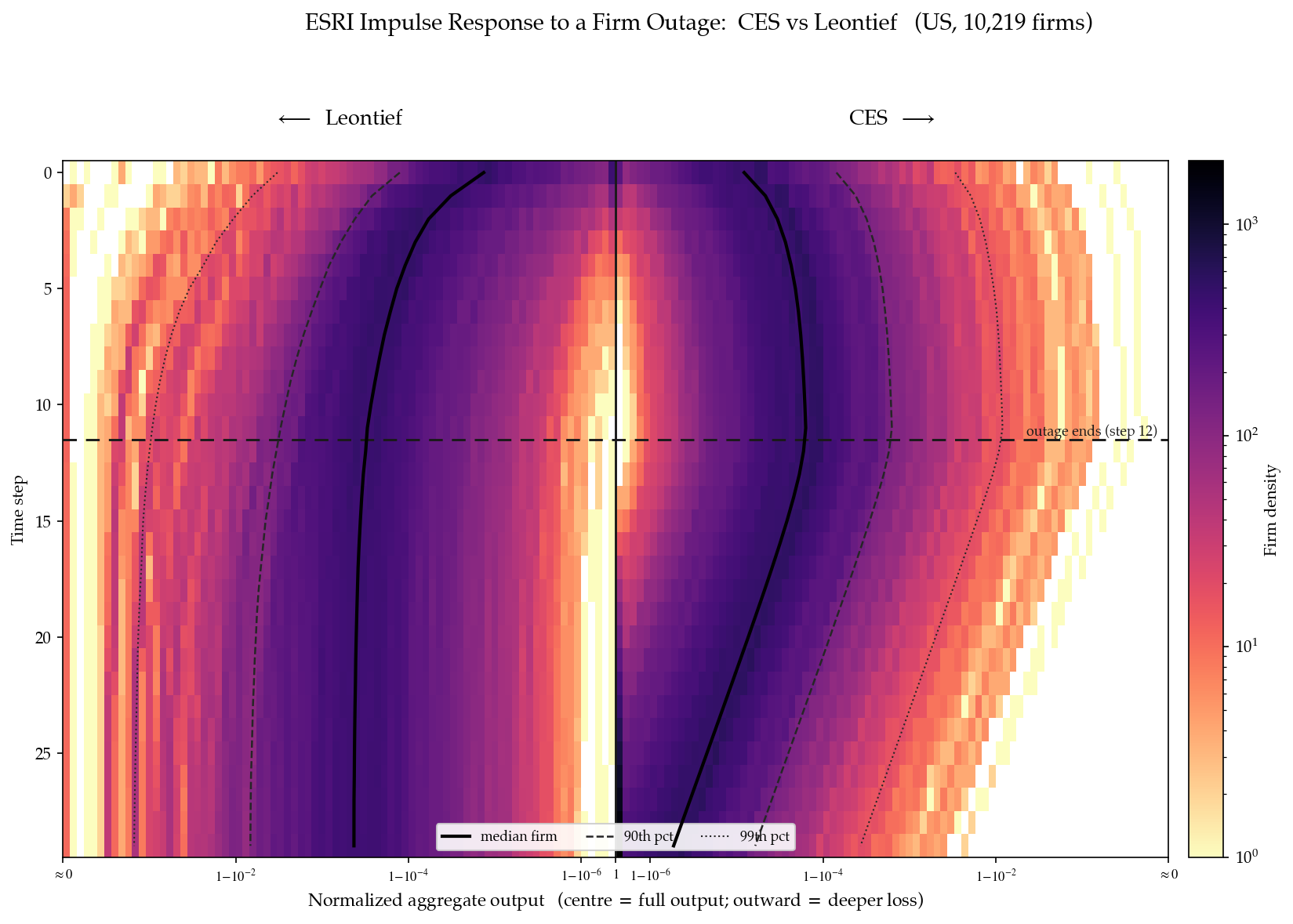}
  \caption{Aggregate-output impulse response to a firm outage, CES versus Leontief, on a US
  \(10^4\) reconstruction. A single firm is forced to zero for a \(12\)-step outage and then restored
  over a horizon of \(30\) steps, with \(s=0.8\). Time runs downward, and normalized aggregate
  output is mirrored about the center. CES fans to the right and Leontief to the left. Color gives
  log firm density. The weighted median, \(90\)th percentile, and \(99\)th percentile are overlaid.
  The demand channel and outage protocol are identical. Only input substitution differs. Under CES,
  the median loss peaks at the outage end and then largely recovers. Under Leontief, the median
  response keeps moving outward and does not return. The upper tail reaches near-total collapse.
  This is the dynamic counterpart to the static CES--Leontief comparison in Section~\ref{sec:esri}.}
  \label{fig:esri_ces_leontief_irf}
\end{figure}

This appendix collects the supplementary figures for the systemic-risk exercise in
Section~\ref{sec:esri}. The first group, Figures~\ref{fig:esri_sales_scatter_ces}
and~\ref{fig:esri_sales_scatter_leontief} with the distances in Table~\ref{tab:esri_sales_country},
extends the sales-based predictor comparison from the United States to all six economies. Because
a firm's sales are its size in the closed economy, the United States panels correspond to the size
scatter in Figure~\ref{fig:esri_vs_size}. The second group, Figures~\ref{fig:esri_sectors}
and~\ref{fig:esri_sectors_scatter}, ranks sectors by average firm-level
systemic risk. The third group, Figures~\ref{fig:esri_dist}--\ref{fig:esri_ces_leontief_irf}, looks inside a single failure: how many firms it reaches, how much
output it destroys, and how the aggregate response differs under CES and Leontief.

\begin{table}[H]\centering
\caption{Sales predicts systemic risk less well when substitution is removed. The table reports
Kuiper and total-variation distances between the actual \(\mathrm{ESRI}\) distribution and the
distribution implied by firm sales, using the best monotone map from sales to \(\mathrm{ESRI}\).
All quantities are Horvitz--Thompson weighted. Lower values mean a closer fit. Total variation is
high throughout because \(\mathrm{ESRI}\) is heavy-tailed and the isotonic implied values are discrete,
so the informative comparison is CES versus Leontief within each economy.}
\label{tab:esri_sales_country}
\begin{tabular}{@{}lcccc@{}}
\toprule
 & \multicolumn{2}{c}{CES (\(\rho=-1\))} & \multicolumn{2}{c}{Leontief} \\
\cmidrule(lr){2-3}\cmidrule(lr){4-5}
Economy & Kuiper & Total variation & Kuiper & Total variation \\
\midrule
United States  & \(0.46\) & \(0.58\) & \(0.47\) & \(0.83\) \\
Japan          & \(0.46\) & \(0.91\) & \(0.48\) & \(0.92\) \\
United Kingdom & \(0.24\) & \(0.91\) & \(0.48\) & \(0.94\) \\
Australia      & \(0.27\) & \(0.78\) & \(0.53\) & \(0.92\) \\
Finland        & \(0.37\) & \(0.84\) & \(0.47\) & \(0.89\) \\
Denmark        & \(0.24\) & \(0.79\) & \(0.66\) & \(0.83\) \\
\bottomrule
\end{tabular}
\end{table}

\section{Firms with Multiple Production Units}
\label{app:factory}

Some shocks are not firm-level shocks. They occur in space. A flood hits a location. A heatwave
hits a region. An earthquake shuts down plants within a radius of its epicenter. For these questions,
a firm-to-firm network is still too coarse. A large firm may operate many factories in different
places, and the relevant unit of exposure is the production site rather than the legal firm. A
geographically resolved production network must therefore say not only which firms trade with one
another, but also where the production units involved in those relationships are located.

This appendix describes how the reconstructed firm network can be extended to factories. The
extension is a post-processing step. It takes the weighted firm-to-firm network of
Section~\ref{sec:model} as fixed and allocates each firm-level edge across the factories owned by the
two firms. The firm-level network is not re-estimated. The firm-pair adjacency, the aggregate weight between each linked firm pair,
the firm sizes, and the sectoral totals are all preserved after aggregation. What is added
is geography. We have not yet implemented this extension empirically. It is included as a design
that can be applied directly once factory-location data are available.

Suppose firm \(i\) owns a finite set of factories
\[
        \mathcal A(i)=\{a_{i,1},\ldots,a_{i,k_i}\},
        \qquad
        k_i\ge1
\]
Each factory \(a\in\mathcal A(i)\) has latitude and longitude coordinates
\[
        (\mathrm{lat}_a,\mathrm{lon}_a)
\]
measured in radians. The starting point is the reconstructed firm-level weighted network. A
firm-level edge \(i\to j\) says that firm \(i\) buys from firm \(j\), with its weight already determined
by the reconstruction. The factory extension decides which factory of \(i\) buys from which factory
of \(j\). This allocation is required to satisfy two conditions. First, when factory edges are summed
back to firms, the original firm-level network must be recovered exactly. Second, within a linked
firm pair, the allocation should be geographically plausible: nearby factories should be more likely
to trade than distant factories, and factories located in dense production regions should carry more
links.

\subsection{Distance-Based Factory Weights}

The basic geographic primitive is the distance between two factories. We measure this distance
using the great-circle formula
\[
d(a,b)
=
2R_E
\arcsin
\sqrt{
\sin^2\!\left(\frac{\mathrm{lat}_a-\mathrm{lat}_b}{2}\right)
+
\cos(\mathrm{lat}_a)\cos(\mathrm{lat}_b)
\sin^2\!\left(\frac{\mathrm{lon}_a-\mathrm{lon}_b}{2}\right)
}
\]
where \(R_E\simeq6371\) km is the Earth's radius. Distance is then translated into an assignment
weight through the exponential kernel
\[
        G(d;\tau_d)=\exp(-d/\tau_d),
        \qquad
        \tau_d>0
\]
The parameter \(\tau_d\) governs the spatial scale of the allocation. When \(\tau_d\) is small, the
kernel gives much greater weight to nearby factory pairs, and the resulting factory network is local.
When \(\tau_d\) is large, distance matters less, and firm-level links are spread more broadly across
space.

Collect all factories into a single index set and define a factory-to-factory matrix \(\mathcal Q\).
Entry \(\mathcal Q_{ab}\) is an unnormalized assignment weight for sending a link from source
factory \(a\) to destination factory \(b\). It is positive only if the firms that own \(a\) and \(b\) are
linked in the firm-level network:
\[
\mathcal Q_{ab}
=
\begin{cases}
G(d(a,b);\tau_d),
& \text{if \(a\in\mathcal A(i)\), \(b\in\mathcal A(j)\), \(i\neq j\), and \(i\to j\) exists,}\\[0.3em]
0,
& \text{otherwise.}
\end{cases}
\]
Factories belonging to the same firm receive zero weight, and so do factories belonging to firms
with no firm-level edge between them. Each positive row of \(\mathcal Q\) is normalized to sum to
one. Thus, conditional on a source factory, the row of \(\mathcal Q\) gives a probability distribution
over feasible destination factories.

The matrix \(\mathcal Q\) is only an assignment-kernel matrix. It is not used to create additional
firm-level capacity. Firm-level weight is preserved edge by edge. For every linked firm pair \(i\to j\), the
allocation of Appendix~\ref{app:link-allocation} records a finite multiset \(\mathcal R_{ij}\) of elements of
\(\mathcal A(i)\times\mathcal A(j)\). Write
\[
        M_{ij}:=|\mathcal R_{ij}|\ge1
\]
for the number of factory-pair representatives. After all representatives have been recorded, factory weights are
defined by
\[
        \widetilde w_{ab}
        =
        \sum_{i\to j}
        \frac{w_{ij}}{M_{ij}}\,
        \#\{(a,b)\in\mathcal R_{ij}\},
\]
where the count is with multiplicity. This representative rule, rather than a dynamic capacity update
in \(\mathcal Q\), is what prevents the factory extension from creating more trade between two firms
than the original firm-level network contains. Without factory output shares, \(\widetilde{\mathbf W}\)
is not imposed to be row-stochastic factory by factory; this is a disaggregation of firm-level flows,
not a separate factory-level money chain.

\subsection{Factory Prominence within a Firm}

Distance governs which destination factory is likely to be chosen once a source factory has been
selected. We also need a rule for which factories within a firm carry that firm's outgoing links. A
factory located near many other factories should play a larger role than an isolated plant, because it
sits in a denser production geography. We capture this by assigning each factory a within-firm
prominence weight.

For factory \(a\in\mathcal A(i)\), define
\[
        L_i(a)
        =
        \sum_{j\neq i}
        \sum_{b\in\mathcal A(j)}
        G(d(a,b);\tau_d)
\]
This score is large when factory \(a\) is close to many factories owned by other firms. Normalizing
within firm \(i\) gives
\[
        \psi_i(a)
        =
        \frac{L_i(a)}
        {\sum_{a'\in\mathcal A(i)}L_i(a')}
\]
The vector \(\psi_i\) is a probability distribution over the factories of firm \(i\). It is held fixed
throughout the allocation. It does not change firm-pair adjacency or aggregate weights after
aggregation. It only determines which factories of the firm are more likely to carry the corresponding
factory edges.

\subsection{Allocating Firm-Level Links across Factories}\label{app:link-allocation}

The allocation proceeds by first ensuring that factories are represented and then assigning the
remaining firm-level links according to prominence and distance. For each firm-level edge \(i\to j\), the procedure records representatives in the multiset \(\mathcal R_{ij}\). An assignment to \(i\to j\) always chooses a source factory \(a\in\mathcal A(i)\) and a destination factory \(b\in\mathcal A(j)\); conditional on \(a\), the destination is drawn from the row \(\mathcal Q_{a\cdot}\) restricted to \(\mathcal A(j)\) and renormalized. Final weights are assigned only after all representatives have been recorded.
For a source factory \(a\in\mathcal A(i)\) and an outgoing firm-level link \(i\to j\), write
\[
        H_{aj}:=\sum_{b\in\mathcal A(j)}G(d(a,b);\tau_d).
\]
Whenever the allocation must attach a source factory \(a\) to one of a set of eligible outgoing links
of firm \(i\), it chooses link \(i\to j\) with probabilities proportional to \(H_{aj}\), using uniform
tie-breaking if all eligible \(H_{aj}\)'s are zero. Conditional on the chosen link, the destination
factory is drawn from the restricted and renormalized row of \(\mathcal Q\) as above. Consider a firm \(i\) with
\(k_i\) factories and \(d_i^{\mathrm O}\) outgoing firm-level links. If \(d_i^{\mathrm O}=0\), there is no
outgoing firm-level link to allocate, and the outgoing-side pass skips firm \(i\). If
\[
        d_i^{\mathrm O}\ge k_i
\]
there are enough outgoing links to give every factory at least one outgoing factory edge. We pass
once through the factories \(a\in\mathcal A(i)\). For each factory \(a\), one not-yet-used outgoing
link \(i\to j\) is chosen by the \(H_{aj}\)-weighted rule above from the remaining outgoing list. A
destination factory \(b\in\mathcal A(j)\) is then drawn from the restricted and renormalized row of
\(\mathcal Q\), and the representative \((a,b)\) is recorded in \(\mathcal R_{ij}\).

If instead
\[
        d_i^{\mathrm O}<k_i
\]
there are more factories than outgoing firm-level links. In that case we still allow every factory to
be active, but some firm-level links must be represented by more than one factory edge. The first pass
constructs a surjection from the factories of firm \(i\) onto its outgoing firm-level links. While any
outgoing link remains unrepresented, the choice is restricted to those links; once every outgoing link
has received one representative, remaining factories are assigned as duplicate representatives with all
outgoing links eligible. Each link choice uses the \(H_{aj}\)-weighted rule above, and, conditional
on that link, the destination is drawn from \(\mathcal Q\) restricted to the relevant destination firm
and renormalized. Each assignment records one element of the corresponding \(\mathcal R_{ij}\). Thus a firm with
many production sites is not forced to leave some plants disconnected merely because its firm-level
out-degree is small. At the same time, the total weight from firm \(i\) to the destination firm remains
exactly the original firm-level weight.

After this initial pass, every factory of every firm has at least one outgoing factory edge, unless the
firm itself has no outgoing firm-level link. This is an outgoing-side representation guarantee:
incoming factory activity is induced by the allocations made by customer firms and is not separately
forced. Let \(r_i\) denote the number of outgoing firm-level links
of firm \(i\) that remain without representatives after this pass. These residual links are allocated using the
prominence vector \(\psi_i\) and the distance matrix \(\mathcal Q\). For each remaining outgoing
firm-level link \(i\to j\), a source factory \(a\in\mathcal A(i)\) is first drawn according to
\(\psi_i\). Conditional on this source, a destination factory \(b\in\mathcal A(j)\) is then drawn from
the row \(\mathcal Q_{a\cdot}\) restricted to \(\mathcal A(j)\) and renormalized. The representative
\((a,b)\) is recorded in \(\mathcal R_{ij}\).

The allocation therefore has two effects. It gives each production unit a role in the extended
network, and it assigns additional links to the production units that are geographically central. At
every stage, however, the allocation is constrained by the original firm-to-firm graph. Distance and
prominence determine how a firm-level relationship is distributed across factories. They do not
create new firm-level relationships.

\subsection{What Is Preserved and What Is Added}

The central property of the extension is aggregation consistency. If the factory network is summed
back to firms, the original firm-level network is recovered. No factory edge can appear between
firms that are not linked at the firm level, because \(\mathcal Q_{ab}=0\) for such pairs from the
start. No linked firm pair can receive more total weight than the original firm-level edge allows,
because the \(M_{ij}\) representatives of \(i\to j\) each receive weight \(w_{ij}/M_{ij}\). Hence
\[
        \sum_{a\in\mathcal A(i)}
        \sum_{b\in\mathcal A(j)}
        \widetilde w_{ab}
        =
        w_{ij}
\]
for every linked firm pair \(i\to j\). A firm-level edge may be represented by several factory edges,
but its total weight between the two firms is unchanged.

Sectoral totals are preserved for the same reason. Each factory inherits the sector label of its firm.
Since the firm-level weights are unchanged after aggregation, sector-to-sector flows are unchanged
as well. The extension therefore leaves all firm-level and sector-level objects from the original
reconstruction intact: firm degrees after aggregation, firm-pair weights, firm sizes, and sectoral
flows.

What changes is the geography inside a firm pair. If firm \(i\) buys from firm \(j\), the extension
decides which factories of \(i\) buy from which factories of \(j\). That decision is governed by
distance and prominence. Nearby factory pairs are more likely to be connected, and factories in
dense production regions carry more of the links. The parameter \(\tau_d\) controls how local the
resulting factory network is. If the factory network is too local, \(\tau_d\) can be increased. If it is
too diffuse, \(\tau_d\) can be decreased.

This is why the extension is formulated as a post-processing step rather than as a new
reconstruction. Re-running the entire reconstruction at the factory level would create two problems.
First, it would tend to generate links between factories owned by the same firm, because factories
of the same firm may look similar in sector, size, and location. Those intra-firm links are not the
object of the exercise. Second, most countries publish firm-size distributions by sector, but not
factory-size distributions by sector. A from-scratch factory reconstruction would therefore have to
assume how each firm's size is split across its factories. The procedure here avoids both problems.
It keeps the firm-level reconstruction fixed and allocates each firm-level edge across factories by a
transparent, distance-based rule.

\end{document}